\pdfoutput=1
\documentclass[11pt,twoside,a4paper,cmspaper,final,collab]{cms-tdr}
\def\svnVersion{1cf8373}\def\svnDate{2023/07/11}\def\cmsCernNoTag{CERN-EP-2022-005}\def\cmsCernDate{\today}\def\cmsMessage{Published in the European Physical Journal C as \href{http://dx.doi.org/10.1140/epjc/s10052-023-11630-8}{\doi{10.1140/epjc/s10052-023-11630-8}.}}
\begin{document}\cmsNoteHeader{GEN-17-002}

\newlength\cmsFigWidth
\ifthenelse{\boolean{cms@external}}{\setlength\cmsFigWidth{0.85\columnwidth}}{\setlength\cmsFigWidth{0.4\textwidth}}
\ifthenelse{\boolean{cms@external}}{\providecommand{\cmsTable}[1]{#1}}{\providecommand{\cmsTable}[1]{\resizebox{\textwidth}{!}{#1}}}
\ifthenelse{\boolean{cms@external}}{\providecommand{\cmsLeft}{upper\xspace}}{\providecommand{\cmsLeft}{left\xspace}}
\ifthenelse{\boolean{cms@external}}{\providecommand{\cmsRight}{lower\xspace}}{\providecommand{\cmsRight}{right\xspace}}
\newlength\cmsTabSkip\setlength{\cmsTabSkip}{2ex}
\newcommand{\MG} {\textsc{mg5}\_a\textsc{mc}}
\newcommand{\eff} {$\sigma_{\text{eff}}$}
\newcommand*{\tmax}{transMAX\xspace}
\newcommand*{\tmin}{transMIN\xspace}
\newcommand*{\tdiff}{transDIFF\xspace}
\newcommand*{\etaphi}{$\eta$--$\phi$}
\newcommand*{\delphi}{\Delta\phi}
\newcommand{\FxFx}{FxFx\xspace}
\newcommand{\ktMLM}{\kt--MLM\xspace}
\newcommand{\RIVET}{\textsc{Rivet}\xspace}
\newcommand{\PROFESSOR}{\textsc{Professor}\xspace}
\newcommand{\PYTHIAviii}{{\PYTHIA~8}\xspace}
\newcommand{\ptsum} {\ensuremath{\pt^{\text{sum}}}\xspace}
\newcommand{\ptmax} {\ensuremath{\pt^{\text{max}}}\xspace}
\newcommand{\ptjet} {\ensuremath{\pt^{\text{jet}}}\xspace}
\newcommand{\Nch} {\ensuremath{N_{\mathrm{ch}}}\xspace}
\newcommand{\dNdeta} {\ensuremath{\rd\Nch/\rd\eta}\xspace}
\newcommand{\dEdeta} {\ensuremath{\rd E/\rd\eta}\xspace}
\newcommand{\topmass}{\ensuremath{m_{\PQt}}\xspace}
\newcommand{\Wmass}{\ensuremath{m_{\PW}}\xspace}
\newcommand{\ndof} {\ensuremath{N_\mathrm{dof}}\xspace}
\newcommand{\ptzero}{\ensuremath{p_\mathrm{T0}}\xspace}
\newcommand{\ptrec}{\ensuremath{p_{\mathrm{T}_{\mathrm{Rec}}}}\xspace}

\cmsNoteHeader{GEN-17-002} 
\title{CMS \texorpdfstring{\PYTHIAviii}{PYTHIA 8} colour reconnection tunes based on underlying-event data}

\date{\today}

\abstract{New sets of parameter tunes for two of the colour reconnection models, quantum chromodynamics-inspired and gluon-move, implemented in the  \PYTHIAviii event generator, are obtained based on the default  CMS \PYTHIAviii underlying-event tune, CP5. Measurements sensitive to the underlying event performed by the CMS experiment at centre-of-mass energies $\sqrt{s}=7$ and 13\TeV, and by the CDF experiment at 1.96\TeV are used to constrain the parameters of colour reconnection models and multiple-parton interactions simultaneously. The new colour reconnection tunes are compared with various measurements  at 1.96, 7, 8, and 13\TeV including measurements of the underlying-event, strange-particle multiplicities, jet substructure observables, jet shapes, and colour flow in top quark pair (\ttbar) events. The new tunes are also used to estimate the uncertainty related to colour reconnection modelling in the top quark mass measurement using  the decay products of \ttbar events in the semileptonic channel at 13\TeV.}

\hypersetup{
pdfauthor={CMS Collaboration},
pdftitle={CMS PYTHIA 8 colour reconnection tunes based on underlying-event data},
pdfsubject={CMS},
pdfkeywords={CMS, software, computing, tuning}}

\maketitle

\section{Introduction}
Monte Carlo (MC) event generators, such as \PYTHIAviii~\cite{pythia8}, are 
indispensable tools for measurements at the LHC proton-proton (pp) collider. 
To provide an accurate description of high-energy collisions, both the hard scattering and the so-called underlying 
event (UE) are computed for each simulated event. In the hard scattering process, two initial partons interact with a large 
exchange of transverse momentum, $\pt > \mathcal{O}(\GeVns{})$ (we use natural units with $c=1$ throughout the paper). 
The UE represents additional activity 
occurring at lower energy scales that accompany the hard scattering.
It consists of multiple-parton interactions (MPIs), 
initial- and final-state radiation (ISR and FSR), and beam-beam remnants (BBR).
According to Quantum Chromodynamics (QCD), strong interactions are affected by colour charges that are carried by quarks and gluons.
All of the coloured partons produced by these components are finally combined to form colourless hadrons
through the hadronisation process.

Particularly relevant for the characterisation of the UE are the MPI, which consist of 
additional 2-to-2 parton-parton interactions occurring within the single collision event.
With increasing collision energy, the interaction probability for partons with small longitudinal 
momentum fractions also increases, which enhances MPI contributions.

The \PYTHIAviii generator regularises the cross sections of the primary hard scattering processes and MPIs with respect
to the perturbative 2-to-2 parton-parton differential cross section through an energy-dependent dampening parameter \ptzero, which
depends on the centre-of-mass energy $\sqrt{s}$.
The energy dependence of the \ptzero parameter in \PYTHIAviii is described with a 
power law function of the form
\begin{linenomath}
\begin{equation}\label{eqmpi}
\ptzero(\sqrt{s})= \ptzero^\text{ref} \left(\frac{\sqrt{s}}{\sqrt{s_0}}\right)^\epsilon,
\end{equation}
\end{linenomath}
where $\ptzero^{\text{ref}}$ is the value of \ptzero at a reference energy $\sqrt{s_0}$, and 
$\epsilon$ is a tunable parameter that determines the energy dependence. 
At a given $\sqrt{s}$, the mean number of additional interactions from MPI depends on \ptzero,
the parton distribution functions (PDFs), and the overlap of the matter
distributions of the two colliding hadrons~\cite{Corke:2011yy}.

\begin{figure*}[!hbtp]
 \centering
 \includegraphics[width=0.98\textwidth]{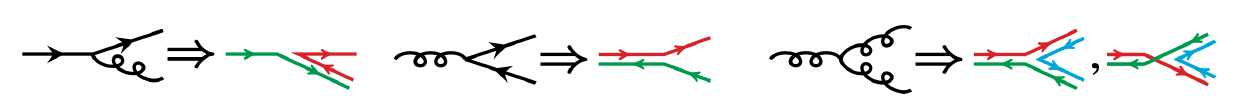}
 \caption{Rules for colour flow for quark-gluon vertices. Figure is taken from Ref.~\cite{Zyla:2020zbs}. 
Quark-gluon vertices are shown in black with Feynman diagrams and 
colour connection lines are shown with coloured lines.}
 \label{fig:lines_and_arrows}
 \end{figure*}

To track the colour information during the development of the parton shower, 
partons are represented and also connected by colour lines. Quarks and antiquarks are represented 
by colour lines with arrows pointing in the direction of the colour flow, and gluons are represented 
by a pair of colour lines with opposite arrows. 
Rules for colour propagation are shown in Fig.~\ref{fig:lines_and_arrows}.
Because each MPI system adds coloured partons to the final state, 
a dense net of colour lines that overlap with the coloured parton fields of the hard scattering 
and with each other is created. 
Parton shower algorithms, in general, use the leading colour (LC) approximation~\cite{Buckley:2011ms,Gustafson:1986db}
in which each successively emitted parton is colour connected only to its parent emitters in the limit of infinite number of colours.
Colour reconnection (CR) models allow colour lines to be 
formed between partons also from different interactions and thus allow different colour topologies 
compared with a simple LC approach.

The CR was first included in minimum-bias (MB) simulations (see Sec.~\ref{sec:UE_MB}) to reproduce the increase
of average transverse momentum $\langle\pt\rangle$ of charged particles
as a function of the measured multiplicity of the charged particles, \Nch, and also to
describe the \dNdeta distribution~\cite{Sjostrand:1987su,Sjostrand:2013cya}.
The pseudorapidity is defined as $\eta=-\ln[\tan(\theta/2)]$, 
where the polar angle $\theta$ is defined with respect to the anticlockwise-beam direction.
Introducing correlations between partons, including those also resulting from MPIs, generally changes the number of charged particles in an event and allows a more realistic simulation of \Nch, and $\langle\pt\rangle$ vs \Nch distributions than in an event scenario without CR~\cite{Sjostrand:2013cya}.

The CR effects are also important for processes occurring at larger scales in pp collisions.
For example, in \ttbar events, the inclusion of CR effects can lead to a significant improvement in
the description of UE variables~\cite{Sirunyan:2018avv}.
The effects of CR may become more prominent in precision measurements, such as the top quark mass
\topmass.
Uncertainties in \topmass related to CR are usually estimated from comparing
the prediction of a given model with and without CR, which might underestimate their
effect~\cite{Argyropoulos:2014zoa}.
A better way to approach the uncertainty estimation would be to consider a
variety of CR models and variations of their parameters~\cite{Sirunyan:2018gqx} that probe
the effects of the underlying soft physics of pp collisions
on the relevant observable.

Various phenomenological models for CR have been developed and are included in \PYTHIAviii. In these models, the general idea is to determine the partonic configuration that reproduces the minimal total string length. In the Lund string fragmentation model~\cite{Andersson:1983ia} used in \PYTHIAviii, the confining colour field between two partons is approximated by a one-dimensional string stretched between the partons according to the colour flow. The fragmentation of a string with a probability given by the fragmentation function produces a set of hadrons. Thus, the colour flow of an event determines the string configuration and therefore hadronic production. 

None of the MPI processes or the CR models are completely determined from first
principles, and they all include free parameters.
A specified set of such parameters that is adjusted to better fit some
aspects of the data is referred to as a ``tune''.
It is possible to derive a tune that describes the data at a particular $\sqrt{s}$. 
However, such a model, without energy dependence, will be biased and cannot provide any reliable 
information about other $\sqrt{s}$. Thus, whenever the collision energy ($\sqrt{s}$) has changed, 
additional constraints on the models must be applied using the information obtained 
from the new measurements. This is not a straightforward procedure since no single tune 
can describe all the data with the same precision.
The default CMS \PYTHIAviii tune CUETP8M1 for 7\TeV~\cite{CMS:2015wcf} was derived using the inputs from the 0.9, 
1.96 and 7\TeV measurements, and it describes the data at 7\TeV quite well. The default CMS \PYTHIAviii 
tune CP5, where CP stands for ``CMS \PYTHIAviii'' for 13\TeV~\cite{Sirunyan:2019dfx} was derived using the inputs from the 1.96, 7 and 13\TeV measurements. The CUETP8M1 describes data at 7\TeV better than CP5, but the overall performance of CP5 is much better 
than CUETP8M1 when 13\TeV data are also included.

{\tolerance=800
This paper presents results from two tunes, which make use of the QCD-inspired~\cite{Christiansen:2015yqa} and the gluon-move~\cite{Argyropoulos:2014zoa} CR models.
The new CR tunes presented are based on the default CMS \PYTHIAviii tune CP5.
Along with the CP5 tune, which is derived from the MPI-based CR model, the performance of the new CR tunes (CP5-CR1 and CP5-CR2 defined below)
is studied using several observables. 
These tunes can be used for the evaluation of the uncertainties due to CR effects, and deepening the understanding of the CR mechanism.
\par}

The paper is organised as follows.
In Section 2, the different colour reconnection models implemented in \PYTHIAviii and used in this study are introduced.
In Section 3, the tuning strategy is explained in detail and the parameters of the new tunes are presented.
Section 4 shows a selection of validation plots related to observables measured at $\sqrt{s}=1.96$, 7, 8, and 13\TeV 
by various experiments compared with the predictions of the new tunes.
In Section 5 a study of the uncertainty in the top quark mass \topmass measurement because of the
CR modelling is presented before summarising the results in Section 6.

\section{Colour reconnection models}
The MPI-based CR model was the only CR model implemented
in \PYTHIAviii until \PYTHIA 8.2, which was released with two additional CR models.
The models implemented in \PYTHIA 8.2, referred to as the ``MPI-based'',
``QCD-inspired'', and ``gluon-move'' CR models, are briefly described
in the following:

\begin{itemize}
\item \textit{\textbf{MPI-based model (CP5):}} The simplest model~\cite{Sjostrand:1987su,Sjostrand:2004pf}
implemented in MC event generators introduces only one tunable parameter.
In this model, the partons are classified according to the MPI system to which they belong.
Each parton interaction is originally a $2\to 2$ scattering. 
For an MPI system with a hardness scale \pt of the $2\to 2$
interaction, a CR probability is defined as:
\begin{linenomath*}
\begin{equation}
P = \frac{\ptrec^2}{(\ptrec^2+\pt^2)},
\label{equation_p}
\end{equation}
\end{linenomath*}
with $\ptrec = r \ptzero$,
where $r$ is a tunable parameter and \ptzero is the energy-dependent dampening 
parameter defined in Eq.(\ref{eqmpi}). The parameter \ptzero avoids a divergence of the partonic cross section at low \pt.
According to Eq.(\ref{equation_p}), MPI systems at high \pt would tend to escape from the interaction point, without being colour reconnected to the hard scattering system.
Colour fields originating from a low-\pt MPI system would instead more likely exchange colour.
Once the systems to be connected are determined, partons of low-\pt systems are added 
to strings defined by the highest \pt system to achieve a minimal total string length.

\item \textit{\textbf{QCD-inspired model (CP5-CR1):}} The QCD-inspired model~\cite{Christiansen:2015yqa} 
implemented in \PYTHIAviii adds the QCD colour rules on top of the minimisation of the string length.
The model constructs all pairs of QCD dipoles allowed to be reconnected by QCD colour rules
that determine the colour compatibility of two strings.
This is done iteratively until none of the allowed reconnection possibilities result in a shortening 
of the total string length.
It uses a simple picture to causally connect the produced strings in spacetime through a
string length measure $\lambda$ to determine favoured reconnections.
The default parametrisation for $\lambda$ is
\begin{linenomath*}
\begin{equation}\label{qcdins}
\lambda = \ln \left(1+\sqrt{2}\frac{E_1}{m_0}\right) + \ln \left(1+\sqrt{2}\frac{E_2}{m_0}\right),
\end{equation}
\end{linenomath*}
where $E_1$ and $E_2$ represent the energies of the coloured partons in the rest frame of the QCD dipole, 
and $m_0$ is a constant with the dimension of energy~\cite{Christiansen:2015yqa}.
In addition, the QCD-inspired model allows us to create junction structures. A junction is a topological structure and is formed when three colour lines meet at a single point. The presence of junctions reduces the number of colour lines that need to be connected to the beam remnant, which in turn can affect the number of particles produced in a collision. Since the QCD-inspired CR model allows for different color topologies beyond LC, it can successfully describe the baryon production measured at the CMS experiment~\cite{Christiansen:2015yqa, Khachatryan:2011tm}, which is not the case for previously available \PYTHIAviii tunes.

\item \textit{\textbf{Gluon-move model (CP5-CR2):}} In this scheme \cite{Argyropoulos:2014zoa},
final-state gluons are identified along with all the colour-connected pairs of partons. 
Then an iterative process starts. 
The difference between string lengths when a final-state gluon belonging to two connected partons is moved to another
connected two-parton system is calculated. 
The gluon is moved to the string for which the move gives
the largest reduction in total string length. 
This procedure can be repeated for all or a fraction of the gluons in the final state,  
which is controlled by the \PYTHIAviii parameter \texttt{ColourReconnection:fracGluon}.

In this scheme, quarks would not be reconnected, \ie they would remain in the
same position without any colour exchange.
To improve this picture, the flip mechanism of the gluon-move model can be included. 
The flip mechanism basically allows reconnection of two different string systems, \ie a quark can connect to a different antiquark.
Junctions (Y-shaped three-quark configurations) are allowed to take part in the flip step as well, but no considerable 
differences are expected due to the limitation of the junction formation in this model. 
The flip mechanism has not been extensively studied and its effect on diffractive 
events is not known. For this reason the flip mechanism is switched off in \PYTHIAviii and not used in this paper. 
The main free parameters of the gluon-move model account for the lower limit of
the string length allowed for colour reconnection,
the fraction of gluons allowed to move, and the lower limit of
the allowed reduction of the string lengths.
\end{itemize}

In addition to these models, the effects of early resonance decay (ERD)~\cite{Argyropoulos:2014zoa} in top quark decays are also studied. 
With this option, top quark decay products are allowed to participate directly in CR. 
Normally the ERD option is switched off in \PYTHIAviii but in Section~\ref{sec:topquark} we investigate the ERD effects. 

Usually, MPI and CR effects are investigated and constrained using fits
to measurements sensitive to the UE in hadron collisions.
The UE measurements have been performed at various collision energies by
ATLAS, CMS, and CDF Collaborations~\cite{CMS:2015zev,Aaboud:2017fwp,CMS:2012zxa, Aad:2010fh, Aaltonen:2015aoa}.
The measurements are typically performed by studying the multiplicity and the scalar \pt sum of the 
charged particles (\ptsum),
measured as a function of the \pt of the leading charged particle in the event, \ptmax.

\begin{figure}[!hbtp]
 \centering
 \includegraphics[width=0.46\textwidth]{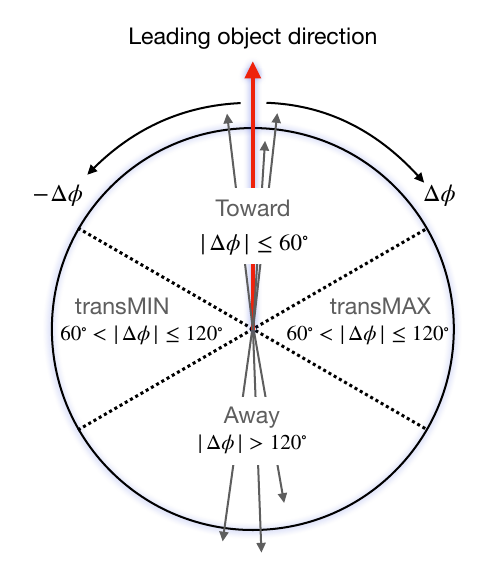}
 \caption{The schematic description of the result of a typical hadron-hadron collision. 
The ``toward'' region contains the ``toward-side'' jet, whereas the ``away'' region may contain an ``away-side'' jet.}
 \label{fig:regions}
 \end{figure}

Different regions of the plane transverse to the direction of the beams are
defined by the direction of the leading charged particle.
A sketch of the different regions is shown in Fig.~\ref{fig:regions}.
A ``toward'' region includes mainly the products of the hard scattering, whereas the ``away'' region
includes the recoiling objects belonging to the hard scattering. 
The two ``transverse'' regions contain the
products of MPIs and are affected by contributions from ISR and FSR.

In Ref.~\cite{CMS:2015zev,Aaboud:2017fwp,Aaltonen:2015aoa}, the transverse region is further subdivided into 
``transMIN'' and ``transMAX'', defined to be the regions with the minimum and maximum number of particles between the two transverse regions.
This is done to disentangle contributions from MPI, ISR, and FSR.
For events with large ISR or FSR, the \tmax\ region contains at least one ``transverse-side'' jet, 
whereas both the \tmax\ and \tmin\ regions contain particles from the MPI and BBR.
Thus, the \tmin\ region is sensitive to MPI and BBR, whereas the difference between \tmax\ and \tmin\ (referred to as the \tdiff\ region) 
is sensitive to ISR and FSR.

The CMS Collaboration showed that a consistent description of the \Nch and the \ptsum distributions 
is not possible using only the \PYTHIAviii hadronisation model without taking into account the CR effects~\cite{CMS:2015wcf}.
In general, the largest difference between the predictions from tunes and the data is observed
in the soft region ($\pt\sim2\text{--}5\GeV$), where CR effects are
expected to be more relevant.

\begin{figure*}[!hbtp]
  \centering
        \includegraphics[width=0.49\textwidth]{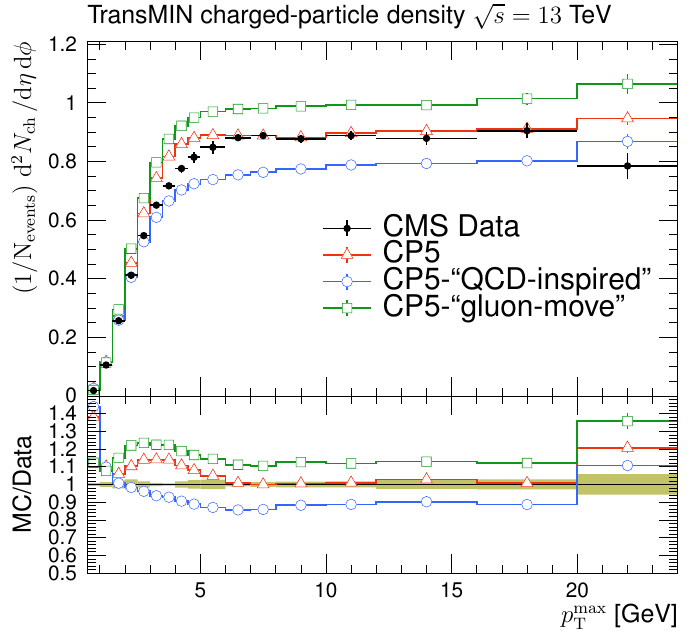}
        \includegraphics[width=0.49\textwidth]{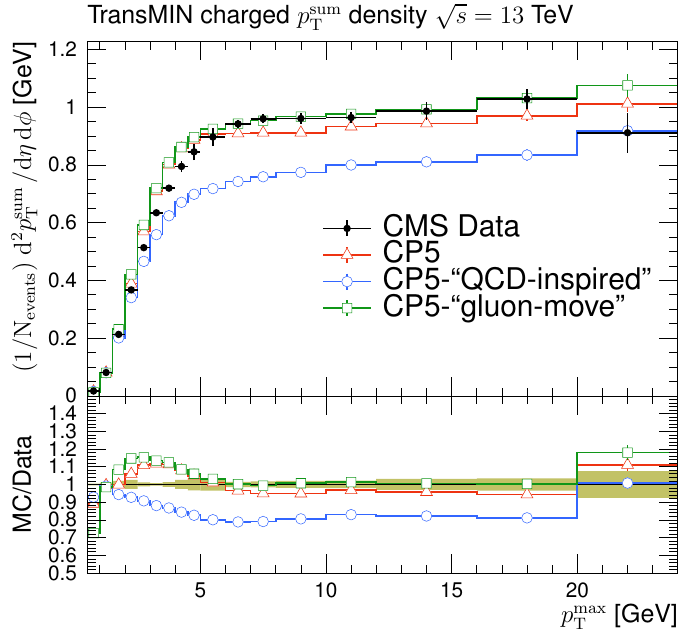}\\
        \includegraphics[width=0.49\textwidth]{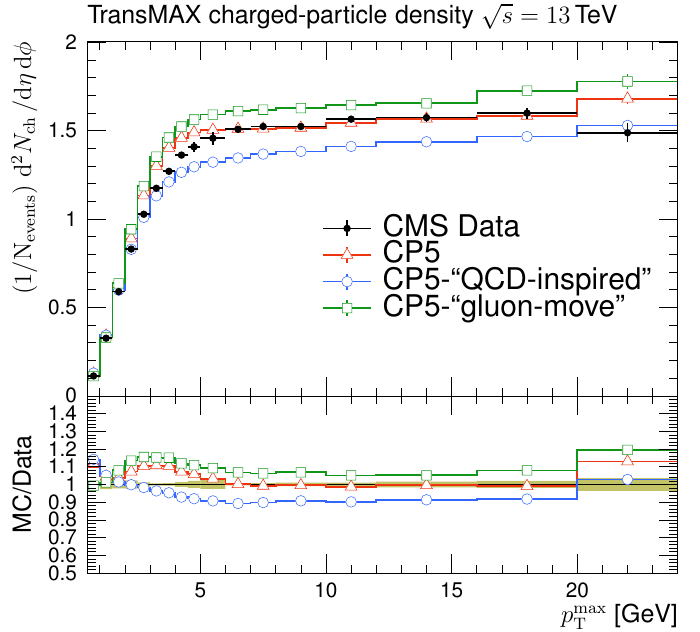}
        \includegraphics[width=0.49\textwidth]{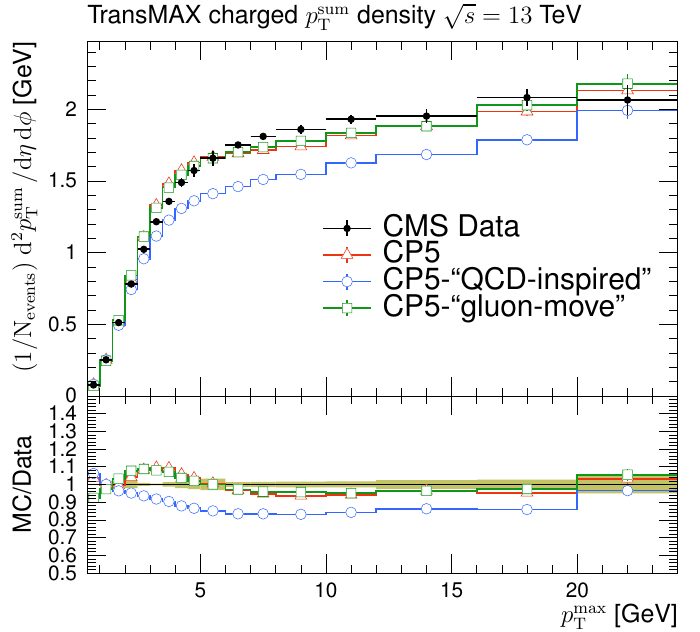}\\
  \caption{
		   The charged-particle (left) and \ptsum densities (right) in the \tmin (upper) and \tmax (lower) regions
		   as functions of the \pt of the leading charged particle, \ptmax, measured by the CMS experiment at $\sqrt{s}=13\TeV$~\cite{CMS:2015zev}.
           The predictions of the tunes CP5, CP5-``QCD-inspired'', and CP5-``gluon-move'' using their default parameter settings in Refs.~\cite{Christiansen:2015yqa,Argyropoulos:2014zoa}, are compared with data.
           The coloured band and error bars on the data points represent the total experimental uncertainty in the data where the model uncertainty is also included.
           The comparisons show that the models do not describe the data and need to be retuned.
		  }
    \label{fig:beforeTuningUE}
\end{figure*}

\begin{figure}[!hbtp]
  \centering
        \includegraphics[width=0.49\textwidth]{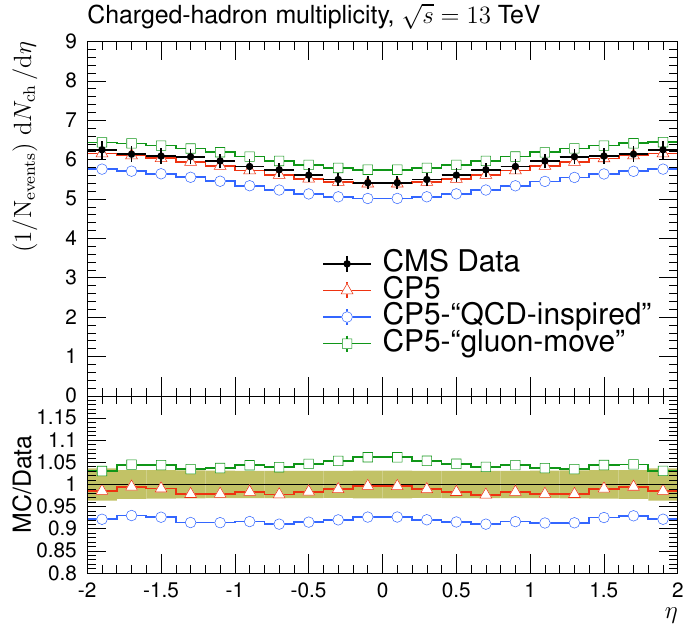}
   \caption{
            The pseudorapidity of charged hadrons, \dNdeta, measured in $\abs{\eta}<2$ by the CMS experiment at $\sqrt{s}=13\TeV$~\cite{CMS:2016gde}. The predictions of the tunes CP5, CP5-``QCD-inspired'', and CP5-``gluon-move'' using their default parameter settings in Refs.~\cite{Christiansen:2015yqa,Argyropoulos:2014zoa}, are compared with data. The coloured band and error bars on the data points represent the total experimental uncertainty in the data where model uncertainty is also included. The comparisons show that the models need to be retuned in order to have a better agreement with the data.
		   }
    \label{fig:beforeTuningMB}
\end{figure}

The new CR models, QCD-inspired and gluon-move, were implemented in \PYTHIA 8.226
after tuning the model parameters to the existing data at $\sqrt{s}=7\TeV$ and at lower centre-of-mass
energies~\cite{Christiansen:2015yqa, Argyropoulos:2014zoa}.
The models were tuned to different data sets starting from different baseline tune settings.
The model predictions, with their default parameter settings in \PYTHIA 8.226 and CP5,
are given in Fig.~\ref{fig:beforeTuningUE} for \Nch and \ptsum densities
measured by the CMS experiment at 13\TeV~\cite{CMS:2015zev} in the \tmin and \tmax regions,
and in Fig.~\ref{fig:beforeTuningMB} for the \dNdeta distribution
measured by CMS at 13\TeV~\cite{CMS:2015zrm}. 
In these figures, the data points, shown in black, are well described by CP5.
The predictions for CP5-``QCD-inspired'' and CP5-``gluon-move'' were obtained by replacing 
the MPI-based CR model in CP5 with the QCD-inspired and gluon-move CR model, respectively.
As mentioned earlier, these models were tuned to data at 7\TeV and at lower centre-of-mass energies. 
The comparisons show that the models must be retuned to describe
the underlying soft physics of pp collisions at 13\TeV.

\section{The new CMS colour reconnection tunes}
\label{sec:Tuning}

{\tolerance=1200

A new set of event tunes, based on UE data from the CMS and CDF experiments, are derived using the QCD-inspired and the gluon-move CR models, as implemented in the
\PYTHIA 8.226 event generator. 
Having tunes for different CR models allows a consistent way of evaluating systematic uncertainties because of colour reconnection effects in specific measurements.
The \RIVET~2.4.0~\cite{Buckley:2010ar} routines used as inputs to the fits, as well as the centre-of-mass energy values and the names of the \RIVET distributions, the $x$-axis ranges (fit ranges), and the relative importance ($R$) of the distributions are displayed in Table~\ref{tab:input_CP5} for the tunes CP5-CR1 and CP5-CR2.
The CP5 tune is used as a baseline for the CR tuning since it is the default \PYTHIAviii
tune for most of the new CMS analyses using data at $\sqrt{s}=13\TeV$ published since 2017, and it has explicitly been tested against a large number of different final states (MB, QCD, top quark, and vector boson + jets) and observables~\cite{Sirunyan:2019dfx}.

\par}

\begin{table*}[!hbtp]
    \centering
   \topcaption{List of input \RIVET routines, centre-of-mass energy values, $\eta$ ranges, names of distributions, fit ranges, and relative importance of the distributions 
   used in the fits to derive the tunes CP5-CR1 and CP5-CR2.}
    \label{tab:input_CP5}
    \cmsTable{
    \begin{tabular}{@{}llllllll}
                                      &           &           &                                   & CP5-CR1    &                 & CP5-CR2     &         \\
         \hline
         \RIVET routine       & $\sqrt{s}$ & $\abs{\eta}$ & Distribution                      & Fit range  & $R$  & Fit range &  $R$      \\
                                      &  [\TeVns{}]     &          &                                   & [\GeVns{}]      &      & [\GeVns{}]     &            \\         
         \hline
         CMS\_2015\_I1384119~\cite{CMS:2016gde}          &  13        & ${<}$2.0   & \Nch versus $\eta$       &            & 1    &           & 1  \\
         CMS\_2015\_PAS\_FSQ\_15\_007~\cite{CMS:2015zev} &  13        & ${<}$2.0   & TransMIN \ptsum & 2--28      & 1    & 3--36     & 0.5 \\
                                      &            &          & TransMAX \ptsum & 2--28      & 1    & 3--36     & 0.5 \\
                                      &            &          & TransMIN \Nch            & 2--28      & 1    & 3--36     & 0.1 \\
                                      &            &          & TransMAX \Nch            & 2--28      & 1    & 3--36     & 0.1 \\ 
         CMS\_2012\_PAS\_FSQ\_12\_020~\cite{Aaboud:2017fwp} &  7         & ${<}$0.8   & TransMAX \Nch            & 3--20      & 1    & 3--20     & 0.1 \\
                                      &            &          & TransMIN \Nch            & 3--20      & 1    & 3--20     & 0.1 \\
                                      &            &          & TransMAX \ptsum & 3--20      & 1    & 3--20     & 0.1 \\
                                      &            &          & TransMIN  \ptsum & 3--20      & 1    & 3--20     & 0.1 \\
         CDF\_2015\_I1388868~\cite{Aaltonen:2015aoa}          &  2         & ${<}$0.8   & TransMIN \Nch            & 2--15      & 1    & 2--15     & 0.1 \\
                                      &            &          & TransMAX \Nch            & 2--15      & 1    & 2--15     & 0.1 \\
                                      &            &          & TransMIN \ptsum & 2--15      & 1    & 2--15     & 0.1 \\
                                      &            &          & TransMAX \ptsum & 2--15      & 1    & 2--15     & 0.1 \\

    \end{tabular}
	}
\end{table*}
{\tolerance=1200
The parameters and their ranges in the fits are shown in Table~\ref{tab:fit_resCR1}.
The minimum and maximum values of the parameters are first taken from \PYTHIAviii, then the ranges of 
the values are further limited using the \PROFESSOR~1.4.0 software~\cite{Buckley:2009bj}. 
The ranges are chosen such that the sampled MC space does not destroy the definition 
of a particular observable in the fits.
\par}

 \begin{table}[!hbtp]
 \centering
 \topcaption{The MPI and CR parameter ranges used in the tuning procedure.}
 \label{tab:fit_resCR1}
 \begin{tabular}{lr}
 \multicolumn{2}{l}{\PYTHIAviii parameter}{Min--Max}\\\hline
 \multicolumn{2}{c}{MPI parameters}\\
 \texttt{MultipartonInteractions:pT0Ref}        & 1.0\NA3.0 \\
 \texttt{MultipartonInteractions:ecmPow}        & 0.0\NA0.3 \\
 \texttt{MultipartonInteractions:coreRadius}    & 0.2\NA0.8 \\
 \texttt{MultipartonInteractions:coreFraction}  & 0.2\NA0.8 \\ [\cmsTabSkip]
 \multicolumn{2}{c}{QCD-inspired model}\\
 \texttt{ColourReconnection:m0}                 & 0.1\NA4.0 \\
 \texttt{ColourReconnection:junctionCorrection} & 0.01\NA10 \\
 \texttt{ColourReconnection:timeDilationPar}    & 0\NA60 \\ [\cmsTabSkip]
 \multicolumn{2}{c}{Gluon-move model}\\
 \texttt{ColourReconnection:m2lambda}           & 0.2\NA8.0 \\
 \texttt{ColourReconnection:fracGluon}          & 0.8\NA1.0
 \end{tabular}
 \end{table}
{\tolerance=1200
Tune CP5 uses the next-to-next-to-leading order (NNLO)  NNPDF31\_nnlo\_as\_0118~\cite{Ball:2017nwa} PDF set, the strong coupling parameter \alpS value of 0.118 for ISR, FSR, and MPI,
and the MPI-based CR model. 
It also uses a double-Gaussian functional form with two tunable parameters, \texttt{coreRadius} and \texttt{coreFraction}, to model the overlap of the matter distribution of the two colliding protons~\cite{Sjostrand:1987su}.
The tune parameters are documented in Ref.~\cite{Sirunyan:2019dfx} and displayed in Table~\ref{tab:fit_res1}.
Also in Figs. 11 and 12 in Ref.~\cite{Sirunyan:2019dfx}, predictions of the CP5 tune are compared with event shape 
observables measured at LEP. The results show that a value of $\alpS^\mathrm{FSR}(m_\PZ)$$\sim$0.120 better describes the data 
compared with higher values of $\alpS^\mathrm{FSR}(m_\PZ)$ which generally overestimates the number of final-state partons. 
As concluded in the Ref.~\cite{Sirunyan:2019dfx}, LEP event shape observables are well described by \MGvATNLO~+~\PYTHIAviii with CP5.
\par}

{\tolerance=1200
The new tunes are obtained by constraining simultaneously the parameters controlling the contributions of the MPI and of each of
the CR models.
The strategy followed to obtain the CP5-CR1 and CP5-CR2 tunes is
similar to that used for the CP5 tune, \ie the same observables sensitive to MPI are considered to constrain the parameters.
These are the \Nch and average \ptsum as functions of the leading
charged particle transverse momentum \ptmax, measured in the \tmin and \tmax regions by the CMS experiment
at $\sqrt{s}=13\TeV$~\cite{CMS:2015zev} and 7\TeV~\cite{CMS:2012zxa} and by the CDF experiment
at 1.96\TeV~\cite{Aaltonen:2015aoa}.
The \Nch as a function of $\eta$, measured
by CMS at $\sqrt{s}=13\TeV$~\cite{CMS:2015zrm} is also used in the fit. 
In Ref.~\cite{CMS:2015zev}, the \tmin and \tmax regions are defined with respect to both the leading 
charged particles and the leading charged-particle jets as reference objects. 
The uncertainty in measurements using leading charged particles as reference objects is lower than the uncertainty in measurements using leading charged-particle jets as reference objects. This is one of the reasons why we choose to use leading charged particle observables instead of leading charged-particle jet observables in the fits. Another reason is that we want to use the same observables that were used to derive CP5, and CP5 was derived using leading charged particle observables in the fits.
As a cross-check, we also derived another version of the CP5-CR1 tune using leading charged-particle 
jet observables, such as \Nch and average \ptsum, as functions of the transverse momentum of the leading 
charged-particle jets in the fits. The results showed that the use of leading charged-particle 
jet observables in the fits makes a very small difference, which is negligible when tune uncertainties 
are taken into account.
As for CP5, the region with \ptmax between 0.5 and 2.0 or 3.0\GeV is excluded depending on the distribution 
from the fit, since this region is affected by diffractive processes 
whose free parameters are not considered in the tuning procedure.
\par}

The MPI-related parameters that are kept free in both the CP5-CR1 and CP5-CR2 tunes are:
 \begin{itemize}
 \item \texttt{MultipartonInteractions:pT0Ref}, the parameter $\ptzero^{\text{ref}}$ included in the regularisation of the partonic QCD cross section 
as described in Eq.(\ref{eqmpi}).
It sets the lower cutoff scale for MPIs;
 \item \texttt{MultipartonInteractions:ecmPow}, the exponent $\epsilon$ of the $\sqrt{s}$ dependence as shown in Eq.(\ref{eqmpi});
 \item \texttt{MultipartonInteractions:coreRadius}, the width of the core when a double-Gaussian matter profile is assumed for
the overlap distribution between the two colliding protons~\cite{Sjostrand:1987su}. A double-Gaussian form identifies an inner,
dense part, which is called core, and an outer, less dense part;
 \item \texttt{MultipartonInteractions:coreFraction}, the fraction of quarks and gluons contained in the core when
a double-Gaussian matter profile is assumed.
 \end{itemize}

 The tunable CR parameters in CP5-CR1 that are considered in the fit are:

{\tolerance=1800

 \begin{itemize}
 \item \texttt{ColourReconnection:m0}, the variable that determines whether a
 possible reconnection is actually favoured in the $\lambda$ measure in Eq.(\ref{qcdins});
 \item \texttt{ColourReconnection:junctionCorrection}, the multiplicative correction for junction
 formation, applied to the \texttt{m0} parameter;
 \item \texttt{ColourReconnection:timeDilationPar}, the parameter controlling the time dilation
 that forbids colour reconnection between strings that are not in causal contact.
 \end{itemize}

\par}

More details on these parameters are reported in Ref.~\cite{pythia8}.
For the CP5-CR1 tune, the parameters related to the hadronisation, \texttt{StringZ:aLund}, \texttt{StringZ:bLund},
\texttt{StringFlav:probQQtoQ}, and \texttt{StringFlav:probStoUD}, proposed in Ref.~\cite{Christiansen:2015yqa},
are also used as fixed inputs to the tune.
The first two of these parameters govern the longitudinal fragmentation function 
used in the Lund string model in \PYTHIAviii, whereas the latter two are the probability 
of diquark over quark fragmentation, and the ratio of strange to light quark production, respectively.

For the optimisation of CP5-CR2, the following parameters are considered:

{\tolerance=1800

 \begin{itemize}
 \item  \texttt{ColourReconnection:m2lambda}, an approximate hadronic mass-square
scale and the parameter used in the calculation of $\lambda$; 
 \item \texttt{ColourReconnection:fracGluon}, the probability that a given gluon will be moved. 
It thus gives the average fraction of gluons being considered.
 \end{itemize}
\par}

The remaining parameters of \PYTHIAviii are kept the same as in the CP5 tune.

 The fits are performed using the \PROFESSOR~1.4.0 software,
 which takes random values for each parameter in the defined multidimensional parameter space,
 and \RIVET, which provides the data points and uncertainties, and produces the individual generator
 predictions for the considered observables.
 About 200 different choices of parameters are considered to build a
 random grid in the parameter space. For each choice of parameters, one million pp
 inelastic scattering events, including contributions from single-diffractive dissociation (SD),
 double-diffractive dissociation (DD), central diffraction (CD), and nondiffractive (ND) processes, are generated.
 The bin-by-bin envelopes of the different MC predictions are checked.
 After building the grid in the parameter space, \PROFESSOR performs an interpolation of the bin
 values for the observables in the parameter space using a third-order
 polynomial function. We verified that the degree of the polynomial used for the
 interpolation does not affect the tune results significantly.
 The function $f^{\scriptsize b}(\mathbf{p})$ models the MC response of each
 bin $b$ of the observable $O$ as a function of the parameter vector $\mathbf{p}$. 
The final step is the
 minimisation of the $\chi^{*2}$ function given by:
\begin{linenomath*}
 \begin{equation}\label{chi2}
 \chi^{*2}(\mathbf{p})=\sum_{O}\sum_{{\scriptsize b}\in O}\frac{(f^{\scriptsize b}(\mathbf{p})-\mathcal{R}_{\scriptsize b})^2}{\Delta_{\scriptsize b}^2},
 \end{equation}
\end{linenomath*}
 where $\mathcal{R}_{\scriptsize b}$ is the data value for each bin $b$, and
 $\Delta_{\scriptsize b}^2$ expresses the total bin uncertainty of the data.

The $\chi^{*2}$ is not a true $\chi^2$ function as explained in the following. 
Treating equally all distributions that are used as inputs to the fit for the CP5-CR2 tune
results in a tune that describes the data poorly; in particular, it underestimates 
the \dNdeta distribution measured in data at $\sqrt{s}=13\TeV$ by about 30\%.
This is because the $\chi^2$ definition treats all bins equally and the importance of \dNdeta
may be lost because of its relatively low precision with respect to other observables.
The \dNdeta distribution is one of the key observables that is sensitive to a number of 
processes and, therefore, increasing the importance of this observable in the fit is reasonable.

In \PROFESSOR, this is done by using weights with a nonstandard $\chi^2$ definition.
To keep the standard properties of a $\chi^2$ fit, we increase the total uncertainties of
the other distributions. 
The total uncertainty in each bin is scaled up by $1/\sqrt{R}$ with $R$ (relative importance) values displayed in Table~\ref{tab:input_CP5}. 
Therefore, the total uncertainty of each bin of \ptsum in the \tmin and \tmax regions at $\sqrt{s}=13\TeV$
is scaled up by $\sqrt{2}$ and that of all other distributions by $\sqrt{10}$.
These scale factors ensure that the distributions are well described after the tuning.
For the CP5-CR1 model, a good description of the input observables is obtained without scaling, meaning that all distributions are considered equally important.

 The experimental uncertainties used in the fit, in general, have bin-to-bin correlations.
 However, some of the bins of the UE distributions used in the fit,
 \eg $\ptmax > 10\GeV$, are dominated by statistical uncertainties, which are uncorrelated between bins.
 In the minimisation procedure, because the correlations between bins are not available for the input measurements,
 the experimental uncertainties are assumed to be uncorrelated between data points.
 
 The parameters obtained from the CP5-CR1 and CP5-CR2 fits, as well as the value of the goodness of the fit
 are shown in Table~\ref{tab:fit_res1}. 
Uncertainties in the parameters of these tunes are discussed in Appendix~\ref{sec:Uncertainties}.
In Ref.~\cite{Sirunyan:2019dfx}, the number of degrees of freedom (\ndof), defined as the sum of the number of bins of fit observables minus the number of fit parameters, for the tune CP5 is given as 63. 
However, this value of \ndof corresponds to the case when only 13\TeV distributions are used. 
The value of \ndof for CP5 consistent with our calculation in this paper is 183. 
The tune CP5 was derived using two additional distributions in the fits; \dNdeta at 13\TeV
with NSD-enhanced selection and SD-enhanced selection. Since these two observables depend on modelling of
single diffraction dissociation, which is not well understood, they are not included 
in the fits for CP5-CR tunes. Therefore, the \ndof values for CP5-CR tunes are lower than the 
\ndof of CP5. The slight difference in the \ndof values between the CP5-CR1 and CP5-CR2 tunes is due to the difference 
in the number of fit parameters used in each tune, which are 7 and 6 respectively.
Although the fit ranges for the CP5-CR tunes differ slightly, as shown in Table~\ref{tab:input_CP5}, the sum of number of bins of fit observables is the same for both tunes.

 \begin{table*}[!hbtp]
 \centering
 \topcaption{The parameters obtained in the fits of the CP5-CR1 and CP5-CR2 tunes,
 compared with that of the CP5 tune. The upper part of the table displays the fixed
 input parameters of the tune, whereas the lower part shows the fitted tune parameters.
 The number of degrees of freedom (\ndof) and the goodness of fit divided by \ndof are also shown.}
\cmsTable{
 \begin{tabular}{lccc}
 \PYTHIAviii parameter                                 &  CP5~\cite{Sirunyan:2019dfx} & CP5-CR1 & CP5-CR2\\
\hline & \\[-2.0ex]
 PDF set & NNPDF3.1 NNLO & NNPDF3.1 NNLO & NNPDF3.1 NNLO                                  \\
 $\alpS(m_\PZ)$                 & 0.118          &  0.118 &  0.118 \\
 \texttt{SpaceShower:rapidityOrder}       &   on & on & on \\
 \texttt{MultipartonInteractions:ecmRef} [\GeVns{}]      &    7000        & 7000 & 7000\\
 $\alpS^\mathrm{ISR}(m_\PZ)$ value/order     &   0.118/NLO         & 0.118/NLO  & 0.118/NLO   \\
 $\alpS^\mathrm{FSR}(m_\PZ)$ value/order     &   0.118/NLO            & 0.118/NLO  & 0.118/NLO   \\
 $\alpS^\mathrm{MPI}(m_\PZ)$ value/order     &  0.118/NLO             &  0.118/NLO & 0.118/NLO   \\
 $\alpS^\mathrm{ME}(m_\PZ)$ value/order     &   0.118/NLO            & 0.118/NLO & 0.118/NLO   \\
 \texttt{StringZ:aLund}                              &\NA  & 0.38    &\NA \\
 \texttt{StringZ:bLund}                              &\NA  & 0.64    &\NA\\
 \texttt{StringFlav:probQQtoQ}                       &\NA  & 0.078   &\NA\\
 \texttt{StringFlav:probStoUD}                       &\NA  & 0.2     &\NA\\
 \texttt{SigmaTotal:zeroAXB}               			 & off 	& off 	  & off \\
 \texttt{BeamRemnants:remnantMode}         			 &\NA 	& 1		  &\NA\\
 \texttt{ColourReconnection:mode}                    &\NA 	& 1       & 2  \\ [\cmsTabSkip]
 \texttt{MultipartonInteractions:pT0Ref} [\GeVns{}]  & 1.410 & 1.375   & 1.454 \\
 \texttt{MultipartonInteractions:ecmPow}             & 0.033 & 0.033   & 0.054 \\
 \texttt{MultipartonInteractions:coreRadius}         & 0.763 & 0.605   & 0.649 \\
 \texttt{MultipartonInteractions:coreFraction}       & 0.630 & 0.445   & 0.489 \\
 \texttt{ColourReconnection:range}                   & 5.176 &\NA     &\NA \\
 \texttt{ColourReconnection:junctionCorrection}      &\NA   & 0.238   &\NA \\
 \texttt{ColourReconnection:timeDilationPar}         &\NA   & 8.580   &\NA\\
 \texttt{ColourReconnection:m0}                      &\NA   & 1.721   &\NA \\
 \texttt{ColourReconnection:m2lambda}                &\NA   &\NA     & 4.917 \\
 \texttt{ColourReconnection:fracGluon}               &\NA   &\NA     & 0.993 \\
  \ndof                                     & 183   & 157     & 158 \\
  $\chi^{*2} / \ndof$   						 & 1.04 & 2.37    & 0.89
 \end{tabular}
}
 \label{tab:fit_res1}
 \end{table*}

A preliminary version of the CP5-CR2 tune was
derived including several jet substructure observables~\cite{Chatrchyan:2012mec,Aad:2011sc,Aad:2016oit}
in the fits. This tune, called CP5-CR2-j, has been used in the MC production in the CMS experiment.
The CP5-CR2 and CP5-CR2-j tunes have very similar predictions in all final states discussed in this paper,
because the tunes differ slightly only in the following parameters, where the listed values are for CP5-CR2-j:
\begin{itemize}
\item \texttt{MultipartonInteractions:ecmPow} = 0.056, 
\item \texttt{MultipartonInteractions:coreRadius} = 0.653, 
\item \texttt{MultipartonInteractions:coreFraction} = 0.439, 
\item \texttt{ColourReconnection:m2lambda} = 4.395, 
\item \texttt{MultipartonInteractions:fracGluon} = 0.990.
\end{itemize}

The CP1 and CP2 are the two tunes in the CPX (X = 1--5) tune family~\cite{Sirunyan:2019dfx} that use an LO PDF set~\cite{Ball:2017nwa}. 
We also derive CR tunes based on the CP1 and CP2 settings to study the effect of using a leading order (LO)
PDF set with alternative CR models, although they are not used in precision measurements.
We find that the predictions of the CR tunes based on CP1 and CP2 for the MB and UE observables are similar to the predictions
of CR tunes based on CP5. 
However, CP1-CR1 (\ie CP1 with the QCD-inspired colour reconnection model) has a different trend in particle multiplicity 
distributions compared with the predictions of other tunes discussed in this study. This different trend of CP1-CR1 cannot be 
attributed to the use of LO PDF set, because both CP1 and CP2 use the same LO PDF set and we do not see a different trend 
with CP2-CR1. 
The different trend observed with CP1-CR1 in the particle multiplicity distributions 
may become a collective effect rather than a single parameter effect, and could be an input for further tuning and development of the QCD-inspired model.
Therefore, in Appendix~\ref{sec:lopdf} of this paper, we present the tune settings of the CR tunes based on CP1 and CP2, along with their predictions 
in the particle multiplicity distributions.

\section{Performance of the tunes}
In Figs.~\ref{fig:CMS_Internal_FSQ_1_CP5}--\ref{fig:dyuetransverse} we show the 
observables measured at centre-of-mass energies of 1.96, 7, 8, and 13\TeV.
The data points are shown in black, and are compared with simulations obtained
from the \PYTHIAviii event generator with the tunes CP5 (red), CP5-CR1 (blue), and CP5-CR2 (green).
For simplicity, the tunes CP5-CR1 and CP5-CR2 will be referred to as CP5-CR when convenient.
The lower panels show the ratios between each MC prediction and the data.

For the plots presented in Figs.~\ref{fig:CMS_Internal_FSQ_1_CP5}--\ref{fig:strangeness}
in Sections~\ref{sec:UE_MB} and~\ref{sec:Par_Mul}, inelastic
events (\ie ND, SD, DD, and CD) are simulated with \PYTHIA 8.226 and compared with data at different centre-of-mass energies.
The rest of the plots are produced with \PYTHIA 8.235.
An update to the description of the elastic scattering component in \PYTHIA 8.235 led to a
slight decrease in the default ND cross section.
The default ND cross section in \PYTHIA 8.226, which is 55.5\unit{mb} at $\sqrt{s}=13\TeV$, is lowered to 55.1\unit{mb} in \PYTHIA 8.235.
Hence, to reproduce the conditions
of \PYTHIA 8.226 in \PYTHIA 8.235 or in a newer version, one should set the ND
cross section manually.

\subsection{Underlying-event and minimum-bias observables}
\label{sec:UE_MB}
MB is a generic term used to describe 
events collected with a loose selection process 
that are dominated by relatively soft particles.
Although these events generally correspond
to inelastic scattering, including ND and SD+DD+CD processes,
these contributions may vary depending on the trigger requirements used in the experiments.
For example, a sample of non-single-diffractive-enhanced (NSD-enhanced) events is selected by suppressing
the SD contribution at the trigger level. 

\begin{figure*}[!hbtp]
  \centering
        \includegraphics[width=0.49\textwidth]{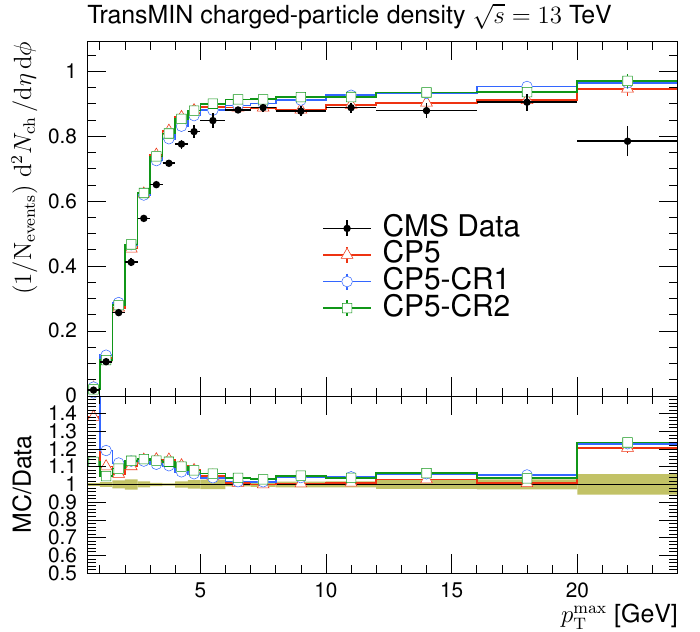}
        \includegraphics[width=0.49\textwidth]{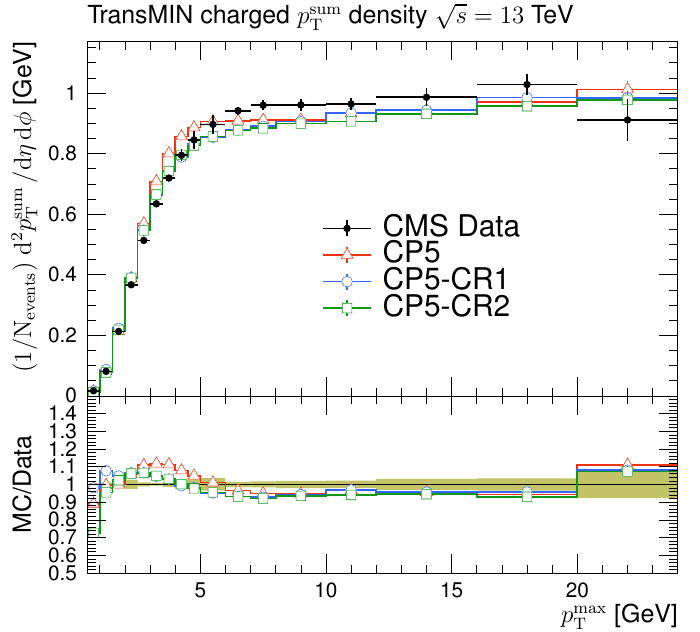}\\
        \includegraphics[width=0.49\textwidth]{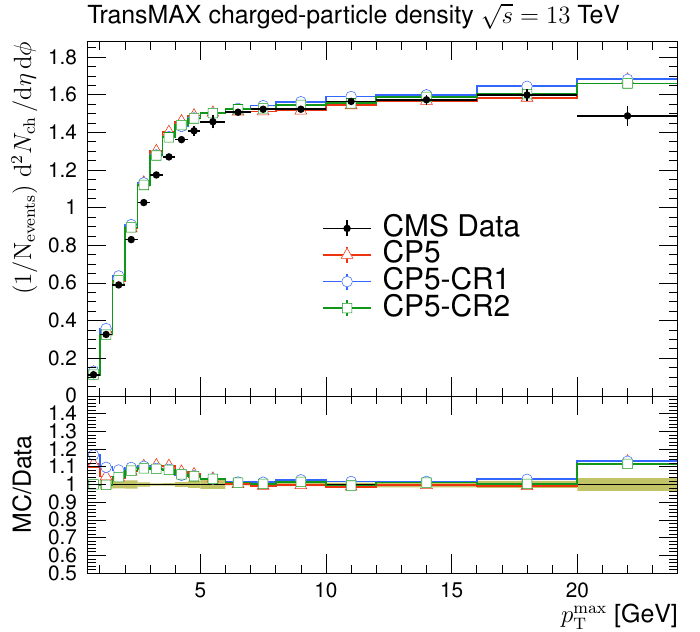}
        \includegraphics[width=0.49\textwidth]{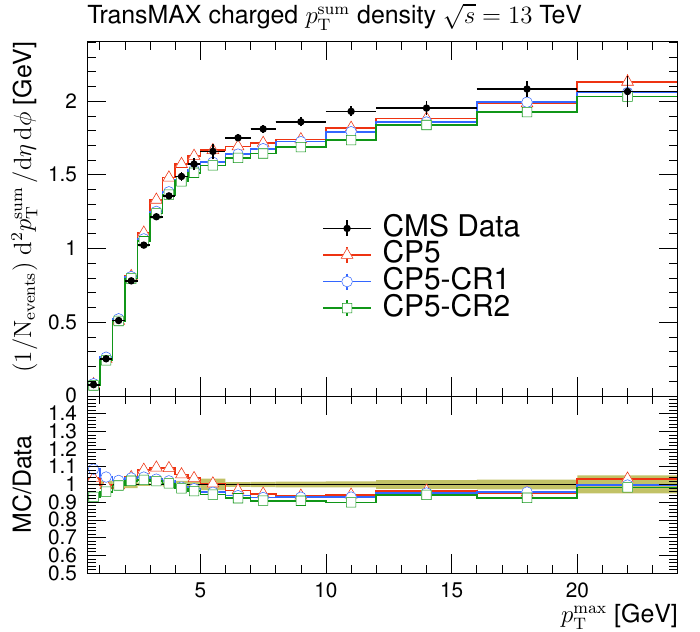}\\
  \caption{
		   The charged-particle (left) and \ptsum (right) densities in the \tmin (upper) and \tmax (lower) regions, 
		   as functions of the \pt of the leading charged particle, \ptmax, measured by the CMS experiment at $\sqrt{s}=13\TeV$~\cite{CMS:2015zev}.
		   The predictions of the CP5 and CP5-CR tunes are compared with data.
		   The coloured band and error bars on the data points represent the total experimental uncertainty in the data.
		   }
    \label{fig:CMS_Internal_FSQ_1_CP5}
\end{figure*}

\begin{figure}[!hbtp]
  \centering
        \includegraphics[width=0.49\textwidth]{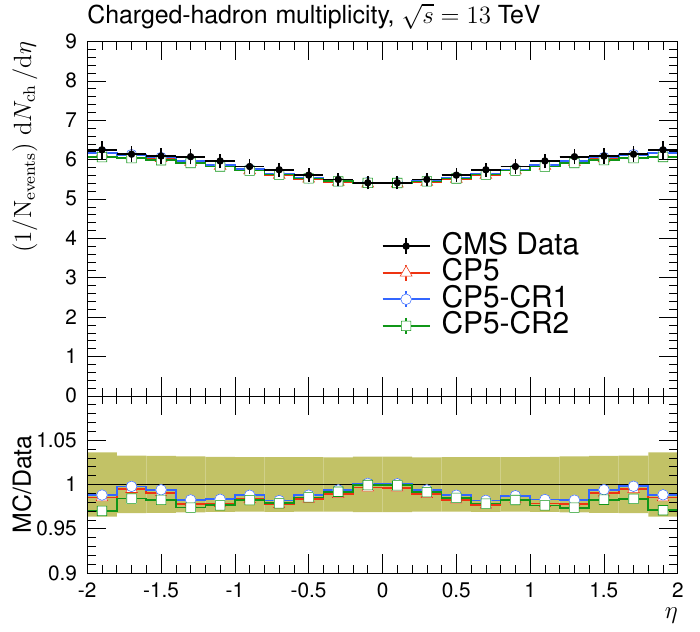}
   \caption{
            The pseudorapidity of charged hadrons, \dNdeta, measured by the CMS experiment at $\sqrt{s}=13\TeV$~\cite{CMS:2015zrm}.
            The predictions of the CP5 and CP5-CR tunes are compared with data.
            The coloured band and error bars on the data points represent the total experimental uncertainty in the data.
			}
    \label{fig:2015_I1384119_FSQ_15_008}
\end{figure}

The UE observables measured by the CMS experiment at $\sqrt{s}=13\TeV$~\cite{CMS:2015zev},
namely \Nch density and the average \ptsum
in the \tmin and \tmax regions are well described by all tunes in the plateau
region as shown in Fig.~\ref{fig:CMS_Internal_FSQ_1_CP5}.
The region up to ${\approx}5$\GeV of $\ptmax$ is highly sensitive to diffractive contributions~\cite{CMS:2020dns}. 
There is a lack of measurements in this region where the tunes, in general, do not perform well. 
Although the optimisation of these components is beyond the scope of this study, we have extended the fit range to ${\approx}$2--3\GeV as long as the data are well described.
The rising part of the spectrum excluding the region up to $\approx$5\GeV of the \Nch density distributions is similarly described by all tunes, whereas in the \ptsum
density distributions the predictions of CP5 differ slightly from the predictions of the CR tunes.
These show that the CP5 tune has a harder \pt spectrum at low $\ptmax$ values.
Through tuning the \Nch and average \ptsum density in the \tmin and \tmax regions, 
a satisfactory agreement is obtained for the same observables in the \tdiff region as well.
Figure~\ref{fig:2015_I1384119_FSQ_15_008} shows the pseudorapidity distribution of 
charged hadrons in inelastic pp collisions measured by the CMS experiment at 
$\sqrt{s}=13\TeV$~\cite{CMS:2015zrm}. This observable is sensitive to the softer 
part of the MPI spectrum and well described by all tunes.

\begin{figure}[!hbtp]
\centering
        \includegraphics[width=0.49\textwidth]{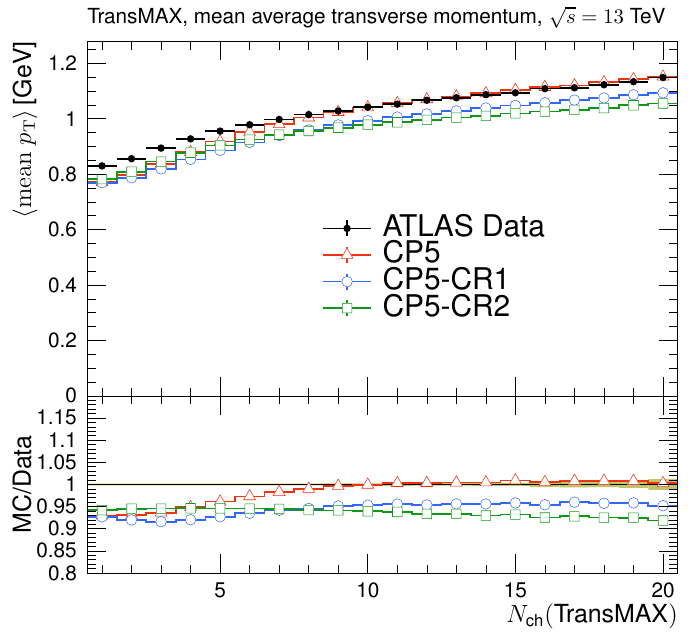}
        \includegraphics[width=0.49\textwidth]{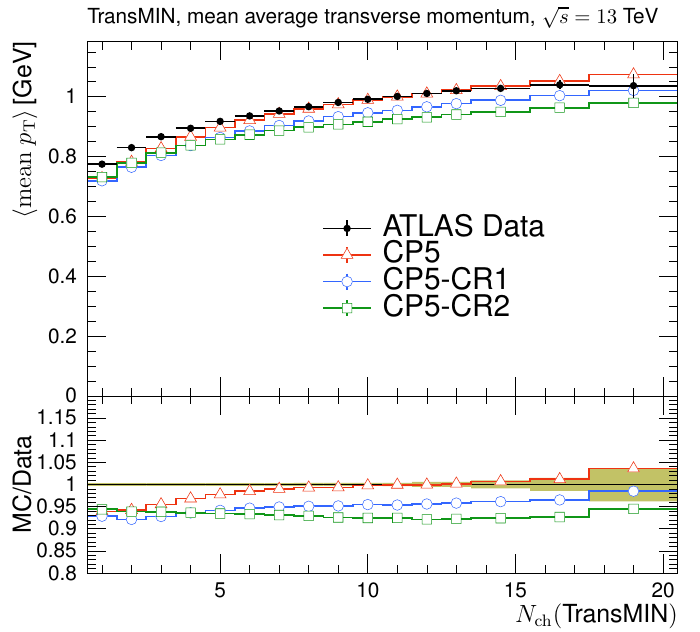}
    \caption{
			 The mean charged-particle average transverse momentum as functions of charged-particle multiplicity in the
			 \tmax (\cmsLeft) and \tmin (\cmsRight) regions, measured by the ATLAS experiment at $\sqrt{s}=13\TeV$~\cite{Aaboud:2017fwp}.
             The predictions of the CP5 and CP5-CR tunes are compared with data.
             The coloured band and error bars on the data points represent the total experimental uncertainty in the data.
			 }
    \label{fig:ATLASCR}
\end{figure}

A crucial test for the performance of UE tunes, and of the
CR simulation in particular, is the description of the average \pt of 
the charged particles as a function of \Nch. Comparisons of the mean average \pt to the measurements by the ATLAS Collaboration
at $\sqrt{s}=13\TeV$ in the \tmax and \tmin regions~\cite{Aaboud:2017fwp}
are displayed in Fig.~\ref{fig:ATLASCR}. 
The tune CP5 describes the central values of the data perfectly for $\Nch>7$,
whereas the CR tunes show an almost constant discrepancy of 5--10\% 
because of the harder \pt spectrum predicted by the tune CP5 for low-\pt particles.
All CR tunes show a reasonable agreement with
the data, confirming the accuracy of the parameters obtained for the new CR models.
The improvement in the tuned CR models and their success in describing the data is seen by
comparing Fig.~\ref{fig:CMS_Internal_FSQ_1_CP5} with Fig.~\ref{fig:beforeTuningUE},
and Fig.~\ref{fig:2015_I1384119_FSQ_15_008} with Fig.~\ref{fig:beforeTuningMB}.
In these figures, CP5 tune predictions are also shown for easier comparison of CR tunes predictions with CP5.

\begin{figure*}[!hbtp]
\centering
    \includegraphics[width=0.49\textwidth]{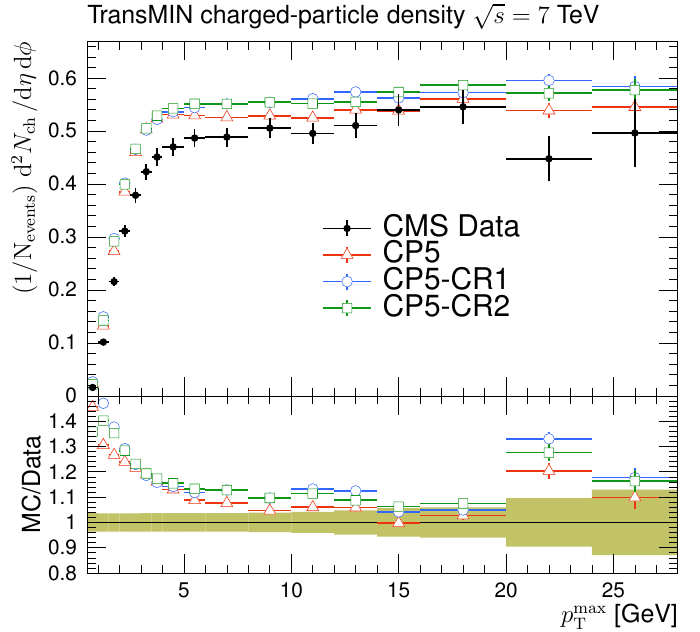}
    \includegraphics[width=0.49\textwidth]{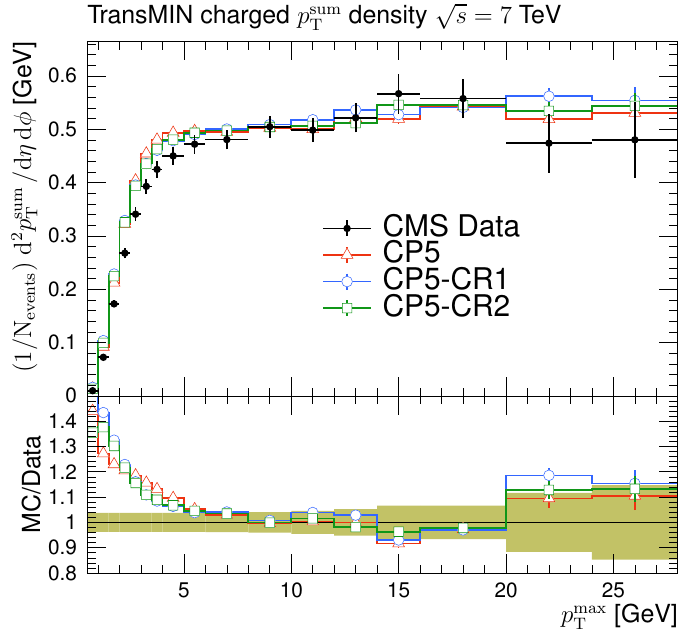}\\
    \includegraphics[width=0.49\textwidth]{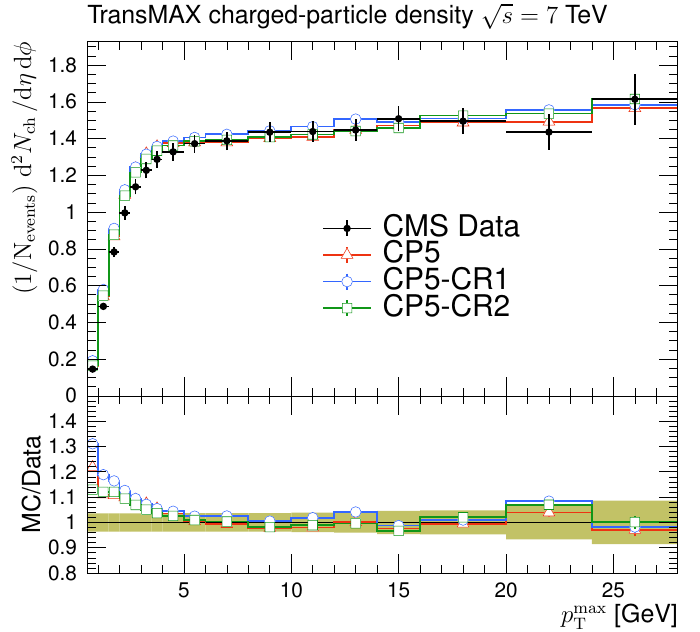}
    \includegraphics[width=0.49\textwidth]{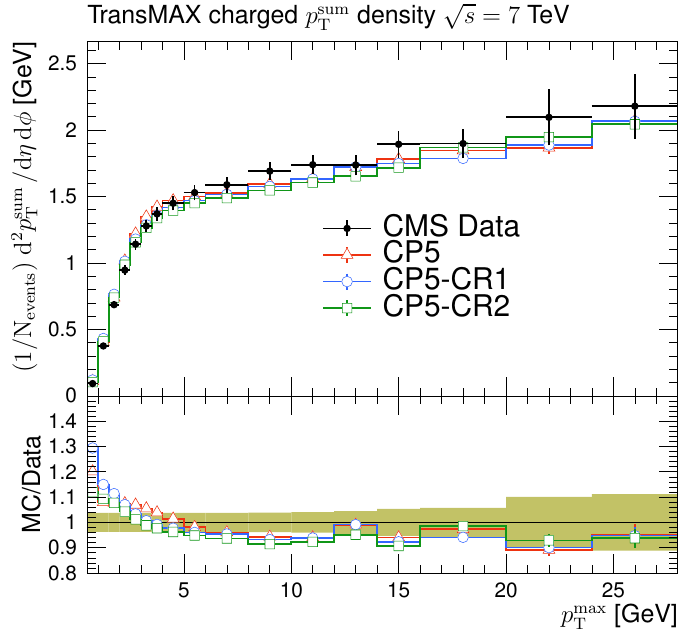}\\
  \caption{
		   The charged-particle (left) and \ptsum (right) densities in the \tmin (upper) and \tmax (lower) regions, 
		   as functions of the \pt of the leading charged particle, \ptmax, measured by the CMS experiment at $\sqrt{s}=7\TeV$~\cite{CMS:2012zxa}.
           The predictions of the CP5 and CP5-CR tunes are compared with data.
           The coloured band and error bars on the data points represent the total experimental uncertainty in the data.
		   }
    \label{fig:CMS_7TeV_CP5}
\end{figure*}

In Fig.~\ref{fig:CMS_7TeV_CP5}, charged-particle and \ptsum densities
measured by the CMS experiment at $\sqrt{s}=7\TeV$~\cite{CMS:2012zxa} in the \tmin and \tmax regions,
as functions of \ptmax, are compared with predictions
from the tunes CP5 and CP5-CR. The data are reasonably well described for $\ptmax>5\GeV$.

\begin{figure*}[!hbtp]
\centering
    \includegraphics[width=0.49\textwidth]{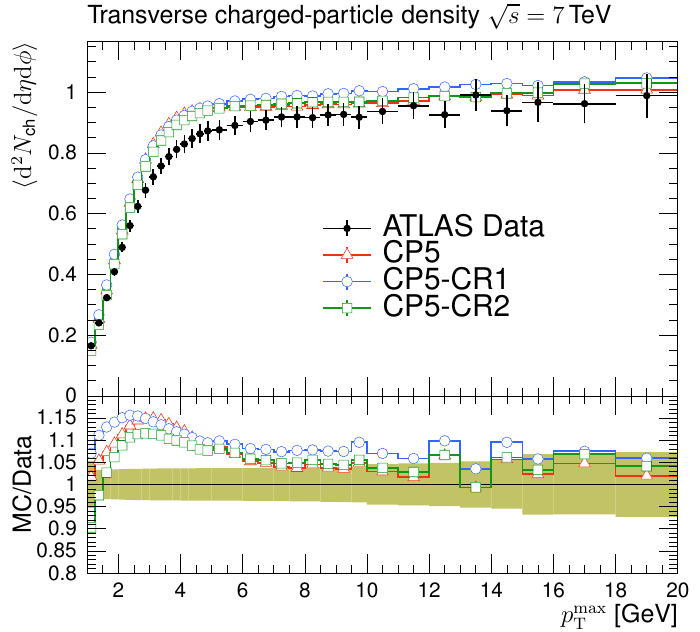}
    \includegraphics[width=0.49\textwidth]{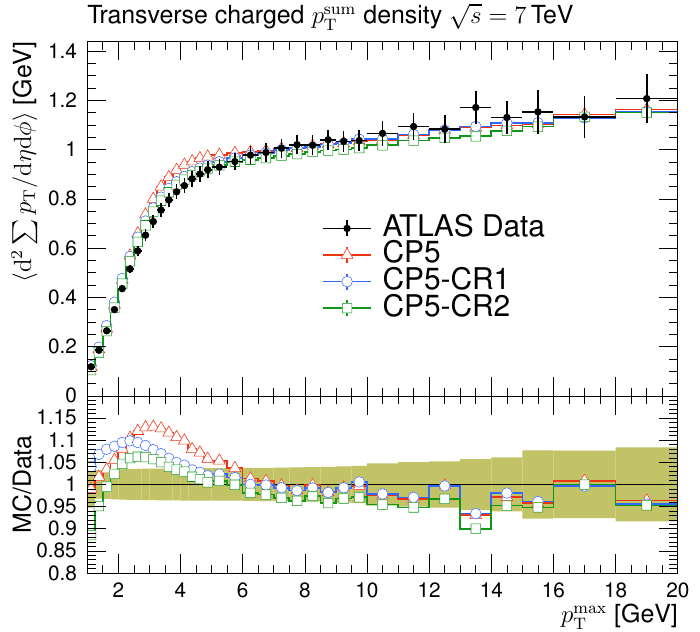}\\
    \includegraphics[width=0.49\textwidth]{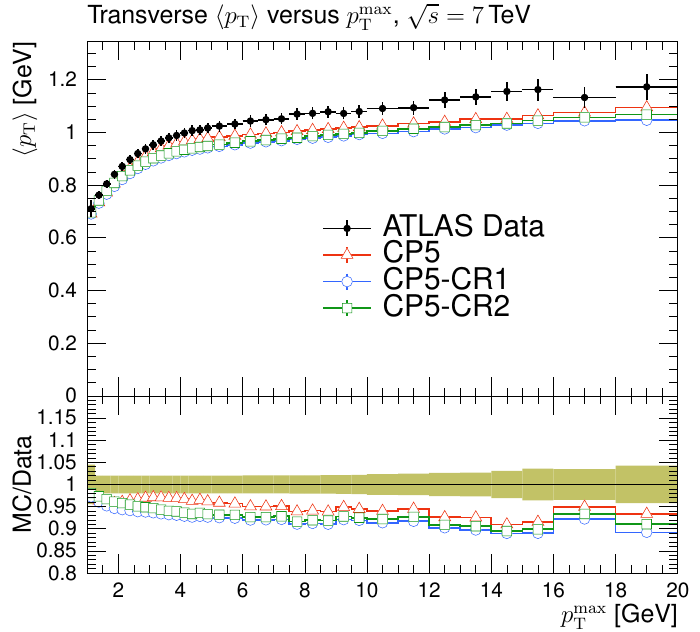}
    \includegraphics[width=0.49\textwidth]{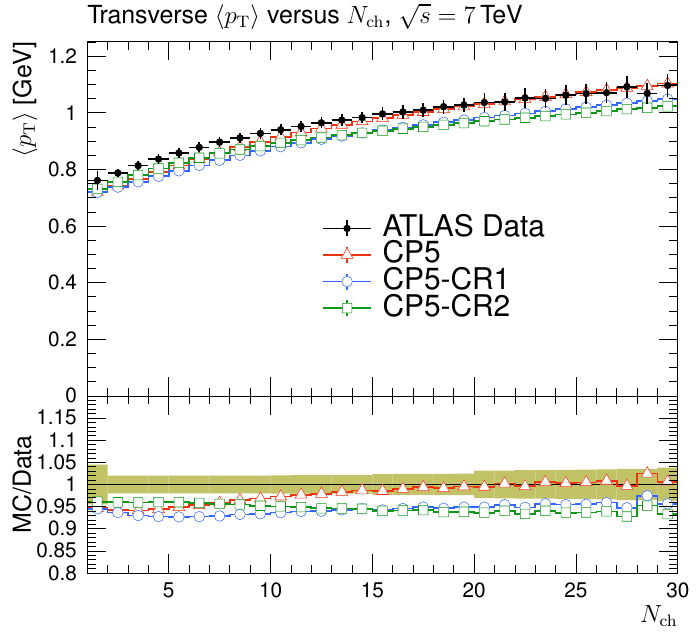}\\
    \caption{
		     The charged-particle (upper left) and \ptsum densities (upper right) in the transverse region, as functions of 
	         the \pt of the leading charged particle, and average transverse momentum in the transverse region as functions 
             of the leading charged particle \pt (lower left) and of the charged particle multiplicity (lower right), 
			 measured by the ATLAS experiment at $\sqrt{s}=7\TeV$~\cite{Aad:2010fh}.
             The predictions of the CP5 and CP5-CR tunes are compared with data.
             The coloured band and error bars on the data points represent the total experimental uncertainty in the data.
			 }
	\label{fig:ATLAS_2010_S8894728_CP5}
\end{figure*}

In Fig.~\ref{fig:ATLAS_2010_S8894728_CP5},
charged particle and \ptsum densities in the transverse region, as functions of \ptmax, 
and the average \pt in the transverse
region as functions of \ptmax and of the \Nch, measured by the ATLAS experiment at $\sqrt{s}=7\TeV$~\cite{Aad:2010fh}, are compared with the predictions
from the tunes CP5 and CP5-CR. The central values of the average \pt
in bins of the leading charged particle \pt and of the \Nch are
consistent with the data points within 10\%.
A similar level of agreement as observed at 13\TeV is achieved by the new tunes at 7\TeV.

\begin{figure}[!hbtp]
\centering
        \includegraphics[width=0.49\textwidth]{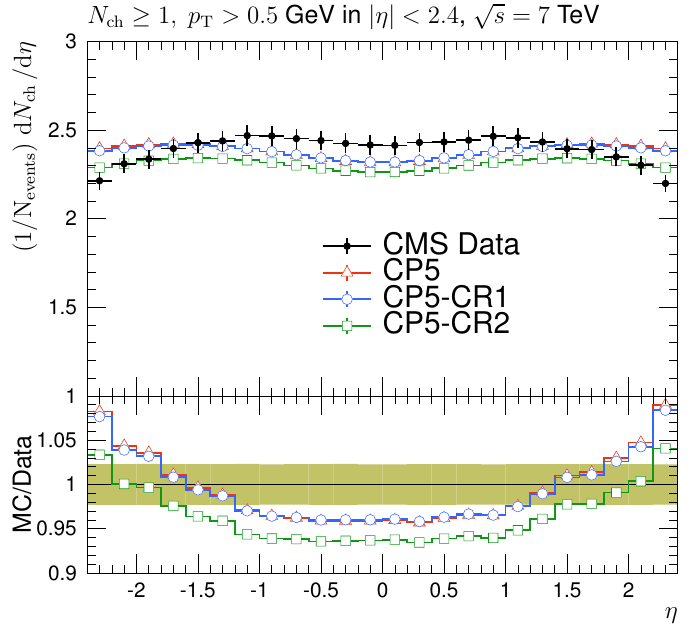}
    \caption{
			 The pseudorapidity of charged particles, \dNdeta, with at least one charged particle in $\abs{\eta}<2.4$, 
			 measured by the CMS experiment at $\sqrt{s}=7\TeV$~\cite{CMS:2011dsa}.
             The predictions of the CP5 and CP5-CR tunes are compared with data.
             The coloured band and error bars on the data points represent the total experimental uncertainty in the data.
			 }
    \label{fig:dNdeta7TeV}
\end{figure}

The performance of the new tunes is also checked at 7\TeV using
inclusive measurements of charged-particle pseudorapidity distributions.
In Fig.~\ref{fig:dNdeta7TeV}, the CMS measurements for \dNdeta at 7\TeV~\cite{CMS:2011dsa} with at least one charged 
particle in $\abs{\eta}<2.4$ are compared with predictions from the tunes CP5 and CP5-CR.
The CP5 and CP5-CR1 have similar predictions, while CP5-CR2 predicts about 4\% less charged 
particles than the first two tunes in all $\eta$ bins of the measurement. Although all tunes provide a reasonable 
description of \dNdeta with deviations up to $\approx$10\%, the data and MC simulation show different trends 
for $\abs{\eta}>1.2$, where the trend for the data is not described well by the tunes.
In the more central region, \ie $\abs{\eta}<1.2$, the shape of the predictions agrees well with the 
data but there is a difference in normalisation. For example, CP5 and CP5-CR1 predict 3--4\% and 
CP5-CR2 predicts about 7\% fewer charged particles in all bins for $\abs{\eta}<1.2$ compared with the data.

\begin{figure*}[!hbtp]
\centering
    \includegraphics[width=0.49\textwidth]{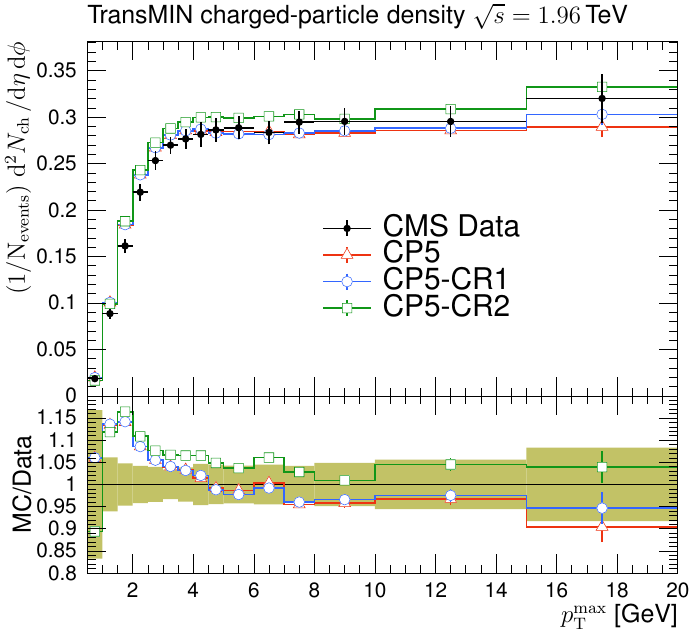}
    \includegraphics[width=0.49\textwidth]{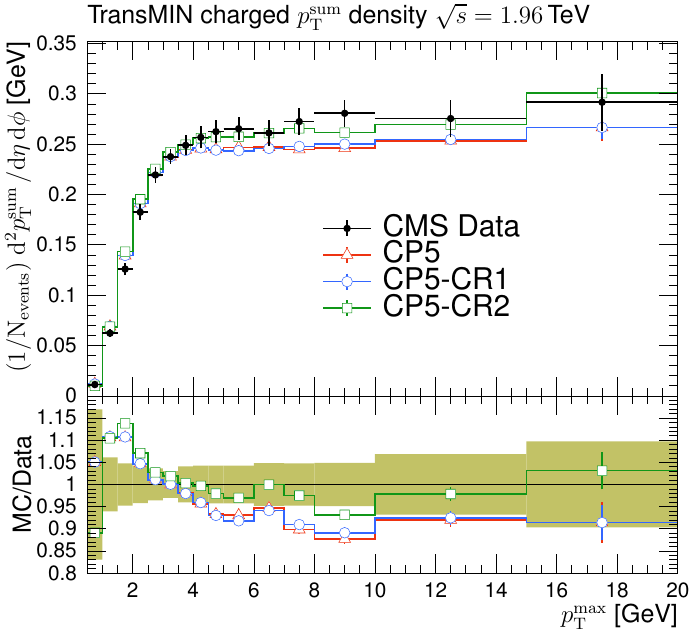}\\
    \includegraphics[width=0.49\textwidth]{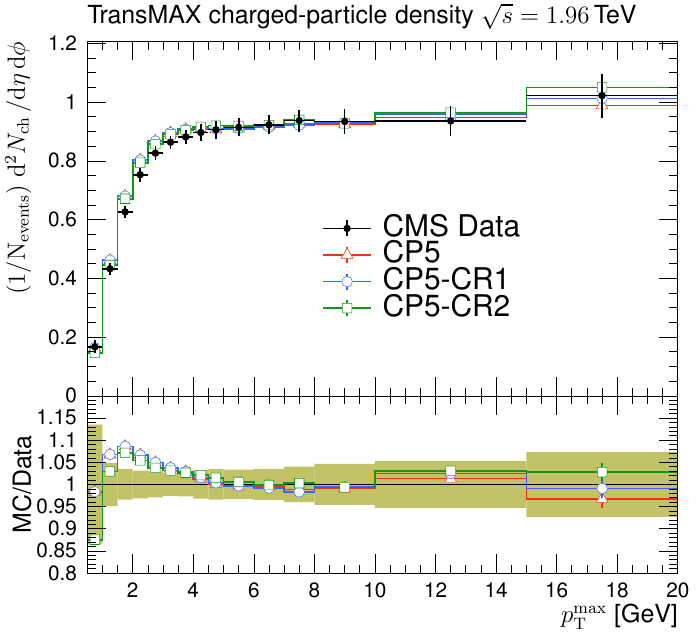}
    \includegraphics[width=0.49\textwidth]{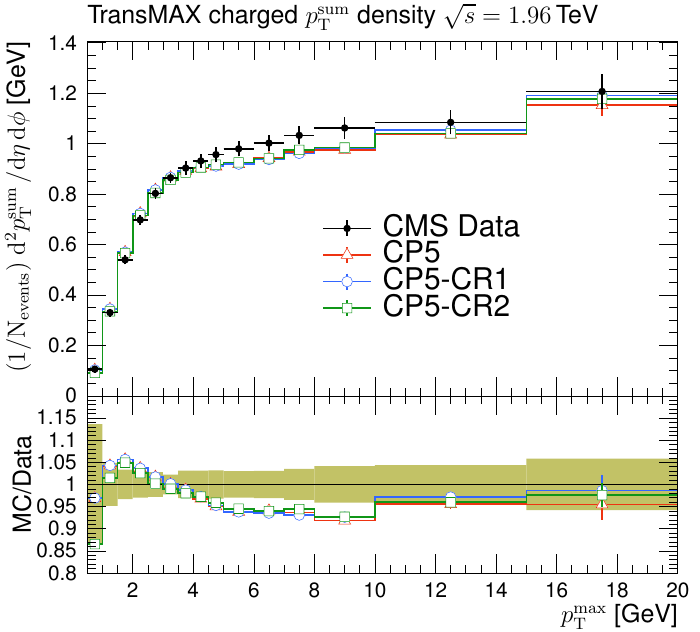}\\
  \caption{
		   The charged-particle (left) and \ptsum densities (right) in the \tmin (upper) and \tmax (lower) regions, 
		   as functions of the \pt of the leading charged particle, \ptmax, measured by the CDF experiment at $\sqrt{s}=1.96\TeV$~\cite{Aaltonen:2015aoa}.
           The predictions of the CP5 and CP5-CR tunes are compared with data.
           The coloured band and error bars on the data points represent the total experimental uncertainty in the data.}
    \label{fig:CDF_1d96TeV_CP5}
\end{figure*}

In Fig.~\ref{fig:CDF_1d96TeV_CP5},
charged-particle and \ptsum densities measured as functions of \ptmax at $\sqrt{s}=1.96\TeV$
by the CDF experiment~\cite{Aaltonen:2015aoa}
in the \tmin and \tmax regions are compared with predictions
from the tunes CP5 and CP5-CR, respectively.
All predictions reproduce the UE observables within $\approx$10\% at $\sqrt{s}=1.96$, 7, and 13\TeV.

\begin{figure}[!hbtp]
\centering
        \includegraphics[width=0.49\textwidth]{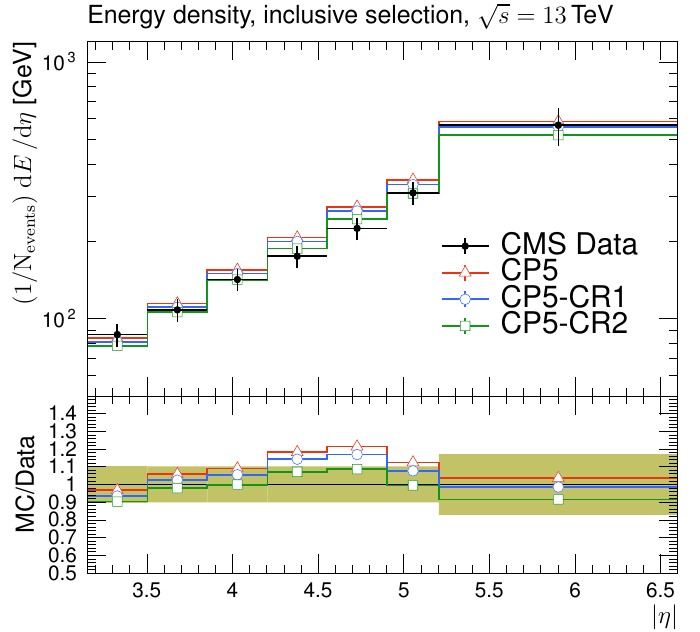}
        \includegraphics[width=0.49\textwidth]{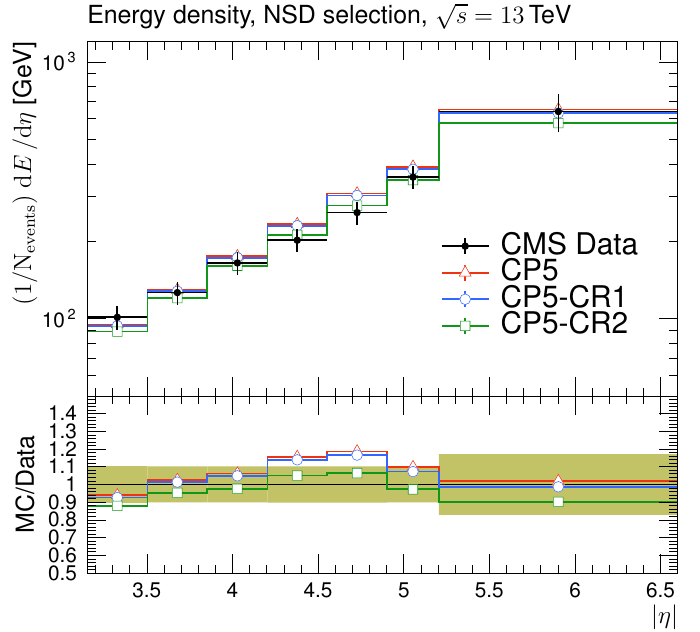}
    \caption{
		     The energy density as a function of pseudorapidity, in two different selections, in MB events (\cmsLeft) and
			 in events with a presence of a hard dijet system (\cmsRight), measured by the CMS experiment at $\sqrt{s}=13\TeV$~\cite{Sirunyan:2018jwn}.
             The predictions of the CP5 and CP5-CR tunes are compared with data.
             The coloured band and error bars on the data points represent the total experimental uncertainty in the data.
			 }
    \label{fig:FWDregion13TeV}
\end{figure}

We compare the new CMS tunes also with MB and UE data
measured at forward pseudorapidities.
The energy density, \dEdeta, measured in MB events and in NSD events
by the CMS experiment at $\sqrt{s}=13\TeV$, is shown in Fig.~\ref{fig:FWDregion13TeV}.
The data are well described by CP5-CR2 within uncertainties and for all measured $\abs{\eta}$ bins. 
The predictions of CP5 and CP5-CR1 overestimate the data in $4.2<\abs{\eta}<4.9$.

\begin{figure}[!hbtp]
\centering
    \includegraphics[width=0.49\textwidth]{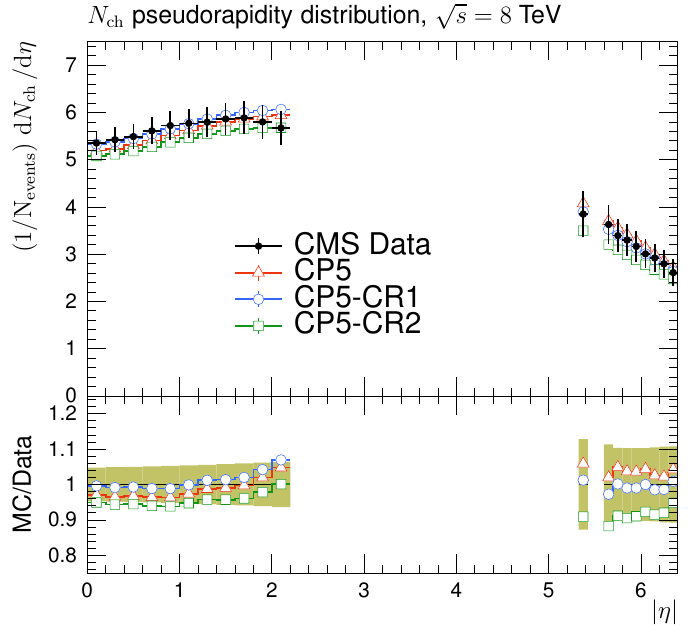}
    \includegraphics[width=0.49\textwidth]{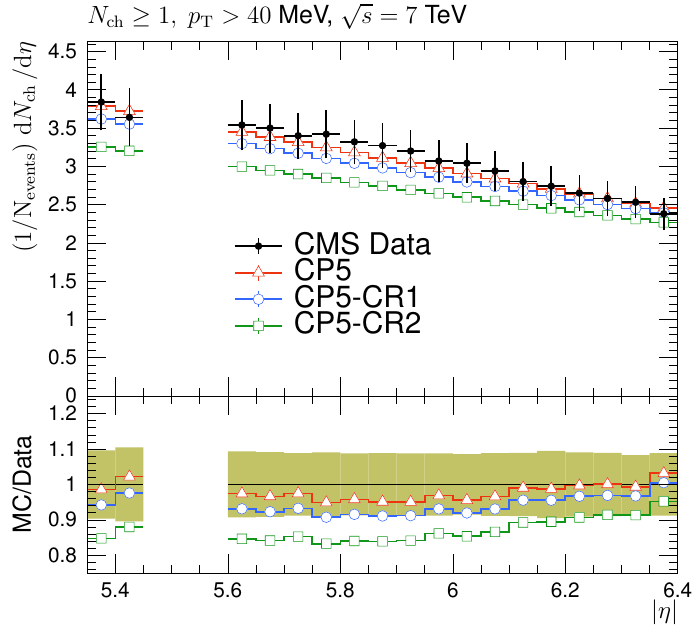}
  \caption{
           The pseudorapidity of charged particles, \dNdeta, measured by the CMS and TOTEM collaborations at 
           $\sqrt{s}=8\TeV$~\cite{CMS:2014kix} (\cmsLeft) and measured by the TOTEM collaboration at $\sqrt{s}=7\TeV$~\cite{TOTEM:2012kvo} 
           (\cmsRight). The predictions of the CP5 and CP5-CR tunes are compared with data. The coloured band and error bars on the data 
           points represent the total experimental uncertainty in the data. For the CMS-TOTEM measurement, at least one charged 
           particle with $\pt>0$ is required in $5.3<\eta<6.5$ or $-6.5<\eta<-5.3$. For the TOTEM measurement, at least one charged 
           particle with $\pt>40\MeV$ is required in $5.3<\abs{\eta}<6.4$.
		  }
    \label{fig:cmstotem}
\end{figure}

The pseudorapidity of charged particles, \dNdeta, in the ranges $\abs{\eta}<2.2$ and $5.3<\abs{\eta}<6.4$ measured by 
the CMS and TOTEM experiments at $\sqrt{s}=8\TeV$~\cite{CMS:2014kix} is presented in Fig.~\ref{fig:cmstotem}. 
The events are required to have at least one charged particle in $5.3<\eta<6.5$ or $-6.5<\eta<-5.3$ with $\pt>0$. 
All tunes describe the data within the uncertainties. Additionally, Fig.\ref{fig:cmstotem} shows the pseudorapidity of 
charged particles, \dNdeta, in $5.3<\abs{\eta}<6.4$ in events with at least one charged particle with $\pt>40\MeV$, measured 
by the TOTEM experiment at $\sqrt{s}=7\TeV$~\cite{TOTEM:2012kvo}. Both CP5 and CP5-CR1 describe the data within the uncertainties, whereas
CP5-CR2 underestimates the data by 15\%. 

\clearpage

\subsection{Particle multiplicities}
\label{sec:Par_Mul}
\begin{figure}[!hbtp]
\centering
         \includegraphics[width=0.49\textwidth]{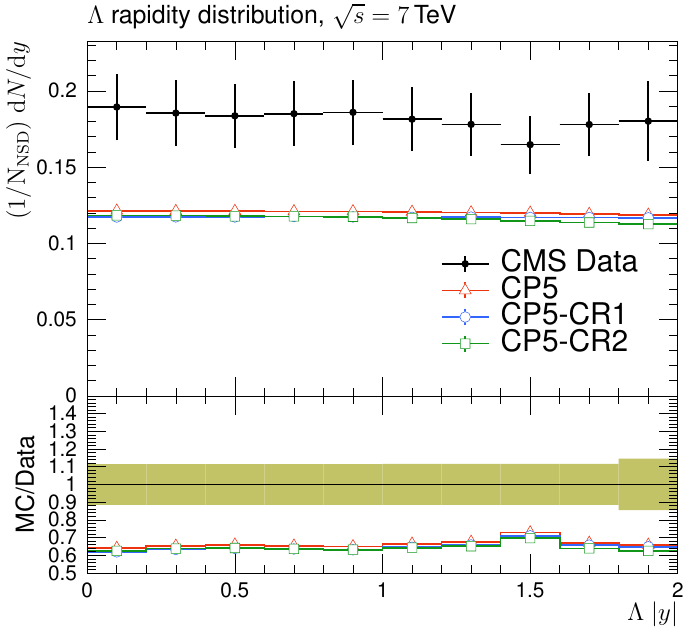}
         \includegraphics[width=0.49\textwidth]{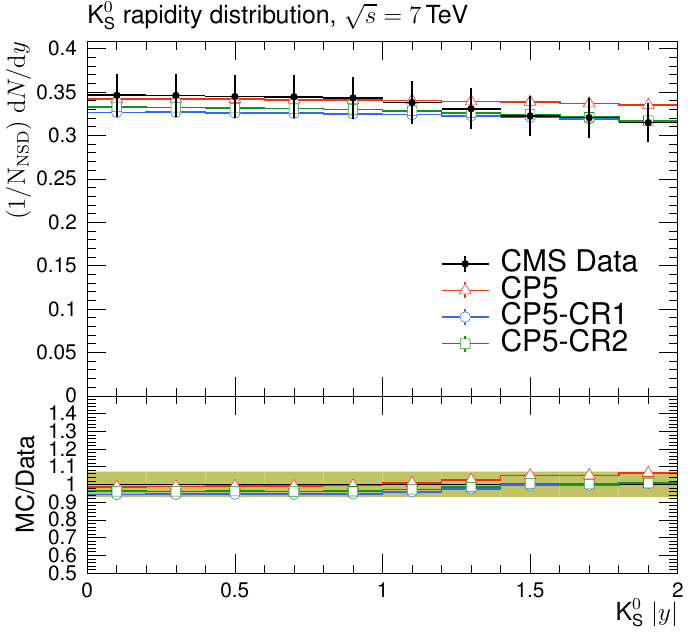}
    \caption{
			 The strange particle production, $\Lambda$ baryons (\cmsLeft) and \PKzS mesons (\cmsRight), as a function of rapidity, 
			 measured by the CMS experiment at $\sqrt{s}=7\TeV$~\cite{Khachatryan:2011tm}.
             The predictions of the CP5 and CP5-CR tunes are compared with data.
             The coloured band and error bars on the data points represent the total experimental uncertainty in the data.
			 }
    \label{fig:strangeness}
\end{figure}

Figure~\ref{fig:strangeness} shows the strange particle production
for $\Lambda$ baryons and \PKzS mesons as a function of rapidity ($y$)
measured by the CMS experiment~\cite{Khachatryan:2011tm} in NSD events at $\sqrt{s}=7\TeV$.
The rapidity is defined as $y=\frac{1}{2}\ln{\frac{E+p_L}{E-p_L}}$, where $E$ is the particle 
energy and $p_L$ is the particle momentum along the anticlockwise-beam direction.
It is shown in Ref.~\cite{Christiansen:2015yqa} that the new CR models
might be beneficial for describing the ratios of strange particle
multiplicities, for example $\Lambda /$\PKzS in pp collisions.
We observe that all CP5 tunes, regardless of the CR model,
describe particle production for \PKzS mesons as a function of rapidity very well. However, they underestimate
particle production for $\Lambda$ versus rapidity by about 30\%. Therefore, 
the ratio $\Lambda /$\PKzS is not perfectly described but this 
could be improved by different hadronisation models~\cite{Bierlich:2016vgw,Bierlich:2017sxk}.
Including these observables, as well as the recent measurements of baryon production from the ALICE and LHCb experiments~\cite{ALICE:2021rzj,LHCb:2019fns}, could be beneficial in future tune derivations. 
This is discussed in Appendix~\ref{sec:junctionCorrection}.

\begin{figure}[tbp]
\centering
    \includegraphics[width=0.49\textwidth]{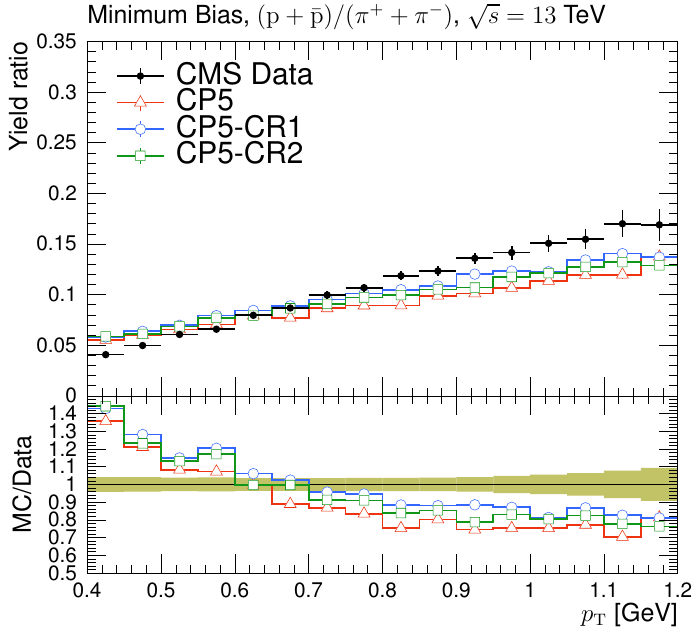}
    \caption{
			 Ratios of particle yields, $\text{p}/\pi$, as a function of transverse momentum in mininum-bias 
			 events, measured by the CMS experiment at $\sqrt{s}=13\TeV$~\cite{Sirunyan:2017zmn}.
			 The predictions of the CP5 and CP5-CR tunes are compared with CMS data.
             The coloured band and error bars on the data points represent the total experimental uncertainty in the data.
			}
    \label{fig:CMS_2017_I1608166}
\end{figure}

The multiplicities of identified particles are also investigated in simulated MB events \sloppy(ND+SD+DD+CD).
Figure~\ref{fig:CMS_2017_I1608166} shows the ratio of proton over pion production, as a function of particle \pt~\cite{Sirunyan:2017zmn}.
All the tunes predict a similar trend, showing that the new CR models do not lead to a significant improvement 
in the description of the ratio of proton to pion production. However, it is known that this observable is strongly correlated 
with event particle multiplicity~\cite{CMS:2012xvn,CMS:2013pdl,Sirunyan:2017zmn} 
and not only CR, since also hadronisation and MPI play a key role in describing the ratios of particle yields. 

\subsection{Jet substructure observables}
\begin{figure}[!hbtp]
\centering
    \includegraphics[width=0.49\textwidth]{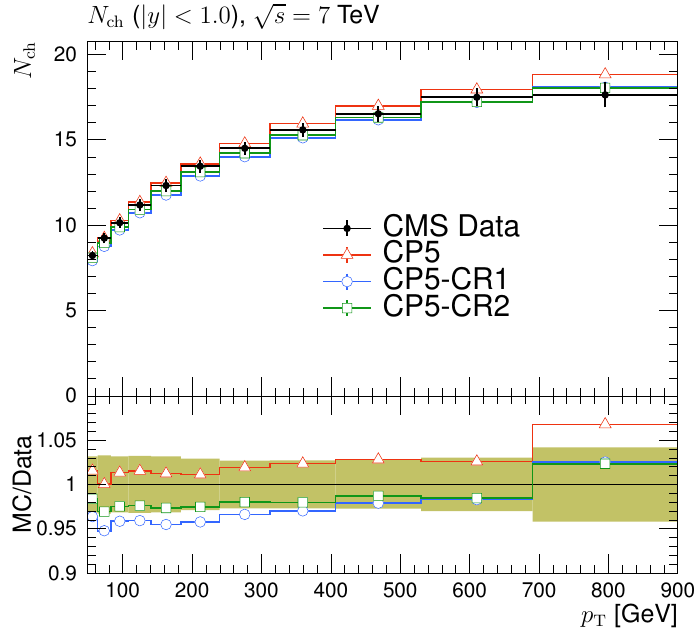}
  \caption{
		   Average charged-hadron multiplicity, as a function of the jet \pt, for jets with rapidity $\abs{y}<1$,
		   measured by the CMS experiment at $\sqrt{s}=7\TeV$~\cite{Chatrchyan:2012mec}.
           The predictions of the CP5 and CP5-CR tunes are compared with data.
           The coloured band and error bars on the data points represent the total experimental uncertainty in the data.
		  }
    \label{fig:ch_jetsSub}
\end{figure}

The number of charged particles contained in jets is an important observable that makes
it possible to distinguish quark-initiated jets from gluon-initiated jets.
The average number of charged hadrons with $\pt>500$\MeV inside the jets
measured by the CMS experiment as a function of the jet \pt is shown in Fig.~\ref{fig:ch_jetsSub}~\cite{Chatrchyan:2012mec}.
The predictions of the CR tunes are comparable, and produce roughly 5\% fewer charged particles
than the CP5 tune. All predictions show a reasonable description of the data.

\begin{figure}[!hbtp]
\centering
    \includegraphics[width=0.49\textwidth]{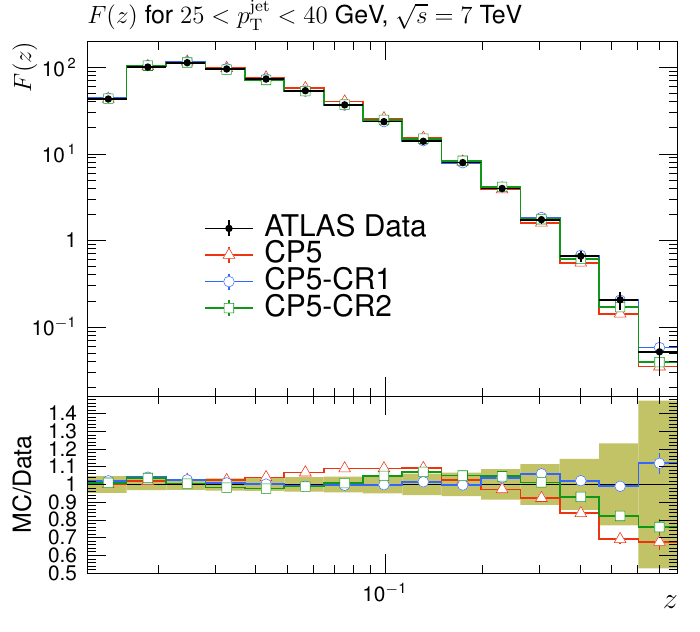}
    \includegraphics[width=0.49\textwidth]{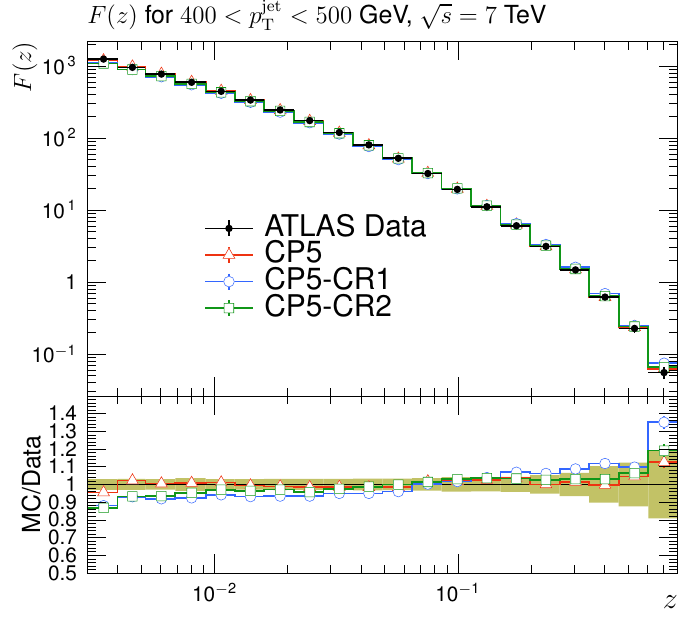}
  \caption{
		   Distributions of $F(z)$ for $25<\ptjet<40\GeV$ (\cmsLeft) 
           and $400<\ptjet<500\GeV$ (\cmsRight)
		   for jets with pseudorapidity $\abs{\eta_\text{jet}}<1.2$, measured by the ATLAS experiment at $\sqrt{s}=7\TeV$~\cite{Aad:2011sc}.
           The predictions of the CP5 and CP5-CR tunes are compared with data.
           The coloured band and error bars on the data points represent the total experimental uncertainty in the data.
		  }
    \label{fig:frag_jetsSub}
\end{figure}

Figure~\ref{fig:frag_jetsSub} presents the distributions of $F(z)=(1/N_\text{jet})(\ddinline{\Nch}{z})$,
where $z$ is the longitudinal momentum fraction, and \Nch is the charged-particle multiplicity in the jet,
measured by the ATLAS experiment at $\sqrt{s}=7\TeV$~\cite{Aad:2011sc}.
The $F(z)$ parameter is related to the fragmentation function
and is presented for $\ptjet=25\text{--}40$\GeV and $\ptjet=400\text{--}500$\GeV.
The CR tunes describe low-\ptjet data better than CP5, and their predictions reasonably agree
with the high-\ptjet data, except for the last bin. The high-\ptjet data
are well described by the CP5 tune within the uncertainties, and its central values agree better with the predictions of the CP5 tune than with those of the CP5-CR tunes.

\subsection{Drell--Yan events}
\begin{figure}[tbp]
\centering
    \includegraphics[width=0.49\textwidth]{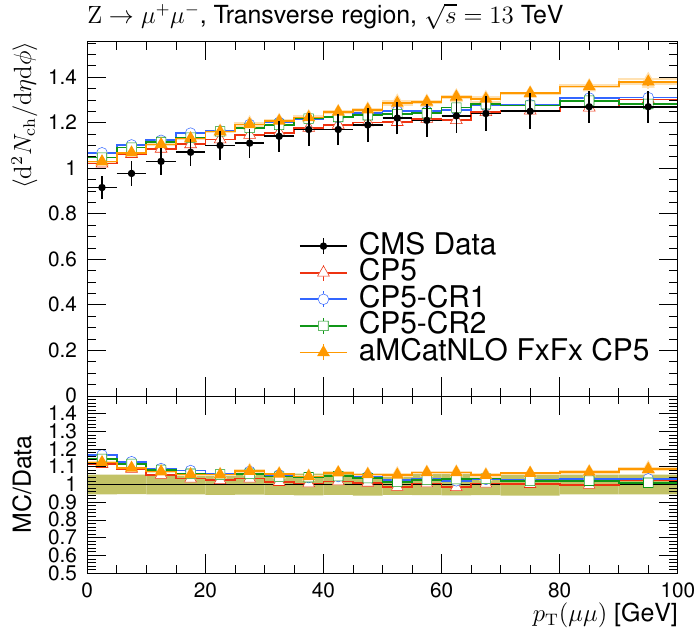}
    \includegraphics[width=0.49\textwidth]{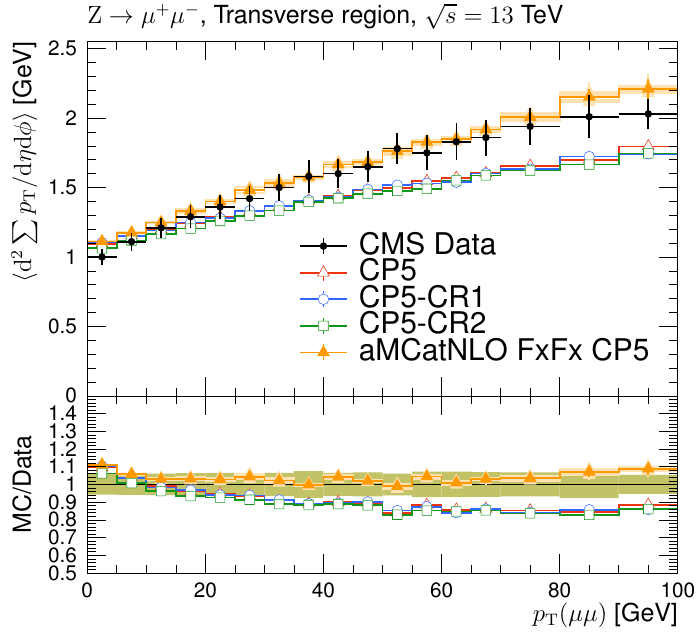}
    \caption{
			 Number of charged particles and \pt flow in the transverse region of DY events, measured by the CMS experiment at $\sqrt{s}=13\TeV$
             in bins of {\PZ} boson \pt~\cite{Sirunyan:2017vio}.
             The plots show the predictions of \PYTHIAviii with the CP5 and CP5-CR tunes, as well as \MGvATNLO with the CP5 tune  compared with data.
             The coloured band and error bars on the data points represent the total experimental uncertainty in the data.
            }
    \label{fig:dyuetransverse}
\end{figure}

Drell--Yan (DY) events~\cite{Drell:1970wh,Christenson:1970um} with the {\PZ} boson decaying to $\PGmp\PGmm$ were generated with
\PYTHIAviii and compared with CMS data at $\sqrt{s}=13\TeV$.
Figure~\ref{fig:dyuetransverse} shows the \Nch and \pt flow as a function
of the {\PZ} boson \pt (in the invariant $\PGmp\PGmm$ mass window of 81--101\GeV) in the region transverse to the boson momentum~\cite{Sirunyan:2017vio},
which is expected to be dominated by the UE.

The CP5 tunes predict up to 15\% too many charged particles at low {\PZ} boson \pt, where additional 
effects, such as the intrinsic transverse momentum of the interacting partons (\ie primordial \kt) 
are expected to play a role.
Higher-order corrections, as implemented in \MGvATNLO v2.4.2~\cite{Alwall:2014hca} with FxFx merging~\cite{Frederix:2012ps}, are necessary to describe the total \pt flow.
The impact of the different CR models is negligible in DY events.

\subsection{Top quark observables}
\label{sec:topquark}
\subsubsection{Jet substructure in \ttbar events}
\label{sec:ttbar_jetshapes}
\begin{figure}[!hbtp]
\centering
    \includegraphics[width=0.49\textwidth]{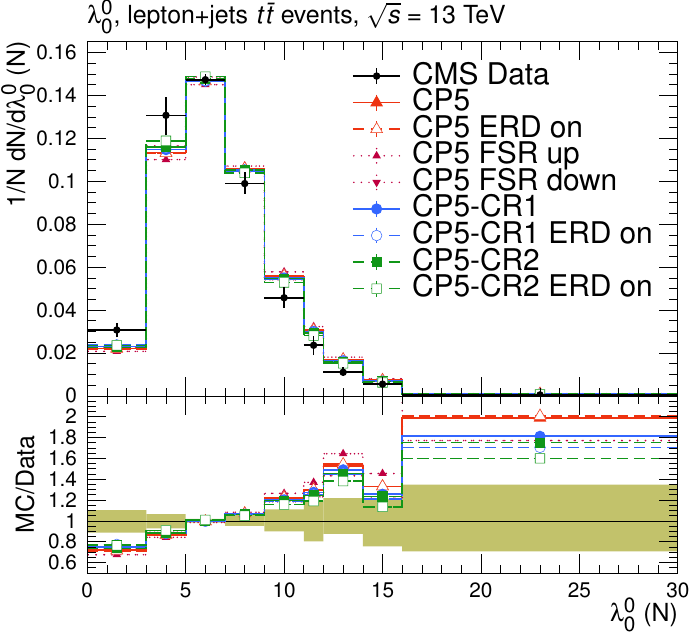}
    \includegraphics[width=0.49\textwidth]{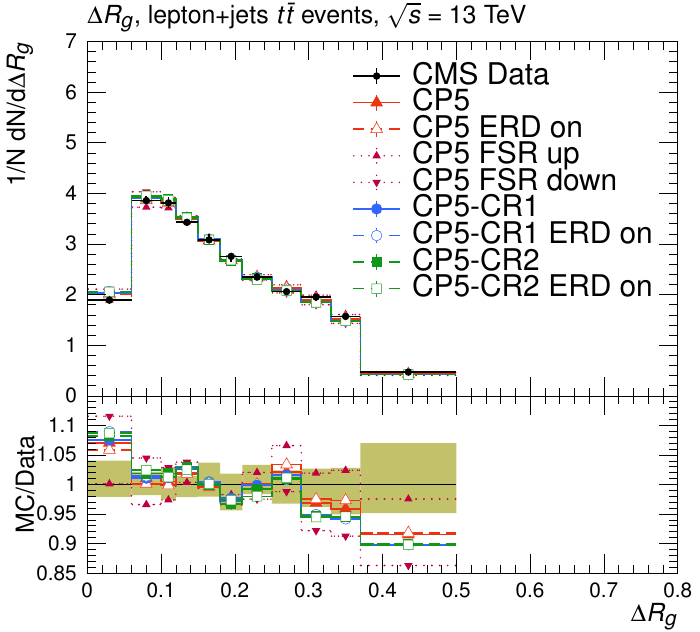}\\
    \caption{Distributions of the particle multiplicity in gluon jets (\cmsLeft) and the angle $\Delta R_g$ between two groomed subjets in inclusive jets (\cmsRight) measured by the CMS experiment in $\ttbar$ events at $\sqrt{s}=13\TeV$~\cite{Sirunyan:2018asm}.
        The coloured band and error bars on the data points represent the total experimental uncertainty in the data.}
    \label{fig:ttbar_jetshapes}
\end{figure}

A study of the UE in \ttbar events~\cite{Sirunyan:2018avv} also estimated the effects of the CR on the top quark decay products
by investigating the differences between predictions
using \PYTHIAviii with the ERD off and on options. In Ref.~~\cite{Sirunyan:2018avv}, in addition to the QCD-inspired and gluon-move models, 
predictions of the rope hadronisation model~\cite{Bierlich:2014xba,Bierlich:2015rha} are also compared with the data. 
In the rope hadronisation model, overlapping strings are treated to act coherently as a ``rope''.
The interactions between overlapping strings are described by an interaction potential inspired by the
phenomenology of superconductors~\cite{Bierlich:2016vgw,Bierlich:2017sxk,Bierlich:2017vhg,Bierlich:2015rha,Bierlich:2014xba,Abramovsky:1988zh,Altsybeev:2015vma,Cea:2014uja}. 
The ERD off and on options allow the CR to take place before or after the top quark decay, respectively. 
In particular, the ERD option allows the top
quark decay products to be colour reconnected with the partons from MPI systems. 
Ref.~\cite{Sirunyan:2018avv} showed that these different models and options produce similar predictions for UE observables in \ttbar events.  
However, some jet-shape
distributions in \ttbar events display a more significant effect~\cite{Sirunyan:2018asm}, \eg in the number of charged particles in jets.
In the following, we investigate how the \PYTHIAviii CR tunes
describe the CMS \ttbar jet substructure data~\cite{Sirunyan:2018asm}.
In the CMS measurement, jets reconstructed using the anti-\kt algorithm~\cite{Cacciari:2008gp}
with a distance parameter of $R=0.4$ as implemented in \textsc{FASTJET} 3.1~\cite{Cacciari:2011ma} are used.
Jets with $\pt>30\GeV$ within $\abs{\eta}<2$ are selected.
Jet pairs ($\mathrm{j_1}$ and $\mathrm{j_2}$) are required to be far from each other in $\eta$-$\phi$ space,
$\Delta R(\mathrm{j_1},\mathrm{j_2}) = \sqrt{\smash[b]{(\eta_\mathrm{j_1}-\eta_\mathrm{j_2})^2 - (\phi_\mathrm{j_1}-\phi\mathrm{j_2})^2}}>0.8$.
Jet substructure observables are calculated from 
jet constituents with $\pt>1\GeV$, \eg in the plateau region of high track finding efficiency and low misidentification rate. 
Here we focus on two variables, (i)
$\lambda_0^0(N)$, which is the number of charged particles with $\pt > 1\GeV$ in the jet, and (ii) 
the separation between two groomed
subjets, $\Delta R_g$, that are shown in Fig.~\ref{fig:ttbar_jetshapes} for gluon jets and inclusive jets, respectively.
A ``groomed jet'' refers to a jet with soft and wide-angle radiation removed by a dedicated grooming algorithm~\cite{Dasgupta:2013ihk,Larkoski:2014wba}. 

The compatibility of data and MC predictions is evaluated using  a measure defined as $\chi^2 = \Delta^T C^{-1} \Delta$, where $\Delta$ is the difference vector between measured and predicted values, and $C$ is the total covariance matrix of the measurement.
Since the measured distribution is normalised to unity, its covariance matrix is singular, \ie not invertible.
To render $C$ invertible, the vector entry and matrix row/column corresponding to one measured bin need to be discarded; we choose to remove the last bin.
The results are displayed in Table~\ref{tab:jetshape_chi2} for all jets inclusively as well as for each jet flavour separately.
We observe that none of the tunes describe the $\lambda_0^0(N)$ data well for all jet flavours.
As concluded in Ref.~\cite{Sirunyan:2018asm}, flavour-dependent improvements in the nonperturbative physics modelling may be required for a better description of the data.
The angle between the groomed subjets, on the other hand, is infrared and collinear safe and can be described very well by an increase in the $\alpS^\mathrm{FSR}(m_\PZ)$, 
which corresponds to a decrease in the FSR renormalisation scale $\mu_R^\mathrm{FSR}$. 
Table~\ref{tab:jetshape_chi2} shows the results obtained by varying $\mu_R^\mathrm{FSR}$ by factors of 0.5 and 2.

\begin{table*}[!hbtp]
    \centering
    \topcaption{The $\chi^2$ values and the numbers of degrees of freedom (\ndof) for the comparison of \ttbar data with the predictions of the different \PYTHIAviii tunes, for the distributions of the charged-particle multiplicity $\lambda_0^0$, the angle between the groomed subjets $\Delta R_g$ at $\sqrt{s}$ = 13\TeV~\cite{Sirunyan:2018asm}, and the pull angle measured in the ATLAS analysis of the colour flow at 8\TeV~\cite{Aad:2015lxa}.
    The FSR up and down entries denote variations of the renormalisation scale in the $\alpS^\mathrm{FSR}(m_\PZ)$ by factors of 0.5 and 2, respectively.
    }
    \label{tab:jetshape_chi2}
    \cmsTable{
    \begin{tabular}{lrrrrrrrrrrrr}
    & \multicolumn{5}{l}{Charged-particle multiplicity $\lambda_0^0$} & \multicolumn{5}{l}{Angle between groomed subjets $\Delta R_g$} & Pull angle & \hspace{-0.75em}$\phi(j_1,j_2)$ \\
    & \multicolumn{4}{c}{$\chi^2$} & \ndof & \multicolumn{4}{c}{$\chi^2$} & \ndof & $\chi^2$ & \ndof \\
    Tune & Incl. & Bottom & Light & Gluon &  & Incl. & Bottom & Light & Gluon &  & & \\
    \hline
    CP5 & 18.4 & 26.6 & 30.7 & 11.8 & 8 & 28.2 & 18.3 & 10.6 & 8.1 & 10 & 4.7 & 3 \\
    CP5 ERD & 19.6 & 28.7 & 32.2 & 12.2 & 8 & 26.9 & 15.0 & 10.7 & 8.7 & 10 & 2.4 & 3 \\  [\cmsTabSkip]
    CP5 FSR up & 28.4 & 43.7 & 33.0 & 14.6 & 8 & 13.3 & 4.2 & 5.8 & 5.7 & 10 & 5.9 & 3 \\
    CP5 FSR down & 15.0 & 19.7 & 44.0 & 11.6 & 8 & 59.6 & 39.2 & 33.1 & 22.6 & 10 & 4.1 & 3 \\  [\cmsTabSkip]
    CP5-CR1 & 14.3 & 28.4 & 29.5 & 4.1 & 8 & 34.6 & 13.4 & 24.4 & 23.6 & 10 & 3.9 & 3 \\
    CP5-CR1 ERD & 11.7 & 24.4 & 27.8 & 3.8 & 8 & 32.7 & 13.4 & 21.1 & 27.0 & 10 & 1.4 & 3 \\  [\cmsTabSkip]
    CP5-CR2 & 14.1 & 23.8 & 38.3 & 8.1 & 8 & 34.3 & 22.3 & 21.3 & 11.7 & 10 & 5.2 & 3 \\
    CP5-CR2 ERD & 11.0 & 16.9 & 38.6 & 7.1 & 8 & 35.3 & 24.8 & 16.1 & 13.1 & 10 & 9.3 & 3 \\
    \end{tabular}
    }
\end{table*}

\subsubsection{Pull angle in \texorpdfstring{\ttbar}{ttbar} events}
\begin{figure}[!hbtp]
\centering
    \includegraphics[width=0.49\textwidth]{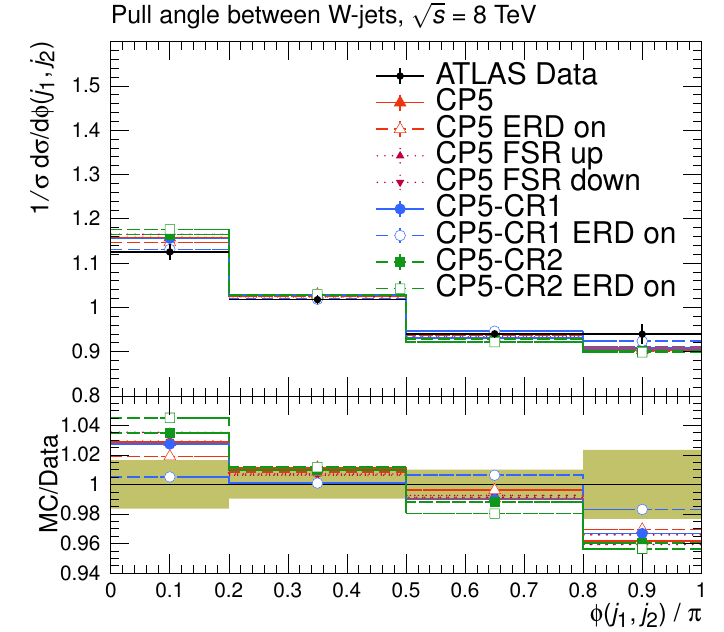}
    \caption{Normalised \ttbar differential cross section for the pull angle between jets from the {\PW} boson in top quark decays, 
        calculated from the charged constituents of the jets, measured by
        the ATLAS experiment using $\sqrt{s}=8\TeV$ data~\cite{Aad:2015lxa} to investigate colour flow.
        The coloured band and error bars on the data points represent the total experimental uncertainty in the data.
        }
    \label{fig:colorflow}
\end{figure}

Figure~\ref{fig:colorflow} displays the normalised \ttbar differential cross section for the jet pull angle~\cite{Gallicchio:2010sw} 
defined using the jets originating from the
decay of a {\PW} boson in \ttbar events, as measured by
the ATLAS experiment~\cite{Aad:2015lxa}.  The observable is shown for the case where only 
the charged constituents of the jet are used in the calculation.
The data are compared with predictions from POWHEG v2~\cite{Frixione:2007nw}+\PYTHIAviii using the CP5 tunes or the corresponding CR tunes.
The $\chi^2$ values are calculated as described in Section~\ref{sec:ttbar_jetshapes} and are shown in the last column of Table~\ref{tab:jetshape_chi2}.
The pull angle is particularly sensitive to the
setting of the ERD option. With ERD turned off, the decay products of the W
boson in \ttbar events are not included in CR, and the predictions
using the tunes with the various CR models are similar to each other.
With ERD enabled, CR now modifies the pull angle between the two
jets, which is observed in Fig.~\ref{fig:colorflow}. The predictions of each tune
also show significant changes when ERD is enabled.
For both the nominal and CP5-CR1 (QCD-inspired) tunes, the prediction with ERD 
improves the description of the data, and the difference between the predictions with
or without ERD is larger for the CP5-CR1-based tune.
We observe the opposite for the CP5-CR2-based (gluon-move)
tunes, for which the choice without ERD is preferable.  
This picture might be different if the flip mechanism had been added in the tuning of 
the gluon-move model. The move step in the gluon-move model is more restrictive because 
it allows only gluons to move between the string end-points. The inclusion of the flip 
mechanism would also allow the string end-points to be mixed with each other and, therefore, 
could further reduce the total string length in an event. 
However, as indicated earlier, the effect of the flip mechanism on diffractive events 
is not well understood and, therefore, this mechanism is not used in this paper.

Overall, the QCD-inspired model
with ERD provides the best description of the jet pull angle.
The differences between the predictions using the different tunes observed here
indicate that the inclusion of observables, such as the jet pull angle and other jet
substructure observables, could be beneficial in future tune derivations.

\section{Uncertainty in the top quark mass due to colour reconnection}

The top quark mass has been measured with high precision
using the 7, 8, and 13\TeV~\ttbar data at the LHC~\cite{ATLAS:2014wva,Aad:2014zea,Aad:2015nba,Aaboud:2016igd,Aaboud:2017mae,Aaboud:2018zbu,Khachatryan:2015hba,Chatrchyan:2012cz,Chatrchyan:2012ea,Chatrchyan:2013xza,Khachatryan:2015hba,CMS:2017mpr,Sirunyan:2018gqx,Sirunyan:2018goh,Sirunyan:2018mlv}.
The most precise value of $\topmass=172.44\pm 0.13\stat \pm 0.47\syst$\GeV
was measured by the CMS Collaboration combining 7 and 8\TeV data~\cite{Khachatryan:2015hba}.
To further improve the precision of \topmass measurements,
a complete analysis of the systematic uncertainties in the measurement is crucial.
One of the dominant systematic uncertainties is due to the modelling of 
CR in top quark decays~\cite{Khachatryan:2015hba}.
The procedure for estimating this uncertainty used for the LHC Run\,1 (years 2009--2013) 
analyses at $\sqrt{s}=7$ and 8\TeV was based
on a comparison of two values of \topmass, calculated by using predictions
with the same UE tune with and without CR effects. 
In Ref.~\cite{Khachatryan:2015hba}, this is done using the tune ``Perugia 2011'' with and 
without CR effects included. The ``Perugia 2011'' tunes family is the updated version of the 
``Perugia (Tevatron)'' tunes family and also takes into account lessons learned from LHC MB 
and UE data at 0.9 and 7\TeV~\cite{Skands:2010ak}.
The new CMS tunes, presented
in Section~2, which use different CR models, can be used to give a better
evaluation of the CR uncertainty. In particular, the uncertainty is now
calculated by comparing results for \topmass values obtained from different realistic CR
models, such as CP5, CP5-CR1, and CP5-CR2.

Additionally, one can also estimate the effects of
the CR on the top quark decay products by investigating the differences between predictions
using \PYTHIAviii with the option ERD off and on, which was done for the UE observables~\cite{Sirunyan:2018avv}. 

A determination of \topmass using a kinematic reconstruction of the decay products in
semileptonic \ttbar events at $\sqrt{s}=13\TeV$ is reported in Ref.~\cite{Sirunyan:2018gqx}.
In these events, one of the {\PW} bosons from the top quark decays into a muon or electron and a neutrino, and the
other into a quark-antiquark pair.
In this analysis, \topmass and the jet energy scale factor were
determined simultaneously through a joint-likelihood fit to the selected events.
The results with the QCD-inspired and gluon-move models were also compared.
The \PYTHIAviii CUETP8M2T4~\cite{CMS:2016kle} UE tune was used, and the
parameters of the CR models were tuned to UE and MB data at $\sqrt{s}=13\TeV$~\cite{Sirunyan:2018gqx}.
They found that the gluon-move model results in a 0.31\GeV shift from the \topmass value obtained with the default simulation. 
This shift, which is larger than the shifts caused by the other CR models, is 
 assumed to be the uncertainty due to the modelling of CR in the measured \topmass. 
It is much larger than the shift, 0.01\GeV, due to the CR modelling in the Run 1 measurement~\cite{Khachatryan:2015hba}.
This is the largest source of uncertainty in the measured \topmass, where the total
uncertainty is 0.62\GeV~\cite{Sirunyan:2018gqx}. Similar studies using single top quark final states are reported 
in Refs.~\cite{CMS:2017mpr,CMS:2021jnp}.

\begin{figure}[!hbtp]
\centering
    \includegraphics[width=0.49\textwidth]{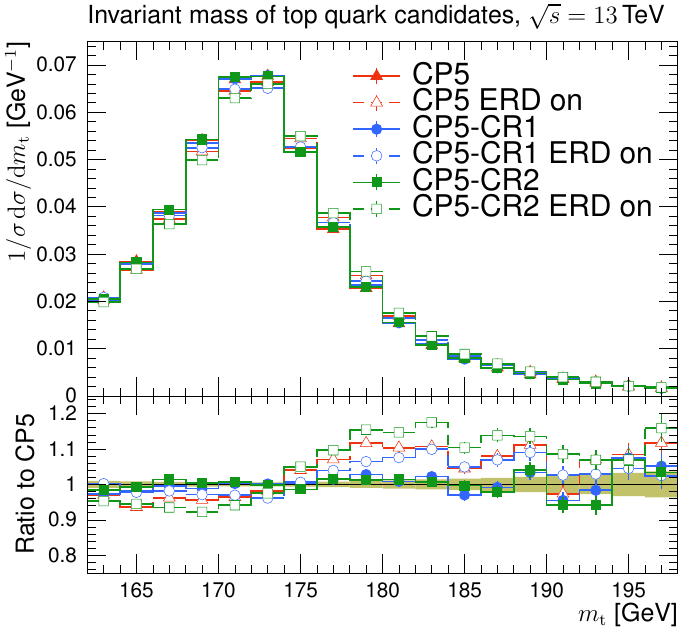}
    \caption{The invariant mass of hadronically decaying top quark candidates for different tune configurations.  The coloured band and vertical bars represent the statistical uncertainty in the predictions.
        }
    \label{fig:topmass}
\end{figure}

We compare the \topmass and {\PW} boson mass values obtained with different tune
configurations based on our new tunes in Table~\ref{tab:mtop_CR}.  Top quark
candidates are constructed by a \RIVET routine in a sample of simulated semileptonic \ttbar
events.  Events must contain exactly one lepton with $\pt>30\GeV$ and $\abs{\eta}<2.1$.  
Leptons are ``dressed'' with the surrounding photons within a cone of $\Delta R = 0.1$ 
and are required to yield an invariant mass window of 5\GeV centred at 80.4\GeV, when combined with a neutrino in the event.  
The events must also contain at least four jets, reconstructed with the anti-\kt algorithm, with $\pt>30\GeV$ within $\abs{\eta}<2.4$.  
At least two of the jets are required to originate from the fragmentation of bottom quarks, and at least two other jets, referred to as light-quark jets, must not originate from bottom quarks.
One jet originating from a bottom quark is combined with the lepton and neutrino to form a leptonically decaying top quark candidate, whereas the other jet originating from a bottom quark is combined with two other jets to form a hadronically decaying top quark candidate.  
The difference in the invariant mass window of the two top quark candidates is required to be less than 20\GeV, and the invariant mass of the two light-quark jets is within a window of 10\GeV centred at 80.4\GeV.  
If more than one combination of jets satisfy these criteria when combined with the lepton and neutrino, then only one combination is chosen based on how similar the invariant masses of the two top quark candidates are to each other and on how close the invariant mass of the light-quark jets is to $80.4\GeV$. The invariant mass of the hadronically decaying top quark candidates constructed in this way for each of the different tune configurations is shown in Fig.~\ref{fig:topmass}.
The top quark and {\PW} boson mass values are obtained from these hadronically decaying top quark candidates by fitting a Gaussian function within an 8\GeV mass window around the corresponding mass peak.
Table~\ref{tab:mtop_CR} also contains the differences from the nominal
\topmass and $\Wmass$ values ($\Delta \topmass$, and $\Delta \Wmass$) and the difference in $\Delta \topmass^\text{hyb}$,
a quantity that was introduced in Ref.~\cite{Khachatryan:2015hba} to incorporate both an in situ jet scale factor determined from the reconstructed \Wmass as well as prior knowledge about the jet energy scale in a hybrid approach to extract \topmass.
Here, $\Delta \topmass^\text{hyb}$ is approximated as $\Delta \topmass-0.5\Delta \Wmass$.
From Table~\ref{tab:mtop_CR}, we observe that the largest deviation from the
predictions of CP5 is CP5-CR2 ERD (0.32\GeV) similar to the largest shift found
in Ref.~\cite{Sirunyan:2018gqx} using CUETP8M2T4.
However, CP5-CR2 ERD is not able to describe the available colour flow data, and can therefore be excluded from the list of modelling uncertainties.

\begin{table*}[!hbt]
\centering
\topcaption{The top quark mass (\topmass) and {\PW} mass (\Wmass) extracted by a fit to the predictions of the different \PYTHIAviii tunes, 
along with the differences from the nominal \topmass value ($\Delta \topmass$), \Wmass value ($\Delta \Wmass$), and $\Delta \topmass^\text{hyb}$ 
which represents an estimation of the \topmass uncertainty considering the shift in \Wmass included with a weight of 0.5.
The uncertainties in the \topmass and \Wmass values correspond to the uncertainty in the fitted \topmass and \Wmass.
}
\label{tab:mtop_CR}
\cmsTable{
\begin{tabular}{llrlrr}
Tune & \topmass [\GeVns{}] & $\Delta \topmass$ [\GeVns{}] & \Wmass [\GeVns{}] & $\Delta \Wmass$ [\GeVns{}] & $\Delta \topmass^\text{hyb}$ [\GeVns{}] \\
\hline
CP5           & $171.93\pm0.02$ & \NA     & $79.76\pm0.02$ & \NA     & \NA     \\
CP5 ERD       & $172.18\pm0.03$ & $ 0.25$ & $80.15\pm0.02$ & $ 0.40$ & $ 0.05$ \\
CP5-CR1       & $171.97\pm0.02$ & $ 0.04$ & $79.74\pm0.02$ & $-0.02$ & $ 0.05$ \\
CP5-CR1 ERD   & $172.01\pm0.03$ & $ 0.08$ & $79.98\pm0.02$ & $ 0.23$ & $-0.04$ \\
CP5-CR2       & $171.91\pm0.02$ & $-0.02$ & $79.85\pm0.02$ & $ 0.10$ & $-0.07$ \\
CP5-CR2 ERD   & $172.32\pm0.03$ & $ 0.39$ & $79.90\pm0.02$ & $ 0.14$ & $ 0.32$ \\
\end{tabular}
}
\end{table*}

\section{Summary and conclusion}
New sets of parameters for two of the colour reconnection (CR) models implemented in the \PYTHIAviii event generator, QCD-inspired
and gluon-move, are obtained,
based on the default CMS \PYTHIAviii tune CP5.
Measurements sensitive to underlying-event (UE) contributions performed at hadron-colliders at $\sqrt{s}=1.96$, 7, and 13\TeV
are used to constrain the parameters for the CR and for the
multiple-parton interactions simultaneously.
Various measurements at 1.96, 7, 8, and 13\TeV are used
to evaluate the performance of the new tunes.
The central values predicted by the new CR tunes for the UE and minimum-bias events describe the data significantly better than the CR models with their default parameters before tuning.
The predictions of the new tunes achieve a reasonable agreement in many UE
observables, including the ones measured at forward pseudorapidities.
However, the models after tuning do not generally perform better 
than the CP5 tune for the observables presented in this study. 
Although the new CR tunes presented in this work are not intended to improve the description of 
the measurements of strange particle multiplicities for $\Lambda$ baryons and \PKzS mesons, 
we test the new tunes against them. We find that the new CR models, when tuned using only 
measurements that are sensitive to the UE, do not provide a better description of the distribution 
of strange particle production as a function of rapidity for $\Lambda$ baryons.
However, we observe that all CP5 tunes, irrespective of the CR model,
describe particle production for \PKzS as a function of rapidity well.
Including these observables in the fits, along with the latest measurements of baryon/meson production, could be beneficial for future tune derivations. 

The predictions of the new tunes for jet shapes and
colour flow measurements done with top quark pair events are also compared with data.
All tunes give similar predictions, but none of the tunes describe the jet shape distributions well.
Some differences are also observed with respect to the colour flow data, which
is particularly sensitive to the early resonance decay option in the CR models.
The differences between the predictions using the different tunes observed here
indicate that the inclusion of observables, such as the jet pull angle and other
jet substructure observables, could be beneficial in tuning studies.
A study of the uncertainty in the top quark mass measurement due to CR
effects is also presented.
The new CR tunes will play a role in the evaluation of systematic uncertainties associated with the modelling of colour reconnection. 

\begin{acknowledgments}
We congratulate our colleagues in the CERN accelerator departments for the excellent performance of the LHC and thank the technical and administrative staffs at CERN and at other CMS institutes for their contributions to the success of the CMS effort. In addition, we gratefully acknowledge the computing centres and personnel of the Worldwide LHC Computing Grid and other centres for delivering so effectively the computing infrastructure essential to our analyses. Finally, we acknowledge the enduring support for the construction and operation of the LHC, the CMS detector, and the supporting computing infrastructure provided by the following funding agencies: BMBWF and FWF (Austria); FNRS and FWO (Belgium); CNPq, CAPES, FAPERJ, FAPERGS, and FAPESP (Brazil); MES and BNSF (Bulgaria); CERN; CAS, MoST, and NSFC (China); MINCIENCIAS (Colombia); MSES and CSF (Croatia); RIF (Cyprus); SENESCYT (Ecuador); MoER, ERC PUT and ERDF (Estonia); Academy of Finland, MEC, and HIP (Finland); CEA and CNRS/IN2P3 (France); BMBF, DFG, and HGF (Germany); GSRI (Greece); NKFIH (Hungary); DAE and DST (India); IPM (Iran); SFI (Ireland); INFN (Italy); MSIP and NRF (Republic of Korea); MES (Latvia); LAS (Lithuania); MOE and UM (Malaysia); BUAP, CINVESTAV, CONACYT, LNS, SEP, and UASLP-FAI (Mexico); MOS (Montenegro); MBIE (New Zealand); PAEC (Pakistan); MES and NSC (Poland); FCT (Portugal); MESTD (Serbia); MCIN/AEI and PCTI (Spain); MOSTR (Sri Lanka); Swiss Funding Agencies (Switzerland); MST (Taipei); MHESI and NSTDA (Thailand); TUBITAK and TENMAK (Turkey); NASU (Ukraine); STFC (United Kingdom); DOE and NSF (USA).

\hyphenation{Rachada-pisek} Individuals have received support from the Marie-Curie programme and the European Research Council and Horizon 2020 Grant, contract Nos.\ 675440, 724704, 752730, 758316, 765710, 824093, 884104, and COST Action CA16108 (European Union); the Leventis Foundation; the Alfred P.\ Sloan Foundation; the Alexander von Humboldt Foundation; the Belgian Federal Science Policy Office; the Fonds pour la Formation \`a la Recherche dans l'Industrie et dans l'Agriculture (FRIA-Belgium); the Agentschap voor Innovatie door Wetenschap en Technologie (IWT-Belgium); the F.R.S.-FNRS and FWO (Belgium) under the ``Excellence of Science -- EOS" -- be.h project n.\ 30820817; the Beijing Municipal Science \& Technology Commission, No. Z191100007219010; the Ministry of Education, Youth and Sports (MEYS) of the Czech Republic; the Hellenic Foundation for Research and Innovation (HFRI), Project Number 2288 (Greece); the Deutsche Forschungsgemeinschaft (DFG), under Germany's Excellence Strategy -- EXC 2121 ``Quantum Universe" -- 390833306, and under project number 400140256 - GRK2497; the Hungarian Academy of Sciences, the New National Excellence Program - \'UNKP, the NKFIH research grants K 124845, K 124850, K 128713, K 128786, K 129058, K 131991, K 133046, K 138136, K 143460, K 143477, 2020-2.2.1-ED-2021-00181, and TKP2021-NKTA-64 (Hungary); the Council of Science and Industrial Research, India; the Latvian Council of Science; the Ministry of Education and Science, project no. 2022/WK/14, and the National Science Center, contracts Opus 2021/41/B/ST2/01369 and 2021/43/B/ST2/01552 (Poland); the Funda\c{c}\~ao para a Ci\^encia e a Tecnologia, grant CEECIND/01334/2018 (Portugal); the National Priorities Research Program by Qatar National Research Fund; MCIN/AEI/10.13039/501100011033, ERDF ``a way of making Europe", and the Programa Estatal de Fomento de la Investigaci{\'o}n Cient{\'i}fica y T{\'e}cnica de Excelencia Mar\'{\i}a de Maeztu, grant MDM-2017-0765 and Programa Severo Ochoa del Principado de Asturias (Spain); the Chulalongkorn Academic into Its 2nd Century Project Advancement Project, and the National Science, Research and Innovation Fund via the Program Management Unit for Human Resources \& Institutional Development, Research and Innovation, grant B05F650021 (Thailand); the Kavli Foundation; the Nvidia Corporation; the SuperMicro Corporation; the Welch Foundation, contract C-1845; and the Weston Havens Foundation (USA).
\end{acknowledgments}

\bibliography{auto_generated}

\numberwithin{table}{section}
\numberwithin{figure}{section}
\appendix
\section{Colour reconnection tunes with a leading-order PDF set}
\label{sec:lopdf}

\begin{figure*}[!hbtp]
\centering
         \includegraphics[width=0.49\textwidth]{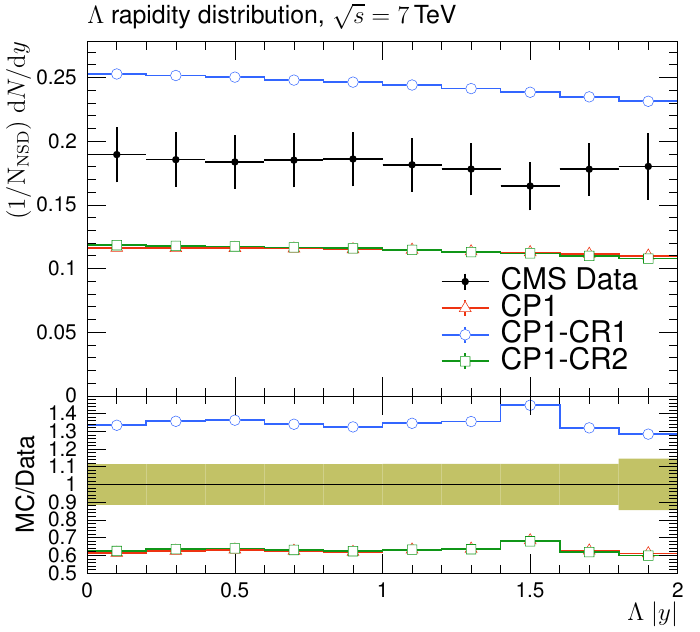}
         \includegraphics[width=0.49\textwidth]{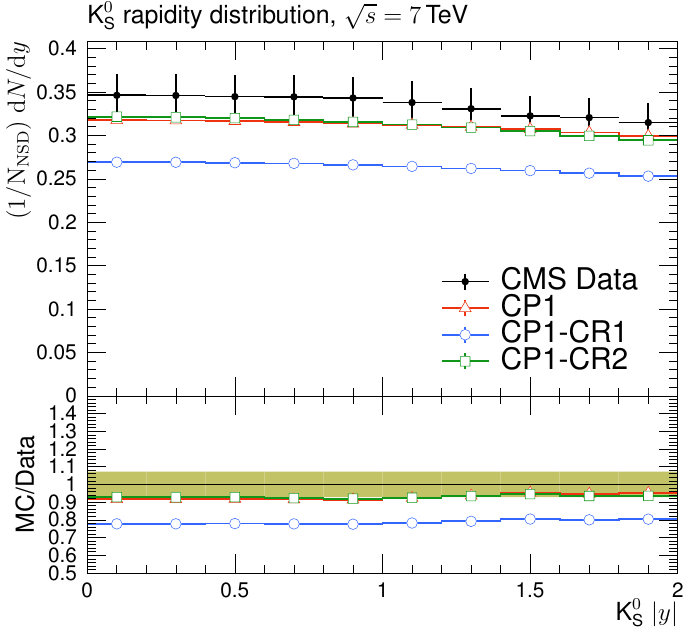}\\
         \includegraphics[width=0.49\textwidth]{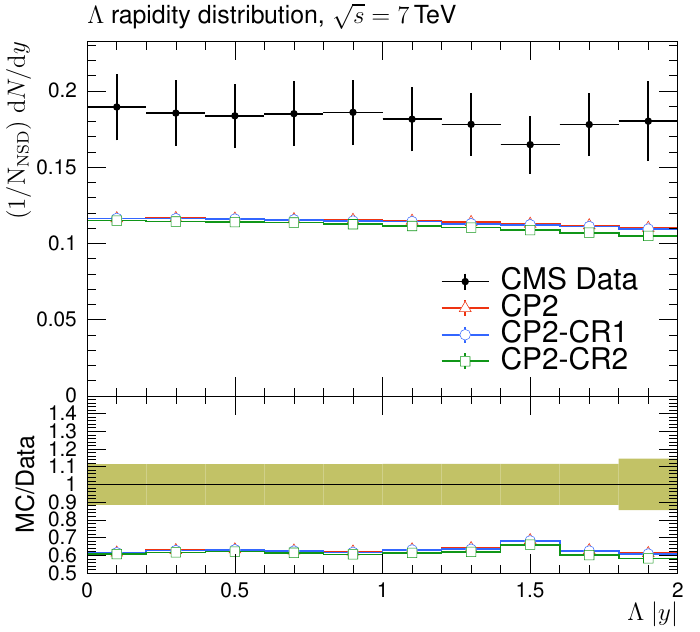}
         \includegraphics[width=0.49\textwidth]{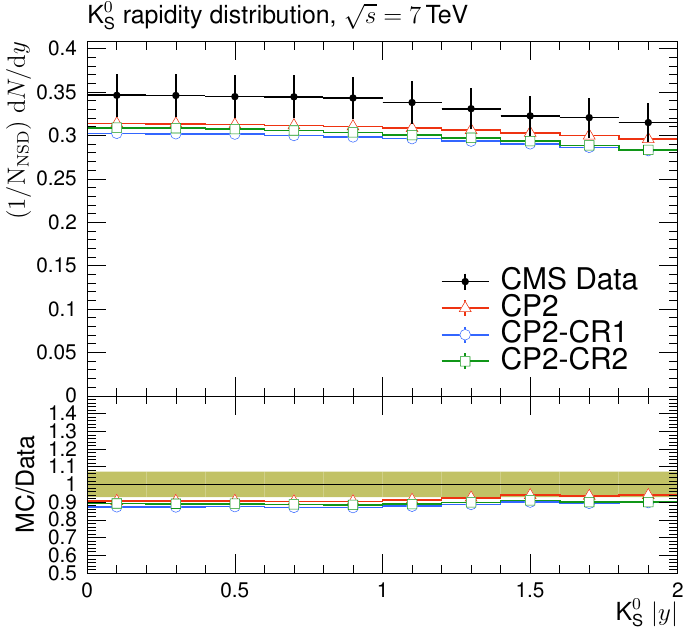}\\
    \caption{
			 The strange particle production, $\Lambda$ baryons (left) and \PKzS mesons (right), as a function of rapidity, 
			 measured by the CMS experiment at $\sqrt{s}=7\TeV$~\cite{Khachatryan:2011tm}.
             The predictions of the CP1 and CP1-CR tunes (upper) and CP2 and CP2-CR tunes (lower) are compared with data.
             The coloured band and error bars on the data points represent the total experimental uncertainty in the data.
            }
    \label{fig:strangeness_CP1}
\end{figure*}

\begin{figure*}[!hbtp]
\centering
    \includegraphics[width=0.49\textwidth]{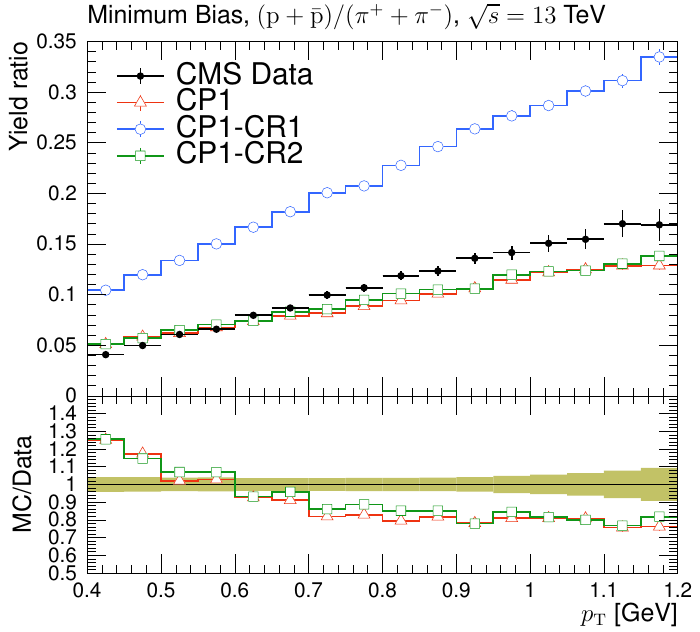}
    \includegraphics[width=0.49\textwidth]{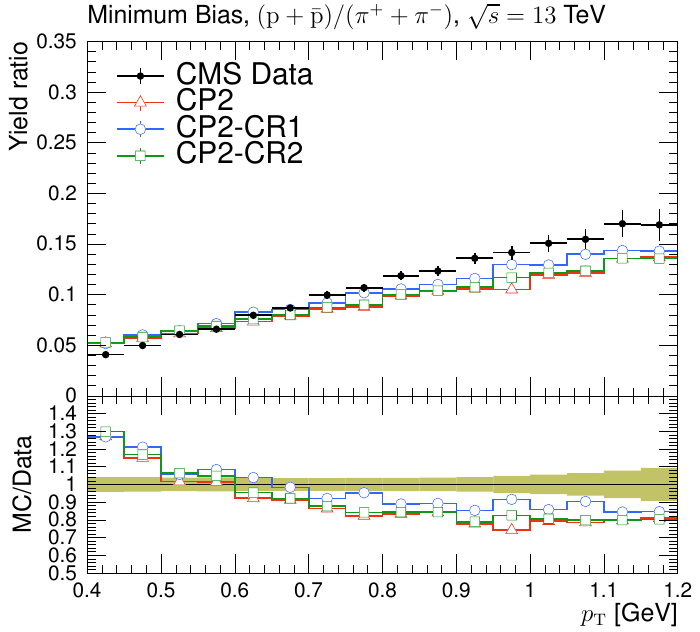}
    \caption{
             Ratios of particle yields, $p/\pi$, as a function of transverse momentum in MB
             events, measured by the CMS experiment at $\sqrt{s}=13\TeV$~\cite{Sirunyan:2017zmn}.
             The predictions of the CP1 and CP1-CR tunes (upper) and CP2 and CP2-CR tunes (lower) are compared with data.
             The coloured band and error bars on the data points represent the total experimental uncertainty in the data.
            }
    \label{fig:CMS_2017_I1608166_CP1}
\end{figure*}

The list of input \RIVET routines used as inputs for the fits, as well as the centre-of-mass energy values, the $\eta$ ranges, the names of the distributions, the $x$-axis ranges, and the $R$ values of the distributions are displayed in Table~\ref{tab:input_CP1} for the tunes CP1-CR1 and CP1-CR2, and in Table~\ref{tab:input_CP2} for the tunes CP2-CR1 and CP2-CR2. 
The baseline tunes CP1 and CP2 use the NNPDF31\_lo\_as\_0130~\cite{Ball:2017nwa} PDF set, with an $\alpS(m_\PZ)$ value of 0.130 for ISR, FSR, and MPI, and the MPI-based CR model.  
The parameters of the tunes are documented in Ref.~\cite{Sirunyan:2019dfx} and displayed in Tables~\ref{tab:CP1_params} and \ref{tab:CP2_params}. 
The parameters obtained from the CP1-CR1, CP1-CR2, CP2-CR1, and CP2-CR2 fits, as well as the value of the goodness of the fits are displayed in Tables~\ref{tab:CP1_params} and \ref{tab:CP2_params}. The predictions of these new CR tunes for particle multiplicities are shown in Figs.~\ref{fig:strangeness_CP1}, and \ref{fig:CMS_2017_I1608166_CP1}. The CR tunes based on CP1 and CP2 describe MB and UE data as well as the CR tunes based on CP5, except for the different trend observed with CP1-CR1 in the particle multiplicity distributions. 

\begin{table*}[!hbtp]
    \centering
   \topcaption{List of input \RIVET routines, centre-of-mass energy values, $\eta$ ranges, names of distributions, fit ranges, and relative importance of the distributions
   used in the fits to derive the tunes CP1-CR1 and CP1-CR2.}
    \label{tab:input_CP1}
    \cmsTable{
    \begin{tabular}{@{}llllllll}
                                      &            &          &                                  &  CP1-CR1  &     & CP1-CR2   &        \\
         \hline
         \RIVET routine       & $\sqrt{s}$ & $\abs{\eta}$ & Distribution                     & Fit range & $R$ & Fit range &  $R$      \\
                                      & [\TeVns{}]      &          &                                  & [\GeVns{}]    &     & [\GeVns{}]    &                \\
         \hline
         CMS\_2015\_I1384119          &  13        & ${<}2.0$   & \Nch versus $\eta$      &           &   1 &           &   1    \\
         CMS\_2015\_PAS\_FSQ\_15\_007 &  13        & ${<}2.0$   & TransMIN \ptsum & 3--36     &   1 & 4--36     & 0.20 \\
                                      &            &          & TransMAX \ptsum & 3--36     &   1 & 4--36     & 0.20 \\
                                      &            &          & TransMIN \Nch           & 3--36     &   1 & 4--36     & 0.20 \\
                                      &            &          & TransMAX \Nch           & 3--36     &   1 & 4--36     & 0.20 \\
         CMS\_2012\_PAS\_FSQ\_12\_020 &  7         & ${<}0.8$   & TransMAX \Nch           & 3--20     &   1 & 3--20     & 0.10 \\
                                      &            &          & TransMIN \Nch           & 3--20     &   1 & 3--20     & 0.10 \\
                                      &            &          & TransMAX \ptsum & 3--20     &   1 & 3--20     & 0.10 \\
                                      &            &          & TransMIN \ptsum & 3--20     &   1 & 3--20     & 0.10 \\
         CDF\_2015\_I1388868          &  2         & ${<}0.8$   & TransMIN \Nch           & 2--15     &   1 & 2--15     & 0.10 \\
                                      &            &          & TransMAX \Nch           & 2--15     &   1 & 2--15     & 0.10 \\
                                      &            &          & TransMIN \ptsum & 2--15     &   1 & 2--15     & 0.10 \\
                                      &            &          & TransMAX \ptsum & 2--15     &   1 & 2--15     & 0.10 \\
    \end{tabular}
    }
\end{table*}

\begin{table*}[!hbtp]
    \centering
   \topcaption{List of input \RIVET routines, centre-of-mass energy values, $\eta$ ranges, names of distributions, fit ranges, and relative importance of the distributions
   used in the fits to derive the tunes CP2-CR1 and CP2-CR2.}
    \label{tab:input_CP2}
    \cmsTable{
    \begin{tabular}{@{}llllllll}
                                      &            &           &                                  & CP2-CR1    &      & CP2-CR2     &    \\
         \hline                                    
         \RIVET routine       & $\sqrt{s}$ & $\abs{\eta}$  & Distribution                     & Fit range  & $R$  & Fit range   & $R$ \\
                                      & [\TeVns{}]      &           &                                  & [\GeVns{}]     &      & [\GeVns{}]      &     \\
         \hline                                    
         CMS\_2015\_I1384119          &  13        & ${<}2.0$    & \Nch versus $\eta$      &            & 0.03 &             & 0.05  \\
         CMS\_2015\_PAS\_FSQ\_15\_007 &  13        & ${<}2.0$    & TransMIN \ptsum & 5--24      & 1    & 5--24       & 1     \\
                                      &            &           & TransMAX \ptsum & 5--24      & 0.17 & 5--24       & 0.25  \\
                                      &            &           & TransMIN \Nch           & 5--24      & 1    & 5--24       & 1     \\
                                      &            &           & TransMAX \Nch           & 5--24      & 0.17 & 5--24       & 0.25  \\
         CMS\_2012\_PAS\_FSQ\_12\_020 &  7         & ${<}0.8$    & TransMAX \Nch           & 5--20      & 0.07 & 5--20       & 0.25  \\
                                      &            &           & TransMIN \Nch           & 5--20      & 1    & 5--20       & 1     \\
                                      &            &           & TransMAX \ptsum & 5--20      & 0.07 & 5--20       & 0.25  \\
                                      &            &           & TransMIN \ptsum & 5--20      & 1    & 5--20       & 1     \\
         CDF\_2015\_I1388868          &  2         & ${<}0.8$    & TransMIN \Nch           & 2--15      & 0.03 & 2--15       & 0.05  \\
                                      &            &           & TransMAX \Nch           & 2--15      & 0.03 & 2--15       & 0.05  \\
                                      &            &           & TransMIN \ptsum & 2--15      & 0.03 & 2--15       & 0.05  \\
                                      &            &           & TransMAX \ptsum & 2--15      & 0.03 & 2--15       & 0.05  \\
    \end{tabular}
    }
\end{table*}

\begin{table*}[!hbtp]
\centering
\topcaption{The parameters obtained in the fits of the CP1-CR1 and CP1-CR2 tunes,
compared with the ones of the tune CP1. The upper part of the table displays the fixed
input parameters of the tune, while the lower part shows the fitted tune parameters.
The number of degrees of freedom (\ndof) and the goodness of fit divided by the number of
degrees of freedom are also shown.}
\cmsTable{
\begin{tabular}{lccc}
\PYTHIAviii parameter                                 &  CP1~\cite{Sirunyan:2019dfx} & CP1-CR1 & CP1-CR2         \\
\hline & \\[-2.0ex]
PDF set & NNPDF3.1 LO & NNPDF3.1 LO & NNPDF3.1 LO                                  \\
$\alpS(m_\PZ)$                 & 0.130          &  0.130 &  0.130 \\
\texttt{SpaceShower:rapidityOrder}       &   off & off & off \\
\texttt{MultipartonInteractions:ecmRef} [\GeVns{}]      &    7000        & 7000 & 7000\\
$\alpS^\mathrm{ISR}(m_\PZ)$ value/order     &   0.1365/LO         & 0.1365/LO  & 0.1365/LO   \\
$\alpS^\mathrm{FSR}(m_\PZ)$ value/order     &   0.1365/LO            & 0.1365/LO  & 0.1365/LO   \\
$\alpS^\mathrm{MPI}(m_\PZ)$ value/order     &  0.130/LO             &  0.130/LO & 0.130/LO   \\
$\alpS^\mathrm{ME}(m_\PZ)$ value/order     &   0.130/LO            & 0.130/LO & 0.130/LO   \\
\texttt{StringZ:aLund}                              &\NA  & 0.38     &\NA \\
\texttt{StringZ:bLund}                              &\NA  & 0.64    &\NA\\
\texttt{StringFlav:probQQtoQ}                       &\NA  & 0.078    &\NA\\
\texttt{StringFlav:probStoUD}                       &\NA  & 0.2      &\NA\\
\texttt{SigmaTotal:zeroAXB}               &  off & off & off \\
\texttt{BeamRemnants:remnantMode}         &\NA & 1 &\NA\\
\texttt{MultipartonInteractions:bProfile} & 2  & 2 & 2 \\
\texttt{ColourReconnection:mode}                     &\NA & 1       & 2    \\  [\cmsTabSkip]
\texttt{MultipartonInteractions:pT0Ref} [\GeVns{}]   & 2.400 & 1.984  & 2.385 \\
\texttt{MultipartonInteractions:ecmPow}              & 0.154 & 0.113  & 0.165 \\
\texttt{MultipartonInteractions:coreRadius}          & 0.544 & 0.746  & 0.587 \\
\texttt{MultipartonInteractions:coreFraction}        & 0.684 & 0.569  & 0.533 \\
\texttt{ColourReconnection:range}                    & 2.633 &\NA    &\NA \\
\texttt{ColourReconnection:junctionCorrection}       &\NA   & 8.382  &\NA \\
\texttt{ColourReconnection:timeDilationPar}          &\NA   & 31.070 &\NA\\
\texttt{ColourReconnection:m0}                       &\NA   & 1.845  &\NA \\
\texttt{ColourReconnection:m2lambda}                 &\NA   &\NA    & 2.769 \\
\texttt{ColourReconnection:fracGluon}                &\NA   &\NA    & 0.979 \\
 \ndof                                      & 183   & 157    & 150 \\
 $\chi^{*2} / \ndof$   						     & 0.89 & 0.73  & 0.20
\end{tabular}
}
\label{tab:CP1_params}
\end{table*}

 \begin{table*}[!hbtp]
 \centering
 \topcaption{The parameters obtained in the fits of the CP2-CR1 and CP2-CR2 tunes,
 compared with the ones of the tune CP2. The upper part of the table displays the fixed
 input parameters of the tune, while the lower part shows the fitted tune parameters.
 The number of degrees of freedom (\ndof) and the goodness of fit divided by the number of
 degrees of freedom are also shown.}
 \cmsTable{
 \begin{tabular}{lccc}
 \PYTHIAviii parameter                       &  CP2~\cite{Sirunyan:2019dfx} & CP2-CR1 & CP2-CR2         \\
\hline & \\[-2.0ex]
 PDF set                                              & NNPDF3.1 LO  & NNPDF3.1 LO & NNPDF3.1 LO   \\
 $\alpS(m_\PZ)$                                      & 0.130        & 0.130 &  0.130 \\
 \texttt{SpaceShower:rapidityOrder}                   & off          & off & off \\
 \texttt{MultipartonInteractions:ecmRef} [\GeVns{}]   & 7000         & 7000 & 7000\\
 $\alpS^\mathrm{ISR}(m_\PZ)$ value/order             & 0.130/LO     & 0.130/LO  & 0.130/LO   \\
 $\alpS^\mathrm{FSR}(m_\PZ)$ value/order             & 0.130/LO     & 0.130/LO  & 0.130/LO   \\
 $\alpS^\mathrm{MPI}(m_\PZ)$ value/order             & 0.130/LO     & 0.130/LO  & 0.130/LO   \\
 $\alpS^\mathrm{ME}(m_\PZ)$ value/order              & 0.130/LO     & 0.130/LO  & 0.130/LO   \\
 \texttt{StringZ:aLund}                               &\NA  & 0.38  &\NA \\
 \texttt{StringZ:bLund}                               &\NA  & 0.64  &\NA\\
 \texttt{StringFlav:probQQtoQ}                        &\NA  & 0.078 &\NA\\
 \texttt{StringFlav:probStoUD}                        &\NA  & 0.2   &\NA\\
 \texttt{SigmaTotal:zeroAXB}                          & off  & off   & off \\
 \texttt{BeamRemnants:remnantMode}                    &\NA  & 1     &\NA\\
 \texttt{MultipartonInteractions:bProfile}            & 2    & 2     & 2 \\
 \texttt{ColourReconnection:mode}                     &\NA  & 1     & 2      \\ [\cmsTabSkip]
 \texttt{MultipartonInteractions:pT0Ref} [\GeVns{}]   & 2.306 & 2.154   & 2.287 \\
 \texttt{MultipartonInteractions:ecmPow}              & 0.139 & 0.119   & 0.146 \\
 \texttt{MultipartonInteractions:coreRadius}          & 0.376 & 0.538   & 0.514 \\
 \texttt{MultipartonInteractions:coreFraction}        & 0.327 & 0.599   & 0.525 \\
 \texttt{ColourReconnection:range}                    & 2.323 &\NA     &\NA \\
 \texttt{ColourReconnection:junctionCorrection}       &\NA   & 0.761   &\NA \\
 \texttt{ColourReconnection:timeDilationPar}          &\NA   & 13.080  &\NA\\
 \texttt{ColourReconnection:m0}                       &\NA   & 1.546   &\NA \\
 \texttt{ColourReconnection:m2lambda}                 &\NA   &\NA     & 6.186 \\
 \texttt{ColourReconnection:fracGluon}                &\NA   &\NA     & 0.978 \\
  \ndof                                      & 183   & 117     & 118 \\
  $\chi^{*2} / \ndof$                          & 0.54  & 0.21    & 0.22
 \end{tabular}
 }
 \label{tab:CP2_params}
 \end{table*}

\section{Parameter ranges and uncertainties in the tunes}
\label{sec:Uncertainties}

\begin{figure*}[!hbtp]
\centering
         \includegraphics[width=0.49\textwidth]{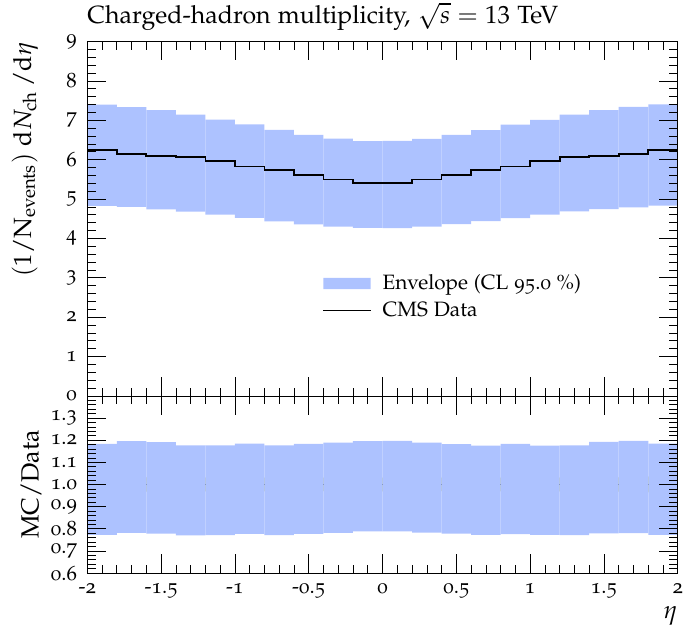}
         \includegraphics[width=0.49\textwidth]{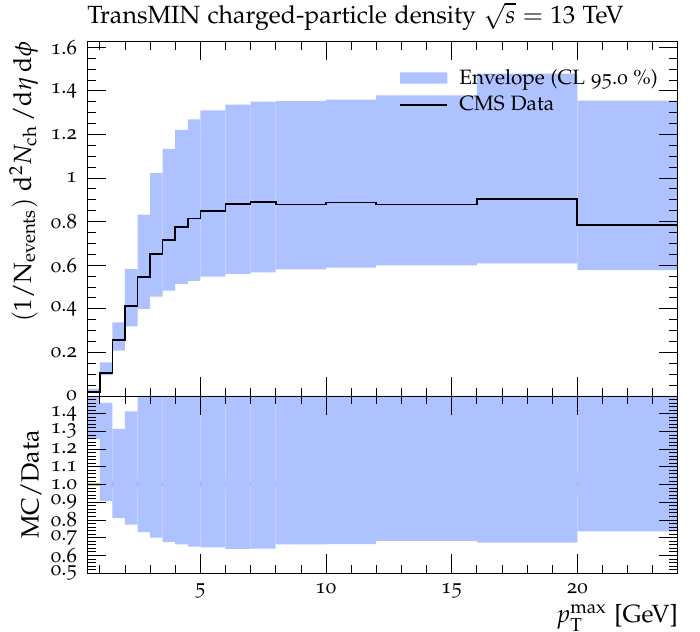}\\
         \includegraphics[width=0.49\textwidth]{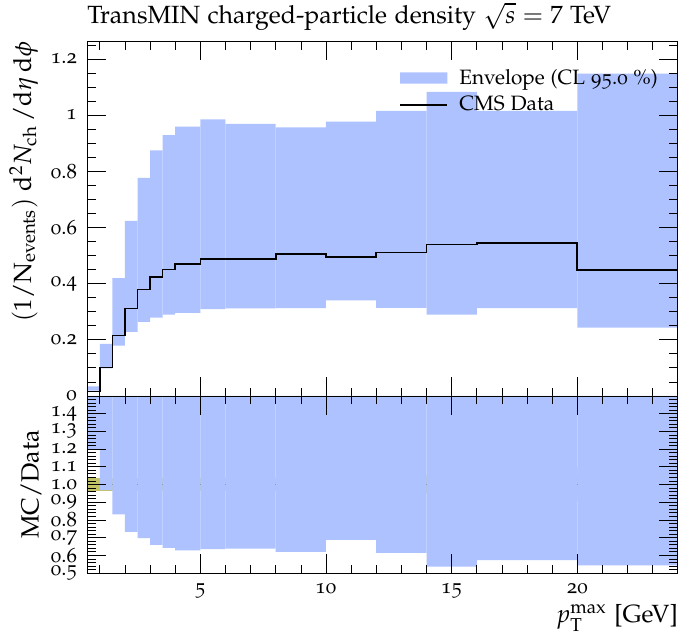}
         \includegraphics[width=0.49\textwidth]{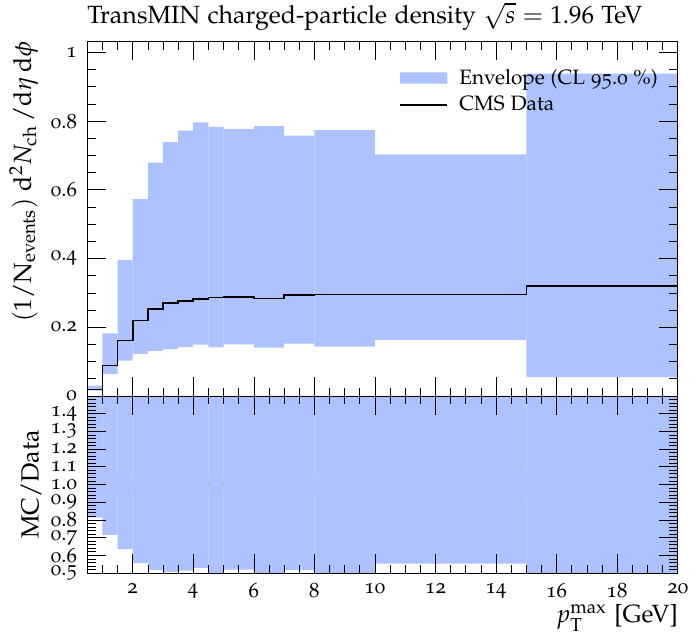}\\
    \caption{
			Sample histograms showing the range of variation available on the observable histograms are given for CP5-CR1.
            }
    \label{fig:envelopesCR1}
\end{figure*}

The parameter ranges are chosen such that the sampled MC space does not destroy the definition of a particular observable in the fits. 
In Fig.~\ref{fig:envelopesCR1}, some sample histograms showing the range of variation available on the observable histograms are given for CP5-CR1. The results are similar for other observables used in the fits as well as for CP5-CR2. 

The CP5-CR1 and CP5-CR2 tunes were developed to evaluate the uncertainty in CP5 that results from different color reconnection models.
The uncertainties in the parameters for these tunes were estimated using eigentunes provided by \PROFESSOR. Eigentunes represent variations
of the tuned parameters in the parameter space along the maximally independent directions. The magnitude of the variation corresponds to
a change in the $\chi^{*2}$ ($\Delta\chi^{*2}$) equal to the $\chi^{*2}$ of the fit. The choice of $\Delta\chi^{*2}$, which is recommended by
the \PROFESSOR Collaboration, is based on empirical grounds since modifying it in equation Eq.(\ref{chi2}) does not yield a statistically meaningful variation.
Such a change, $\Delta\chi^{*2} = \chi^{*2}$, is considered reasonable for reflecting the combined statistical and systematic uncertainty in the model parameters,
and results in variations similar in magnitude to the uncertainties in the fitted data points. However, this approach may result in uncertainties that do not fully
encompass the data in every bin. If the uncertainties in the fitted data points are uncorrelated, their magnitudes will depend on the bin widths.
For the data used in the fit, the uncertainties are mostly correlated between bins. However, for UE observables with high \ptmax ($\ptmax \gtrsim 10\GeV$) 
statistical uncertainties, which are uncorrelated between bins, dominate. This creates some dependence of the eigentunes on the bin widths of the data used in the fit,
and leads to uncertainties in the tunes that are much larger and more asymmetric than those on the data points when all eigentunes are added in quadrature.
 
The number of eigentunes is equal to twice the number of free parameters used in the fit. For the QCD-inspired and gluon-move models, 
there are 14 and 12 eigentunes, respectively. However, using all 12 or 14 eigentunes to calculate the tune uncertainty for a given observable 
is computationally inefficient. Therefore, the ``up'' and ``down'' tune settings are calculated by comparing the positive and negative 
differences between each eigentune and the central prediction of the nominal tune for each bin of the observable. The upper and lower bounds 
of the uncertainty in each bin are defined by adding the positive differences in quadrature and taking the square root, and similarly for the 
negative differences. These ``up'' and ``down'' variations are then fit using the same procedure as in Section~\ref{sec:Tuning} to obtain new 
parameter sets that can be used to estimate the uncertainties in the nominal tune. 

\begin{figure}[!hbtp]
  \centering
        \includegraphics[width=0.49\textwidth]{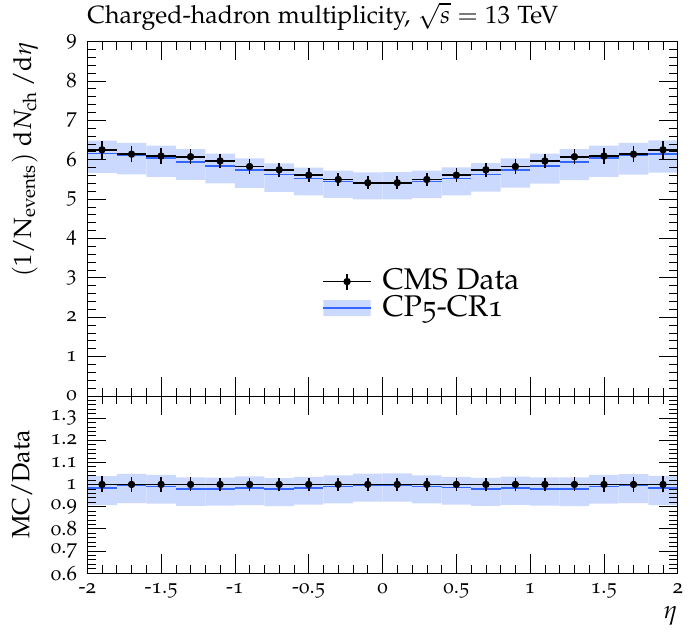}
   \caption{
            The pseudorapidity of charged hadrons, \dNdeta, measured by the CMS experiment at $\sqrt{s}=13\TeV$~\cite{CMS:2015zrm}.
            The prediction of the CP5-CR1 tune is compared with data. The coloured band represents the tune uncertainties.
			}
    \label{fig:CP5CR1uncerdN}
\end{figure}

\begin{figure*}[!hbtp]
  \centering
        \includegraphics[width=0.49\textwidth]{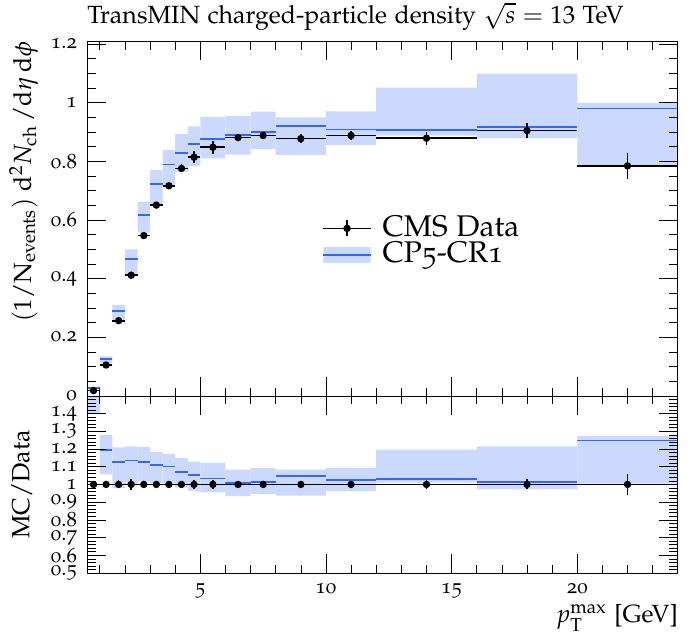}
        \includegraphics[width=0.49\textwidth]{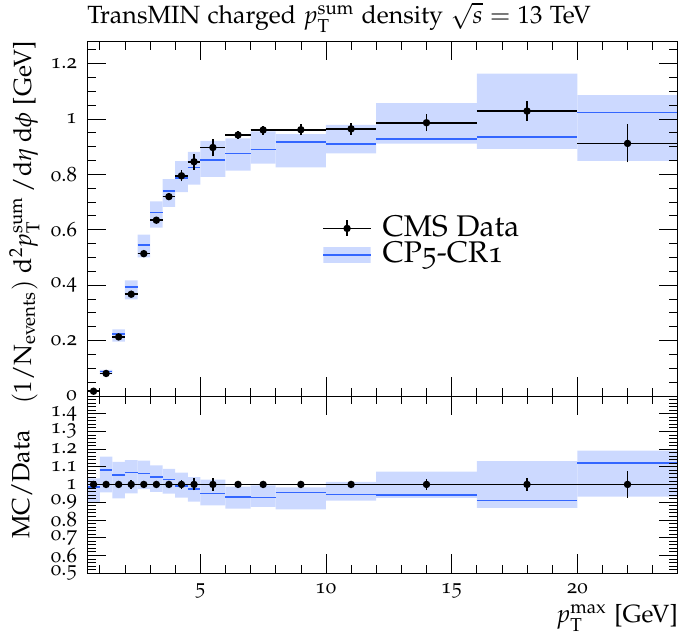}\\
        \includegraphics[width=0.49\textwidth]{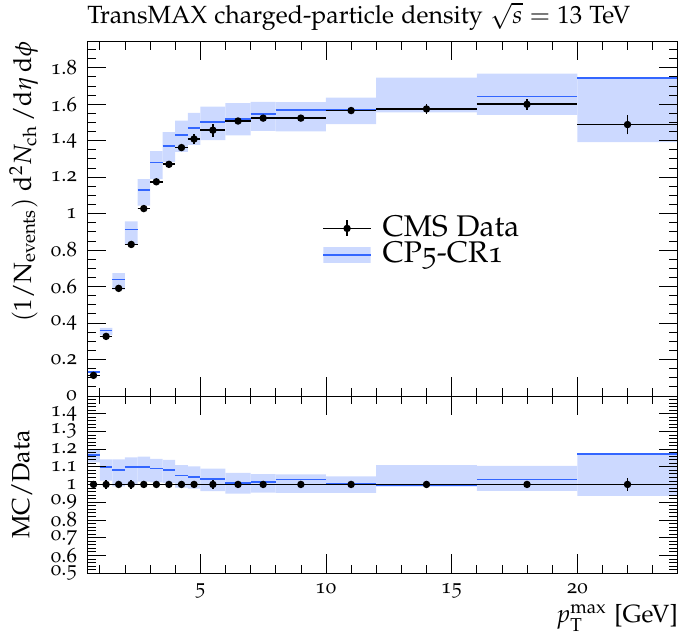}
        \includegraphics[width=0.49\textwidth]{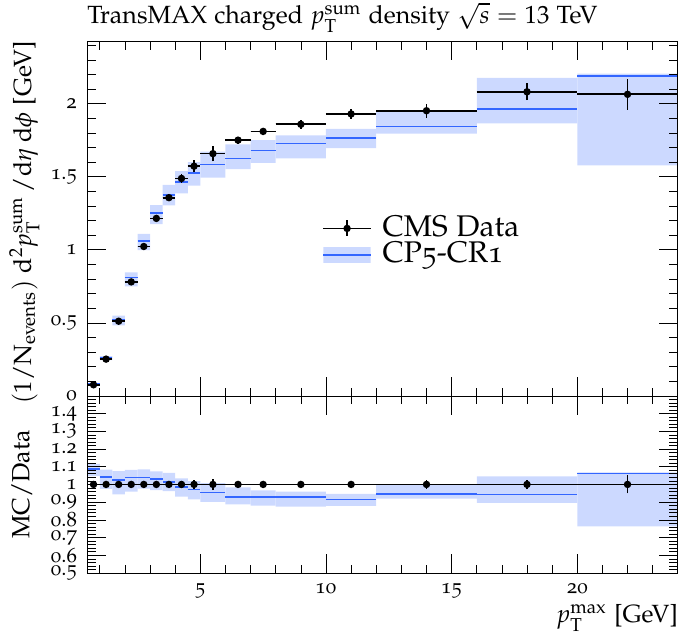}\\
  \caption{
		   The charged-particle (left) and \ptsum densities (right) in the \tmin (upper) and \tmax (lower) regions
		   as functions of the \pt of the leading charged particle, \ptmax, measured by the CMS experiment at $\sqrt{s}=13\TeV$~\cite{CMS:2015zev}.
           The predictions of the tunes CP5-CR1 are compared with data. The coloured band represents the tune uncertainties.
		  }
    \label{fig:CP5CR1uncertainty}
\end{figure*}

\begin{figure}[!hbtp]
  \centering
        \includegraphics[width=0.49\textwidth]{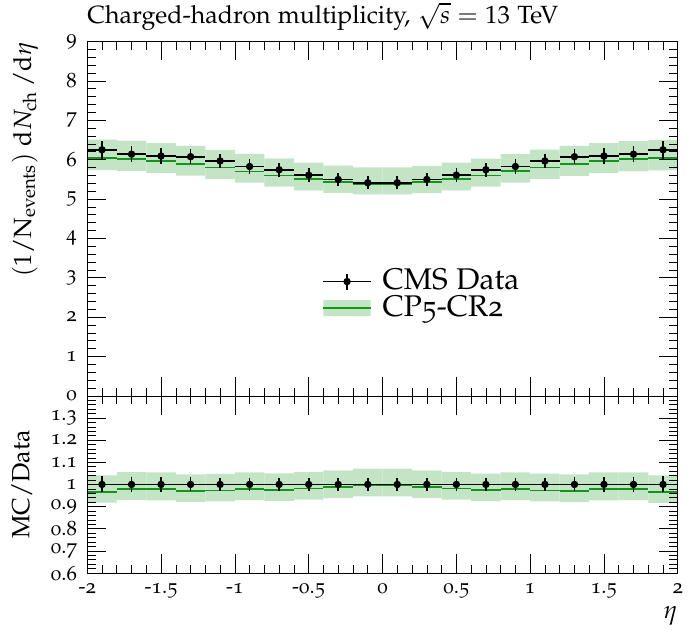}
   \caption{
            The pseudorapidity of charged hadrons, \dNdeta, measured by the CMS experiment at $\sqrt{s}=13\TeV$~\cite{CMS:2015zrm}.
            The prediction of the CP5-CR2 tune is compared with data. The coloured band represents the tune uncertainties.
			}
    \label{fig:CP5CR2uncerdN}
\end{figure}

\begin{figure*}[!hbtp]
  \centering
        \includegraphics[width=0.49\textwidth]{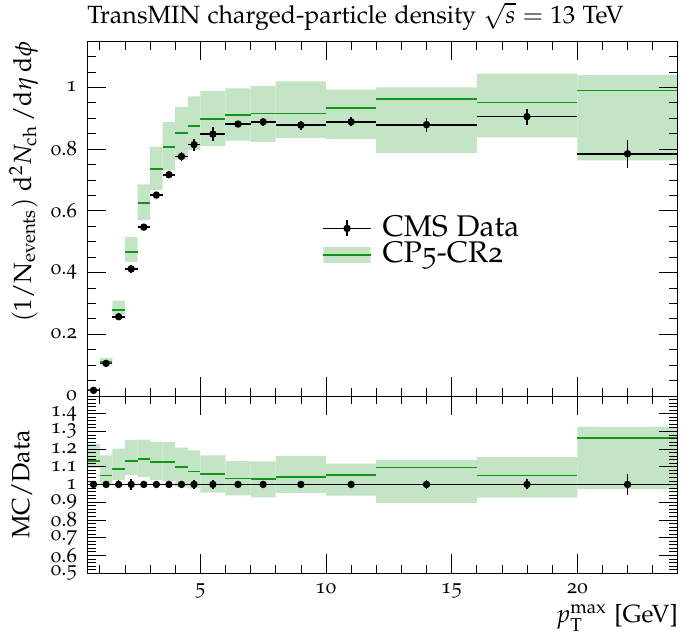}
        \includegraphics[width=0.49\textwidth]{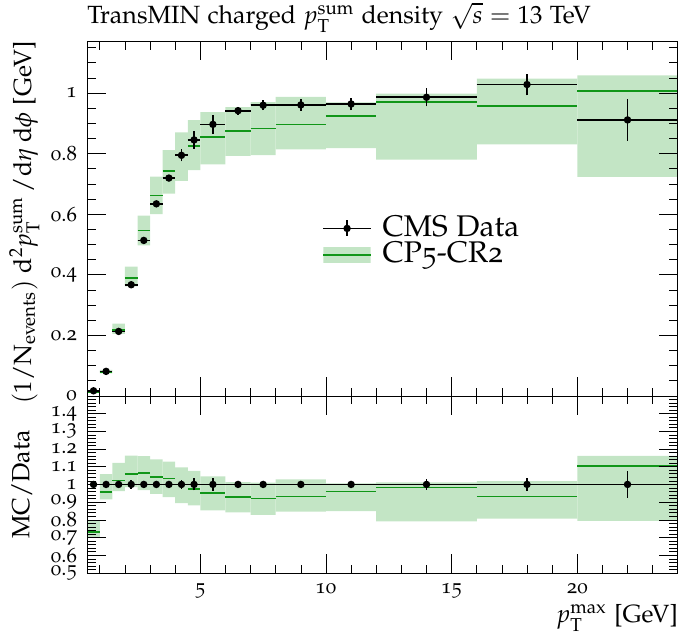}\\
        \includegraphics[width=0.49\textwidth]{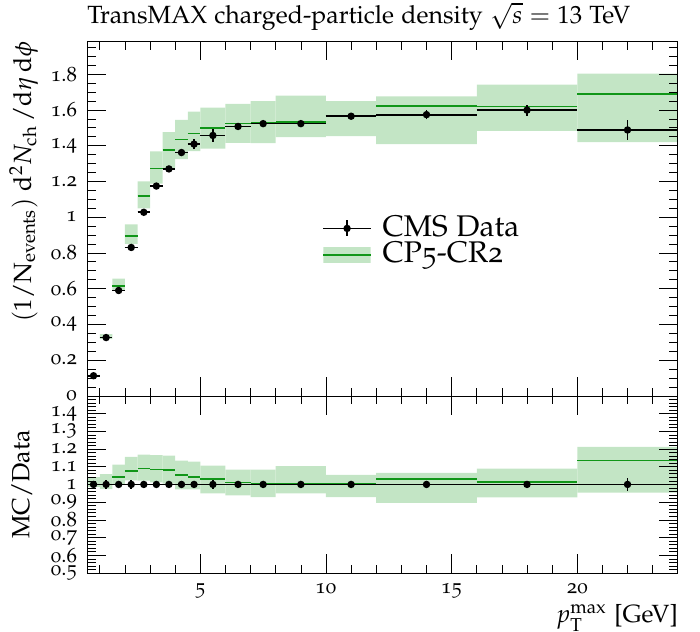}
        \includegraphics[width=0.49\textwidth]{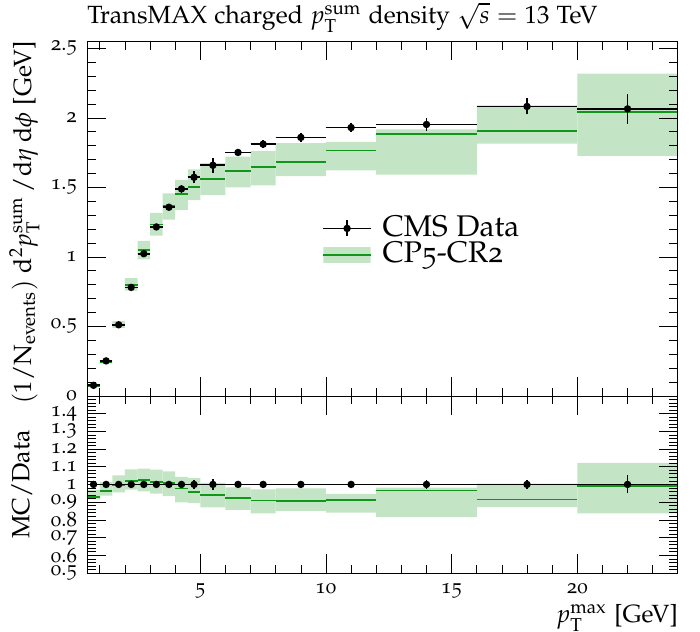}\\
  \caption{
		   The charged-particle (left) and \ptsum densities (right) in the \tmin (upper) and \tmax (lower) regions
		   as functions of the \pt of the leading charged particle, \ptmax, measured by the CMS experiment at $\sqrt{s}=13\TeV$~\cite{CMS:2015zev}.
           The predictions of the tunes CP5-CR2 are compared with data. The coloured band represents the tune uncertainties.
		  }
    \label{fig:CP5CR2uncertainty}
\end{figure*}

The predictions of the CP5-CR1 and CP5-CR2 tunes are compared with observables at 13\TeV in Figs.~\ref{fig:CP5CR1uncerdN}--\ref{fig:CP5CR2uncertainty}. 
The shaded bands in these figures correspond to the envelope of the predictions of the eigentunes of each tune. The parameters of the 
``up'' and ``down'' tunes for CP5-CR1 and CP5-CR2 are given in Tables~\ref{tab:CP5CR1Uncertainties} and~\ref{tab:CP5CR2Uncertainties}, respectively.

\begin{table}[h]
\topcaption{Parameters of the ``up'' and ``down'' variations of the CP5-CR1 tune.}
\centering
\begin{tabular}{l c c }
& \multicolumn{2}{c}{CP5-CR1}  \\
& Down & Up \\
\hline
\texttt{MultipartonInteractions:pT0Ref} [\GeVns{}]      & 1.568 & 1.328  \\
\texttt{MultipartonInteractions:ecmPow}                 & 0.011 & 0.045  \\
\texttt{MultipartonInteractions:coreRadius}             & 0.466 & 0.643  \\
\texttt{MultipartonInteractions:coreFraction}           & 0.516 & 0.579  \\
\texttt{ColourReconnection:junctionCorrection}          & 0.219 & 0.261  \\
\texttt{ColourReconnection:timeDilationPar}             & 9.328 & 11.95  \\
\texttt{ColourReconnection:m0}                          & 1.786 & 1.529  \\
\end{tabular}
\label{tab:CP5CR1Uncertainties}
\end{table}

\begin{table}[h]
\topcaption{Parameters of the ``up'' and ``down'' variations of the CP5-CR2 tune.}
\centering
\begin{tabular}{l c c }
& \multicolumn{2}{c}{CP5-CR2}  \\
& Down & Up \\
\hline
\texttt{MultipartonInteractions:pT0Ref} [\GeVns{}]      & 1.565 & 1.361  \\
\texttt{MultipartonInteractions:ecmPow}                 & 0.057 & 0.075  \\
\texttt{MultipartonInteractions:coreRadius}             & 0.576 & 0.760  \\
\texttt{MultipartonInteractions:coreFraction}           & 0.704 & 0.623  \\
\texttt{ColourReconnection:m2lambda}                    & 3.202 & 2.837  \\
\texttt{ColourReconnection:fracGluon}                   & 0.979 & 0.988  \\
\end{tabular}
\label{tab:CP5CR2Uncertainties}
\end{table}

\section{The \texttt{ColourReconnection:junctionCorrection} parameter}
\label{sec:junctionCorrection}

The QCD-inspired model implemented in \PYTHIAviii allows for the creation of string junctions when three color lines meet at a single point. The presence of these junctions can affect the number of particles produced in a collision and result in the production of additional gluons and quark-antiquark pairs. The \texttt{junctionCorrection} parameter in \PYTHIAviii controls the strength of this effect and can impact various observables, including the charged particle pseudorapidity distribution. 

\begin{figure}[!hbtp]
  \centering
        \includegraphics[width=0.49\textwidth]{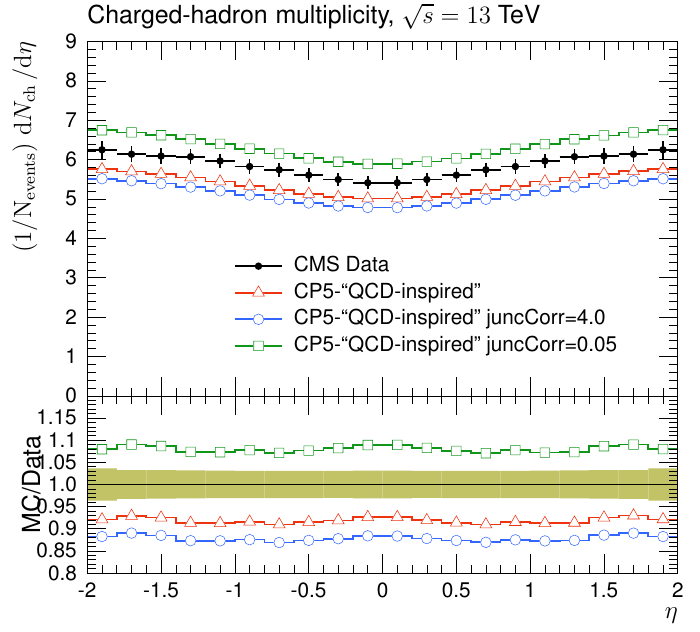}
   \caption{
            The pseudorapidity of charged hadrons, \dNdeta, measured in $\abs{\eta}<2$ by the CMS experiment at $\sqrt{s}=13\TeV$~\cite{CMS:2016gde}. 
            The red line shows the prediction of CP5-``QCD-inspired,'' where the default value of the \texttt{ColourReconnection:junctionCorrection} parameter is 1.2. 
            The predictions shown with the blue and green lines use 4.0 and 0.05 for the \texttt{ColourReconnection:junctionCorrection} parameter, respectively. 
            These comparisons demonstrate the sensitivity of the \texttt{ColourReconnection:junctionCorrection} parameter to this observable. 
           }
    \label{fig:juncCorrMB}
\end{figure}

\begin{figure*}[!hbtp]
  \centering
        \includegraphics[width=0.49\textwidth]{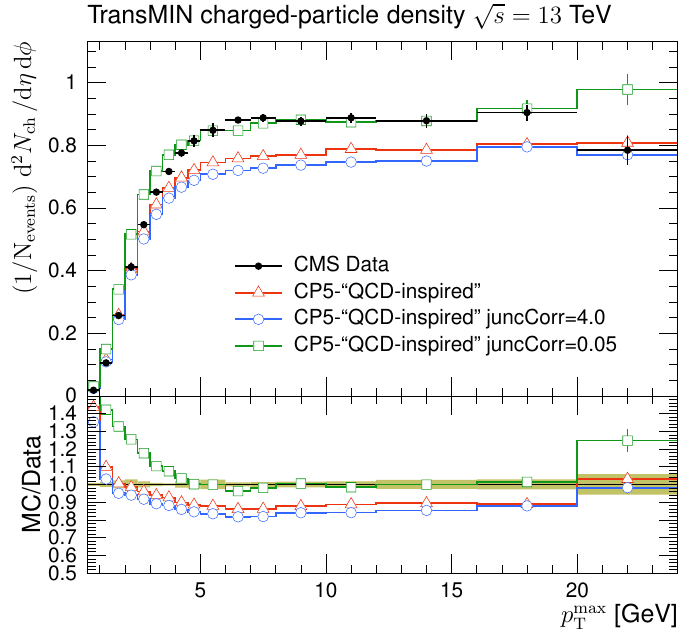}
        \includegraphics[width=0.49\textwidth]{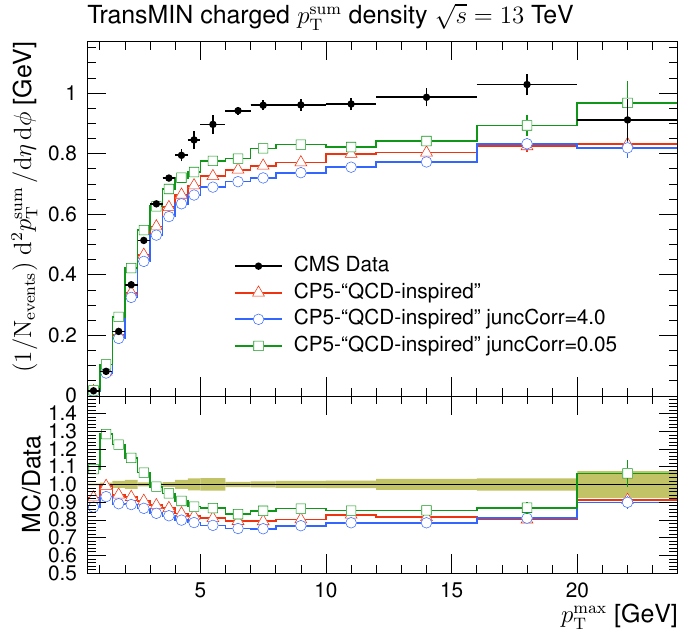}\\
        \includegraphics[width=0.49\textwidth]{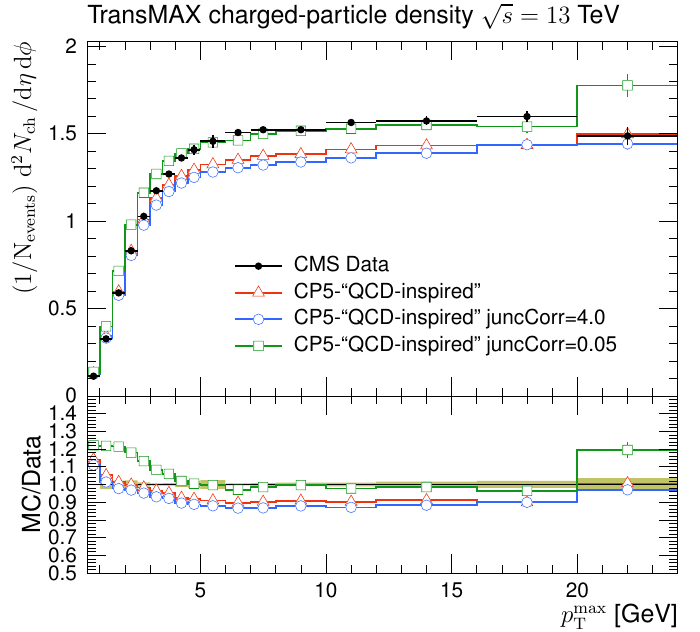}
        \includegraphics[width=0.49\textwidth]{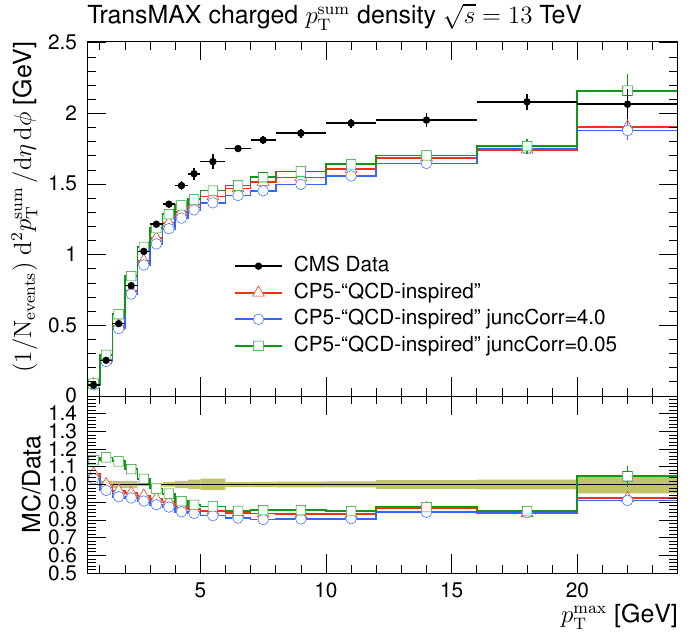}\\
  \caption{
		   The charged-particle (left) and \ptsum densities (right) in the \tmin (upper) and \tmax (lower) regions
		   as functions of the \pt of the leading charged particle, \ptmax, measured by the CMS experiment at $\sqrt{s}=13\TeV$~\cite{CMS:2015zev}.
           The red line shows the prediction of CP5-``QCD-inspired,'' where the default value of the \texttt{ColourReconnection:junctionCorrection} parameter is 1.2. 
           The predictions shown with the blue and green lines use 4.0 and 0.05 for the \texttt{ColourReconnection:junctionCorrection} parameter, respectively. 
           These comparisons demonstrate the sensitivity of the \texttt{ColourReconnection:junctionCorrection} parameter to these observables.
		  }
    \label{fig:juncCorrUE}
\end{figure*}

\begin{figure*}[!hbtp]
  \centering
        \includegraphics[width=0.49\textwidth]{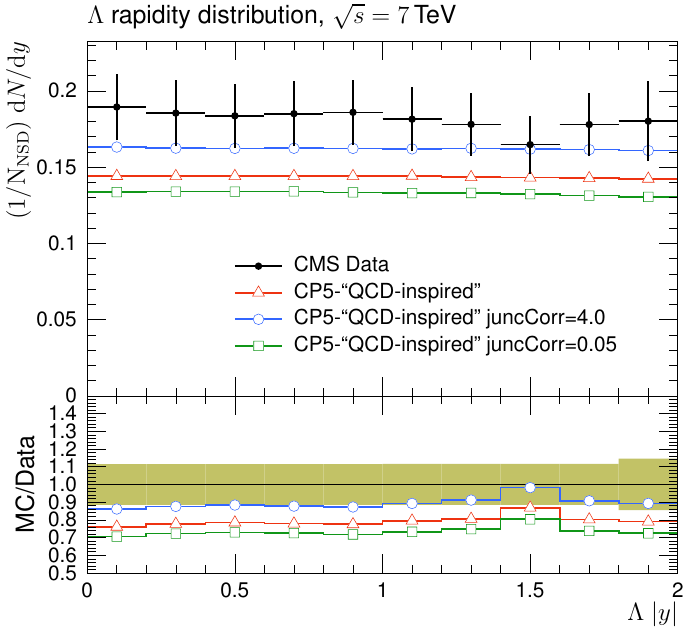}
        \includegraphics[width=0.49\textwidth]{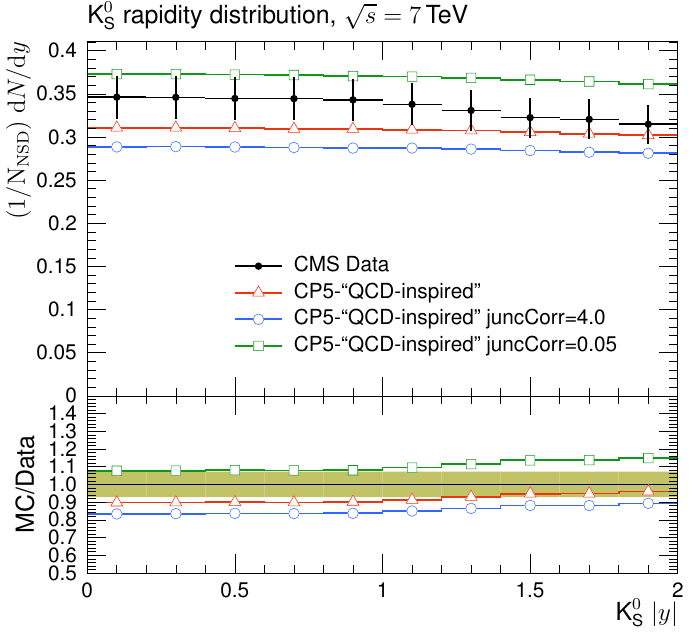}\\
  \caption{
           The strange particle production, $\Lambda$ baryons (upper) and \PKzS mesons (lower), as a function of rapidity,
           measured by the CMS experiment at $\sqrt{s}=7\TeV$~\cite{Khachatryan:2011tm}.
           The red line shows the prediction of CP5-``QCD-inspired,'' where the default value of the \texttt{ColourReconnection:junctionCorrection} parameter is 1.2. 
           The predictions shown with the blue and green lines use 4.0 and 0.05 for the \texttt{ColourReconnection:junctionCorrection} parameter, respectively. 
           These comparisons demonstrate the sensitivity of the \texttt{ColourReconnection:junctionCorrection} parameter to these observables.
		  }
    \label{fig:juncCorrStrange}
\end{figure*}

The model predictions, with their default parameter settings in \PYTHIA 8.226 and CP5,
are given in Fig.~\ref{fig:juncCorrMB} for the \dNdeta distribution measured by the CMS experiment at 13\TeV~\cite{CMS:2015zrm}, and 
in Fig.~\ref{fig:juncCorrUE} for \Nch and \ptsum densities measured by CMS at 13\TeV~\cite{CMS:2015zev} in the \tmin and \tmax regions.
The predictions for CP5-``QCD-inspired'' were obtained by replacing the MPI-based CR model in CP5 with the QCD-inspired model, where the default value 
of the \texttt{junctionCorrection} parameter is 1.2. The other predictions presented in the figures were obtained by setting the 
\texttt{junctionCorrection} parameter to 4.0 and to 0.05, respectively. These values were chosen arbitrarily to test how the prediction 
changes when a relatively high or low value is set for the \texttt{junctionCorrection} parameter. 
These comparisons demonstrate the sensitivity of the \texttt{junctionCorrection} parameter to these observables.
According to Ref.~\cite{Christiansen:2015yqa}, the \texttt{junctionCorrection} parameter is most sensitive to the baryon/meson ratio in pp collisions.
The sensitivity of the \texttt{junctionCorrection} parameter to the production of $\Lambda$ baryons and \PKzS mesons,
measured by the CMS experiment at $\sqrt{s}=7\TeV$~\cite{Khachatryan:2011tm}, is shown in Fig.~\ref{fig:juncCorrStrange}.

We also derived a new version of CP5-CR1 by including the rapidity distributions of $\Lambda$ baryons and \PKzS mesons, as well as some recent baryon and meson measurements from ALICE and LHCb experiments~\cite{ALICE:2021rzj,LHCb:2019fns}.
The new tune, named CP5-CR2-v2, resulted in a significant improvement in the description of the $\Lambda$ rapidity distribution and reasonable agreement with the data for $\Lambda$/\PKzS, but it was not able to reproduce the \dNdeta at 13 TeV. The values of the parameters obtained in the fits of the CP5-CR1-v2 are presented in Table~\ref{tab:CP5-CR1-v2}. The remaining parameters of \PYTHIAviii are kept the same as in the CP5 tune.

\begin{table}[!hbtp]
\centering
\topcaption{The values of the parameters obtained in the fits of the CP5-CR1-v2.}
\label{tab:CP5-CR1-v2}
\begin{tabular}{lr}
\multicolumn{2}{l}{\PYTHIAviii parameter}{CP5-CR1-v2}\\\hline
\texttt{MultipartonInteractions:pT0Ref}        & 1.260 \\
\texttt{MultipartonInteractions:ecmPow}        & 0.042 \\
\texttt{MultipartonInteractions:coreRadius}    & 0.750 \\
\texttt{MultipartonInteractions:coreFraction}  & 0.567 \\
\texttt{ColourReconnection:m0}                 & 1.888 \\
\texttt{ColourReconnection:junctionCorrection} & 1.427 \\
\texttt{ColourReconnection:timeDilationPar}    & 4.990 \\ 
\end{tabular}
\end{table}
\cleardoublepage \section{The CMS Collaboration \label{app:collab}}\begin{sloppypar}\hyphenpenalty=5000\widowpenalty=500\clubpenalty=5000
\cmsinstitute{Yerevan Physics Institute, Yerevan, Armenia}
{\tolerance=6000
A.~Tumasyan\cmsAuthorMark{1}\cmsorcid{0009-0000-0684-6742}
\par}
\cmsinstitute{Institut f\"{u}r Hochenergiephysik, Vienna, Austria}
{\tolerance=6000
W.~Adam\cmsorcid{0000-0001-9099-4341}, J.W.~Andrejkovic, T.~Bergauer\cmsorcid{0000-0002-5786-0293}, S.~Chatterjee\cmsorcid{0000-0003-2660-0349}, K.~Damanakis\cmsorcid{0000-0001-5389-2872}, M.~Dragicevic\cmsorcid{0000-0003-1967-6783}, A.~Escalante~Del~Valle\cmsorcid{0000-0002-9702-6359}, R.~Fr\"{u}hwirth\cmsAuthorMark{2}\cmsorcid{0000-0002-0054-3369}, M.~Jeitler\cmsAuthorMark{2}\cmsorcid{0000-0002-5141-9560}, N.~Krammer\cmsorcid{0000-0002-0548-0985}, L.~Lechner\cmsorcid{0000-0002-3065-1141}, D.~Liko\cmsorcid{0000-0002-3380-473X}, I.~Mikulec\cmsorcid{0000-0003-0385-2746}, P.~Paulitsch, F.M.~Pitters, J.~Schieck\cmsAuthorMark{2}\cmsorcid{0000-0002-1058-8093}, R.~Sch\"{o}fbeck\cmsorcid{0000-0002-2332-8784}, D.~Schwarz\cmsorcid{0000-0002-3821-7331}, S.~Templ\cmsorcid{0000-0003-3137-5692}, W.~Waltenberger\cmsorcid{0000-0002-6215-7228}, C.-E.~Wulz\cmsAuthorMark{2}\cmsorcid{0000-0001-9226-5812}
% ...
\par}
\vskip\cmsinstskip
\dag:~Deceased\\
$^{1}$Also at Yerevan State University, Yerevan, Armenia\\
$^{2}$Also at TU Wien, Vienna, Austria\\
% ...
\end{sloppypar}
\end{document}